\documentclass[acmlarge]{acmart}

\usepackage{soul,color}
\usepackage[ruled,linesnumbered]{algorithm2e}
\usepackage{amsfonts}
\usepackage{amsmath}
\usepackage{amsthm}
\usepackage{multirow,multicol}
\usepackage{pifont}
\usepackage{subfig}
\usepackage{enumitem}
\usepackage{listings}
\usepackage{graphicx}
\usepackage{xcolor}
\usepackage{ulem}
\soulregister\cite7
\soulregister\ref7
\soulregister\pageref7
\soulregister\figurename7
\soulregister\subsection7
\soulregister\subsubsection7
\soulregister\tablename7
\soulregister\systemname7
\soulregister\bfseries7
\soulregister\textbf7

\SetAlFnt{\small}
\SetAlCapFnt{\small}
\SetAlCapNameFnt{\small}
\SetAlCapHSkip{0pt}
\IncMargin{-\parindent}

\setcopyright{acmcopyright}
\copyrightyear{0}
\acmYear{0}
\acmDOI{0}

\acmJournal{HEALTH}
\acmVolume{0}
\acmNumber{0}
\acmArticle{0}
\acmMonth{0}

\definecolor{ccolor}{RGB}{37, 80, 44}
\renewcommand{\figurename}{Fig.}
\newcommand{\shortname}{\textit{DietGlance}\xspace}

\begin{document}

\title{\shortname: Dietary Monitoring and Personalized Analysis at a Glance with Knowledge-Empowered AI Assistant}

\author{\href{https://orcid.org/0000-0003-4857-7143}{Zhihan Jiang}}
\authornote{Co-first authors.}
\email{zj2445@cumc.columbia.edu}
\orcid{0000-0003-4857-7143}
\affiliation{
    \institution{The University of Hong Kong}
    \city{Hong Kong}
    \country{China}
}
\affiliation{
    \institution{Columbia University}
    \city{New York}
    \country{USA}
}

\author{\href{https://orcid.org/0000-0003-2496-3429}{Running Zhao}}
\authornotemark[1]
\email{rnzhao@connect.hku.hk}
\affiliation{
    \institution{The University of Hong Kong}
    \city{Hong Kong}
    \country{China}
}

\author{\href{https://orcid.org/0000-0003-1632-6921}{Lin Lin}}
\email{linlin00wa@gmail.com}
\affiliation{
    \institution{The University of Hong Kong}
    \city{Hong Kong}
    \country{China}
}

\author{\href{https://orcid.org/0000-0002-9302-0793}{Yue Yu}}
\email{yue.yu@connect.ust.hk}
\affiliation{
    \institution{The Hong Kong University of Science and Technology}
    \city{Hong Kong}
    \country{China}
}

\author{\href{https://orcid.org/0000-0002-4223-3502}{Handi Chen}}
\email{hdchen@connect.hku.hk}
\affiliation{
    \institution{The University of Hong Kong}
    \city{Hong Kong}
    \country{China}
}

\author{\href{https://orcid.org/0000-0003-3650-7332}{Xinchen Zhang}}
\email{u3008407@connect.hku.hk}
\affiliation{
    \institution{The University of Hong Kong}
    \city{Hong Kong}
    \country{China}
}

\author{\href{https://orcid.org/0000-0001-5930-3899}{Xuhai Xu}}
\email{orson@google.com}
\affiliation{
    \institution{Google}
    \city{New York}
    \country{USA}
}

\author{\href{https://orcid.org/0000-0001-6267-9440}{Yifang Wang}}
\email{yifang.wang@fsu.edu}
\affiliation{
    \institution{Florida State University}
    \city{Tallahassee}
    \country{USA}
}

\author{\href{https://orcid.org/0000-0002-9847-7784}{Xiaojuan Ma}}
\authornote{Co-corresponding authors.}
\email{mxj@cse.ust.hk}
\affiliation{
    \institution{Hong Kong University of Science and Technology}
    \city{Hong Kong}
    \country{China}
}

\author{\href{https://orcid.org/0000-0002-3454-8731}{Edith C.H. Ngai}}
\authornotemark[2]
\email{chngai@eee.hku.hk}
\affiliation{
    \institution{The University of Hong Kong}
    \city{Hong Kong}
    \country{China}
}
% \author{Anonymous author(s)}

\begin{abstract}
Growing awareness of wellness has prompted people to consider whether their dietary patterns align with their health and fitness goals. 
In response, researchers have introduced various wearable dietary monitoring systems and dietary assessment approaches. 
However, these solutions are either limited to identifying foods with simple ingredients or insufficient in providing an analysis of individual dietary behaviors with domain-specific knowledge. 
In this paper, we present \shortname, a system that automatically monitors dietary behaviors in daily routines and delivers personalized analysis from knowledge sources. 
\shortname first detects ingestive episodes from multimodal inputs using eyeglasses, capturing privacy-preserving meal images of various dishes being consumed. 
Based on the inferred food items and consumed quantities from these images, \shortname further provides nutritional analysis and personalized dietary suggestions, empowered by the retrieval-augmented generation module on a reliable nutrition library.
A short-term user study (N=33) and a four-week longitudinal study (N=16) demonstrate the usability and effectiveness of \shortname, offering insights and implications for future AI-assisted dietary monitoring and personalized healthcare intervention systems using eyewear.
\end{abstract}




\maketitle

\section{Introduction}
Dietary behaviors and human health are inextricably linked \cite{stylianou2021small, marshall2022survey}. For instance, excessive consumption of processed foods high in sugars, unhealthy fats, and sodium increases the risk of chronic diseases \cite{micha2017association}. Even small dietary improvements can significantly benefit health \cite{grummon2023simple}.
Thus, many individuals have sought dietary monitoring and analysis methods to refine their dietary choices.
By offering insights into nutritional intake, these practices enable individuals to make informed decisions on diet planning, correct nutritional imbalances, and achieve health goals such as weight management or disease prevention \cite{adams2020perspective}.
The dominant dietary monitoring and analysis method nowadays is food journaling (on paper, websites, or mobile apps~\cite{shin2022mydj}), such as MyFitnessPal~\cite{myfitnesspal} and Boohee~\cite{boohee}. 
These tools allow users to log their food consumption, track nutrients, reflect on their dietary habits, and seek dietary recommendations from nutrition experts.
However, these methods face several limitations.

 \begin{figure*}[!t]
          \centering
          \subfloat[]{
            \centering
            \includegraphics[width=.17\textwidth]{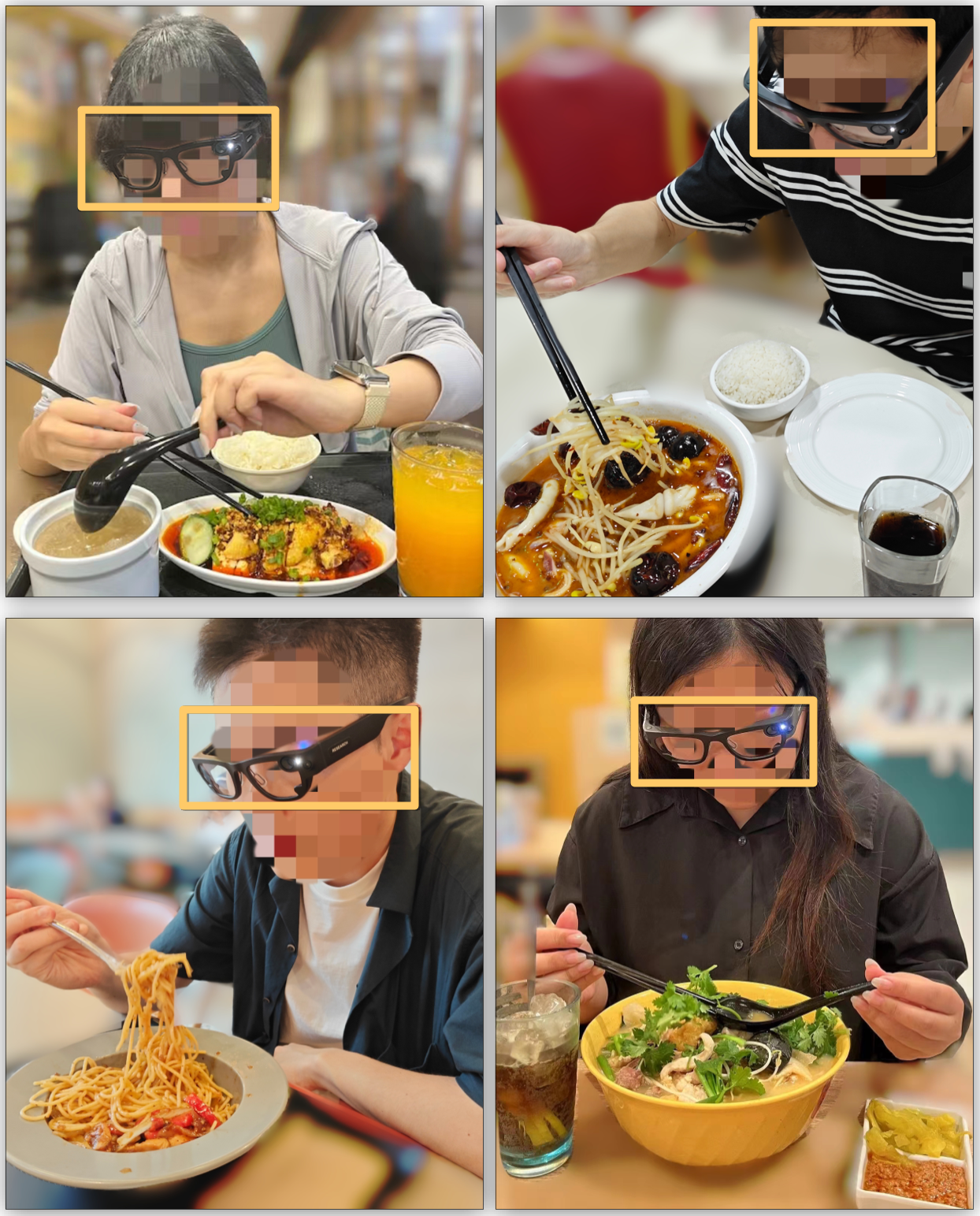}
            \label{fig:teaser1}
          }
          \subfloat[]{    
                  \centering
                  \includegraphics[width=.46\textwidth]{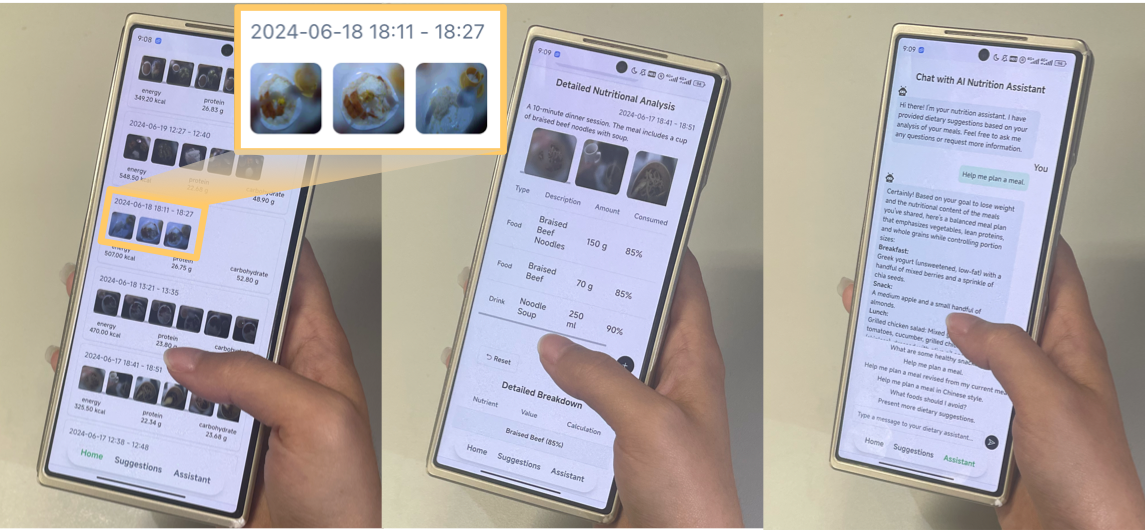}
              \label{fig:teaser2}
          }
          \subfloat[]{    
                  \centering
                  \includegraphics[width=.306\textwidth]{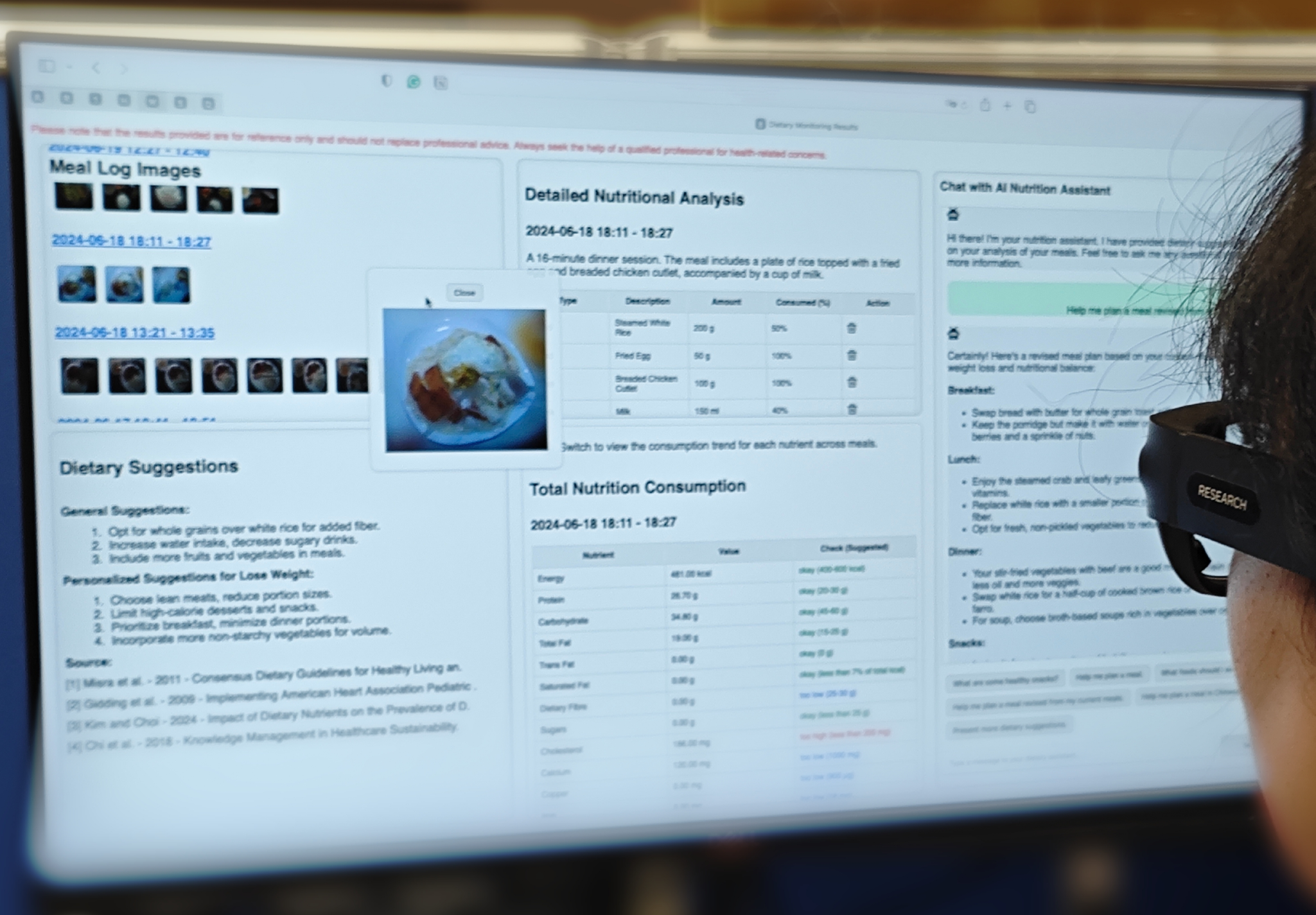}
              \label{fig:teaser5}
          }
          \vspace{-0.1cm}
          \caption{The flow of \shortname. (a) \shortname uses Aria Glasses to detect ingestive episodes and collects data in real-world, unconstrained environments. Users can view their dietary analysis results using the (b) mobile (from left to right: meal log image presentation, detailed nutritional analysis, and personalized suggestions) or (c) desktop interface.}
          \label{fig:teaser}
           \vspace{-0.1cm}
      \end{figure*}

First, manual food journaling is labor-intensive and time-consuming~\cite{selfrepotdrawback,passler2011food}. Most existing tools, like mobile apps, require users to input information or capture images manually, hindering long-term tracking sustainability~\cite{schoeller1995limitations,cordeiro2015barriers, merck2016multimodality}. Second, comprehensive nutritional analysis remains insufficiently addressed in current systems, which is essential for evaluating dietary intake and making informed decisions. 
Mobile apps offer basic nutritional insights with pre-built databases but often lack data on customized or home-cooked meals, leading to vague and inaccurate analysis. 
They also lack real-time, personalized suggestions tailored to individual contexts.
Therefore, a holistic automatic dietary monitoring and analysis system supporting food journaling, nutritional analysis, and dietary suggestions is in great need.

To automate food journaling, researchers have explored eyewear-based devices for detecting ingestive episodes \cite{bedri2020fitbyte, shin2022mydj, piezoelectricglasses}.
Eyeglasses are both socially acceptable and conveniently positioned near the eating region, making them ideal for this purpose \cite{shin2022mydj,bedri2020fitbyte}.
Contemporary commercial eyewear devices (e.g., Vision Pro \cite{apple2024visionpro} and Meta smart glasses \cite{waisberg2024meta}), which integrate advanced sensing capabilities for diverse applications, highlight the potential of eyewear for dietary data collection in daily contexts.
However, existing systems are only capable of handling simple dishes and idealized dining environments, which restricts their applicability in real-world settings. Approaches for more generic and sophisticated nutritional analysis remain largely underexplored. 

Recent advances in Large Language Models (LLMs) offer new opportunities.
Their extensive knowledge and contextual understanding enable effective management of the complexity and diversity of food items, dietary habits, and nutritional needs across various populations \cite{szymanski2024integrating, lo2024dietary}. 
For example, Szymanski et al. \cite{szymanski2024integrating} collaborated with registered dietitians to validate the GPT-4-generated food product explanation based on product names, nutrition labels, and dietary goals. Lo et al. \cite{lo2024dietary} applied GPT-4V to dietary assessment, highlighting its effectiveness in food detection under challenging conditions.
However, directly using LLM may generate plausible-sounding but incorrect or inaccurate results (i.e., hallucination). Moreover, it may show limited and generic health guidelines that are not personalized to individuals' needs, overlook comprehensive dietary patterns, and miss relevant nutrition information \cite{szymanski2024integrating, llmnutritionlimitation1, llmnutritionlimitation2}. 
There is significant potential in enhancing LLMs by domain knowledge for comprehensive nutritional analysis, enabling more reliable and personalized recommendations. Such a system can facilitate the early detection of nutritional issues and support informed dietary decisions to improve personal health.

    \begin{figure}[!t]
          \centering
          \includegraphics[width=\textwidth]{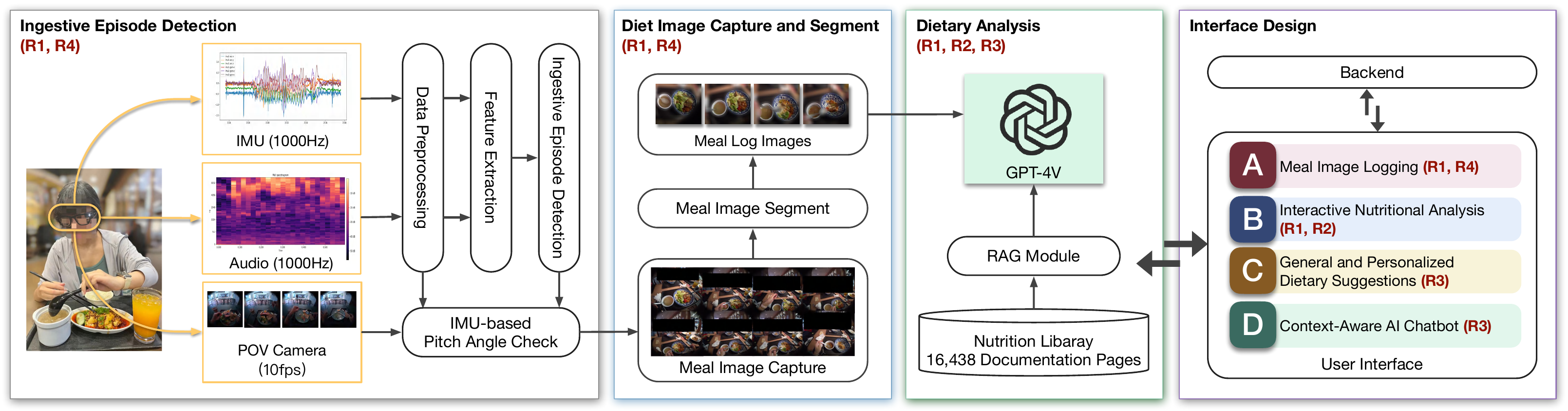}
          \vspace{-0.5cm}
          \caption{System Overview. }
          \label{fig:pip}
           \vspace{-0.1cm}
      \end{figure}
      
To address these gaps, we first conducted a needs-finding study with 45 participants to deepen our understanding of user needs and validate limitations identified in prior research. Consequently, we formulated four key design requirements for a holistic dietary monitoring and analysis system: supporting automated, accurate diet identification and logging, facilitating nutritional analysis based on reliable sources, providing personalized dietary suggestions with domain-specific knowledge, while protecting privacy.
Following these design requirements, we introduced \shortname (\figurename{\ref{fig:pip}}), an AI-assisted system for dietary monitoring and analysis using eyewear, which automatically identifies various food and drink items, and further provides knowledge-supported nutritional analysis and personalized suggestions.
Developed using Aria Glasses \cite{somasundaram2023project}, we proposed a multimodal sensing framework for ingestive episode detection and key-moment meal image capture in free-living conditions.
Based on the consumed type and amount of food and drink identified from processed meal images using vision foundation models, a Retrieval-Augmented Generation (RAG) module on a nutrition library is built to empower LLM in providing nutrition analysis and personalized dietary suggestions with knowledge sources incorporating individual profiles and meal logs. The system's user interface allows users to review personalized dietary suggestions with nutritional analysis, interactively correct errors, add information, and access additional insights. 

To evaluate \shortname, we first conducted a short-term study with 33 participants to explore their perceptions and behaviors (Study I.1). Using real-world data from their daily routine, we quantitatively evaluated the system's performance on diet identification (Study I.2), nutritional analysis (Study I.3), and conducted comparative analysis for the RAG module (Study I.4), involving crowd workers and domain experts. The qualitative and quantitative results demonstrate the system's accuracy in diet logging and its effectiveness in providing nutritional analysis and personalized suggestions.
Additionally, a four-week longitudinal study (Study II) with 16 participants demonstrated the system's positive impact on their dietary behaviors in real-world scenarios.
In summary, this paper presents the following main contributions:

\begin{enumerate}
    \item \shortname, a holistic AI-assisted system for dietary monitoring and analysis using smart glasses in daily contexts. \shortname takes a novel multimodal sensing approach with an LLM augmented by domain-specific knowledge to support accurate automated diet identification and logging, comprehensive nutritional analysis, and personalized dietary suggestions in real-world uncontrolled environments.
    \item A short-term user study (N=33) demonstrating \shortname's effectiveness and usability. With the real-world data collected in this study, we further quantitatively evaluate the system's performance on diet identification, nutritional analysis, and dietary suggestions with crowd workers and domain experts.
    \item A four-week longitudinal study (N=16) highlighting the positive long-term effects of \shortname on dietary behaviors. The quantitative analysis of nutrition consumption trends and qualitative analysis of participants' feedback demonstrate that using \shortname increased the understanding of their diet, raised their awareness of regular meal timing, promoted healthier dietary habits, and reduced consumption of unhealthy foods.
    \item Through real-world implementation and studies, we contribute practical insights into the design and implementation of AI-assisted dietary monitoring and analysis systems using eyewear. These findings offer implications for future AI-assisted dietary monitoring and personalized healthcare intervention systems.
\end{enumerate}

\section{Related Work}
\label{sec:related_work}
\subsection{Ingestive Episode Detection}
\label{dietarydetection}
Ingestive episode detection is essential for automatic food journaling \cite{tang2025video}. 
Various wearable sensors have been explored for automated tracking. Wrist-mounted devices can track hand movements during eating \cite{wrist1,wrist2,wrist3}, but these can be distracted by other gestures and require use on the dominant hand. More reliable approaches involve head- or neck-mounted devices like necklace \cite{necklace_proximity2}, headbands \cite{headband}, caps \cite{cap}, and ear-mounted sensors \cite{EarBit,EarSAVAS,Auracle}, though they are often uncomfortable and socially intrusive. 
In contrast, eyeglasses offer a more acceptable alternative, with their proximity to the mouth allowing for better detection of eating episodes. Previous work using eyeglasses includes sensors such as electromyography \cite{emgglasses}, piezoelectric \cite{piezoelectricglasses}, load cells \cite{loadcellglasses}, and acoustic sensors \cite{acousticglasses} for the detection of ingestive behavior. 
Among these sensors, the Inertial Measurement Unit (IMU) is a widely adopted choice, primarily because it is low-cost, ubiquitous in consumer electronics, and highly effective at capturing the specific 3D motion and orientation of the head during ingestive behaviors.
Rahman et al. \cite{imuglasses} utilized the IMU in Google Glass for detection. IMU sensors combine accelerometers, gyroscopes, and occasionally magnetometers to measure the specific force, angular rate, and sometimes magnetic field of objects, enabling the tracking of motion and orientation in three-dimensional space \cite{bedri2020fitbyte}.
Systems like FitByte \cite{bedri2020fitbyte} integrated IMU and proximity sensors, while MyDJ \cite{shin2022mydj} utilized piezoelectric and accelerometer sensors. Similarly, other work has explored using deep learning with optical tracking sensors in smart glasses to monitor eating behavior \cite{stankoski2024controlled}.

However, prior studies were limited to simple dishes or controlled settings \cite{shin2022mydj, chen2023first} and often relied on crowd-sourced efforts \cite{bedri2020fitbyte}. \shortname introduces a multimodal sensing framework to record the meal throughout dining in uncontrolled environments of the real world, allowing comprehensive nutritional analysis and personalized diet suggestions.

\subsection{Diet Identification and Nutritional Analysis}
While the prior works in Sec.~\ref{dietarydetection} focused primarily on detecting ingestive episodes, their methods for diet identification were often simple and used in controlled settings \cite{shin2022mydj, chen2023first, bedri2020fitbyte}. This section, therefore, reviews more dedicated approaches for diet identification and nutritional analysis.
Traditional expert- or crowd-based methods include uploading meal photos for expert evaluation \cite{expertnutritionestimate} or crowdsourced nutrient estimation \cite{crowdsourcingnutritionestimate}, which are labor-intensive and time-consuming, limiting the system's accessibility and scalability. 
Additionally, online searches or forums often provide irrelevant information, reducing their effectiveness.
With advances in computer vision, some solutions perform food segmentation, volume estimation, and database-based nutrient calculations sequentially \cite{foodseg1,foodseg2,foodseg3,foodseg4}, while end-to-end neural networks directly estimate nutritional content from meal images \cite{end2endfood1,end2endfood2,end2endfood3}. 
Although they have achieved promising results with pre-built nutrition estimation datasets, the generalization of models limits their applicability to open-world settings, and vision models alone lack comprehensive knowledge of diet or nutrition \cite{foodreview}.   

The rise of LLMs offers new potential. They excel in knowledge integration and cross-domain generalization, attributed to their immense parameters and diverse, extensive training data \cite{llmemergence}. 
Lo et al. \cite{lo2024dietary} demonstrated that GPT-4V can identify food items with high accuracy, deduce portion sizes of eating consumption at a comparable performance to dietitians' estimates, and estimate nutritional components aligning with the USDA National Nutrient Database \cite{haytowitz2018usda} under a controlled study protocol that evaluated GPT-4V on pre-collected meal images.
Building on this, our \shortname system employs GPT-4V with crafted prompts and user profiles to estimate diet type and amount. Unlike prior work, we proposed a multimodal sensing framework combining IMU, audio, and images for automatic diet identification and nutritional analysis under completely free-living conditions. Insights from our user studies highlight opportunities to further refine dietary monitoring and analysis systems.

\subsection{Personalized Dietary Suggestions}
The use of machine learning models for personalized nutrition and dietary recommendation has been widely explored, as documented in several systematic reviews \cite{varshney2023personalized, tsolakidis2024artificial}, marking a shift away from the traditional ``one-diet-fits-all'' era \cite{roman2024personalized}.
For example, Mitchell et al. \cite{g2021reflection} developed GlucoGoalie, which combines machine learning and a rule-based expert system to generate dietary goals for individuals with type 2 diabetes. 
To enhance flexibility and generalization, researchers have turned to LLMs for dietary guidance, leveraging their extensive knowledge and reasoning capabilities.
Chatelan et al. \cite{dietaryguidance1} investigated the ability of ChatGPT to provide nutritional and dietary guidance for patients with type 2 diabetes and hemodialysis, and their findings show that the generated diet recipes are in accordance with the Diabetes Plate Method. Ataguba et al. \cite{ataguba2025exploring} explored using LLM for personalized recipe generation and weight-loss management. Further, Yang et al. \cite{dietaryguidance2} introduced ChatDiet, an LLM-powered framework for personalized nutrition-oriented food recommender chatbots, and evaluated it on a case study including personal information, where the generated food recommendation dialogue demonstrates the explainability and personalization. 
However, researchers noted that LLMs remain susceptible to hallucinations. A study by Szymanski et al. \cite{szymanski2024integrating} involved registered dietitians in assessing the capabilities of LLMs in providing nutritionally accurate and personalized information. The results showed that the outputs often failed to align with professional standards and contained falsehoods, which could be misleading.
Similarly, in \cite{llmnutritionlimitation1}, researchers found that while the nutritional advice proposed by ChatGPT is generally accurate, it has the potential to produce unsafe diets containing allergens.

Therefore, different from existing works that directly prompt LLMs for dietary suggestions, we built the RAG module with reliable external dietary and nutritional knowledge to enhance LLM's capabilities for nutritional analysis and personalized dietary suggestions. We also developed a context-aware chatbot that allows users to enhance personalization through iterative conversations.
This approach provides accurate and trustworthy results, reducing the need for costly expert involvement while maintaining high-quality recommendations.

\subsection{Leveraging RAG to Enhance LLMs for Accurate Knowledge Retrieval}
LLMs are widely used for information extraction and summarization due to their vast knowledge and reasoning capabilities. However, directly prompting LLMs can lead to plausible-sounding but inaccurate answers, known as hallucinations, and a lack of domain-specific knowledge. To address these issues, RAG enhances LLMs by integrating a retrievable memory that incorporates knowledge from external sources \cite{llmhallucination,ragsurvey}. 
For example, Fok et al. \cite{Qlarify} use RAG to expand abstracts with additional information from full papers, while Zulfkar et al. \cite{Memoro} develop Memoro to infer the user's memory needs in conversation and present suggestions queried from memories using a RAG module. 
The Arizona Water Chatbot \cite{Arizona_Water_Chatbot} employs RAG to retrieve water-related information from reputable sources, improving decision-making.
Ren et al. \cite{Memolet} explore memory retrieval and generation refinement using RAG for enhanced conversational AI, and Yang et al. \cite{yang2024aqua} develop AQuA that combines software UI elements associated with questions as the query and generates answers using RAG-powered GPT-4 from official documentation and tutorial resources.
Recent work has specifically explored RAG in specialized health domains to address the accuracy limitations of baseline LLMs. For instance, in \cite{bingol2025accuracy},
RAG was used to improve the accuracy of dietary principle determination in patients with Chronic Kidney Disease (CKD), showing that RAG generated more accurate and stable results compared to LLMs alone. Similarly, Bano et al. \cite{bano2024utilizing} found that the augmented model generated more concise and easily comprehensible responses, which was critical for improving patient health literacy.
This recent work has shown the promise of RAG in specialized, offline nutritional applications. In this work, we integrate the RAG module within an end-to-end system that automatically detects and analyzes meals in completely free-living conditions, and investigate the performance with and without RAG.

\section{Existing Practices of Dietary Monitoring and Analysis }

To validate limitations identified in prior research and better reveal users' needs, concerns, and hurdles encountered, we conducted a needs-finding study with 45 participants. 
We then derived four design requirements to motivate our system design.

\subsection{Participants and Procedure}

We recruited 45 participants with diverse demographic and socio-economic backgrounds by word-of-mouth and online advertisement, with inclusion criteria of being adults (aged  > 18), fluent in Chinese or English, and without eating disorders. Eligibility regarding eating disorders was determined based on self-identification during the initial online screening process.
The participants were geographically located across three regions: 26 in Mainland China, 17 in Hong Kong SAR, and 2 in Japan.
Their ages range from 18 to 62 ($M=33.24, SD=10.62$),
including 22 males, 21 females, 1 non-binary, and 1 preferred not to disclose.
Participants had diverse socio-economic backgrounds, including twenty university students, six senior researchers, four engineers, one high school graduate, one government employee, one freelancer, one bank employee, two high school teachers, two manufacturers, two workers, one clerk, and four retirees.
We included both experienced (exp., 23, 51\%) and inexperienced (inexp., 22, 49\%) participants. The exp. group consisted of individuals who had prior experience using dietary monitoring tools, such as mobile apps, online tools, and paper journals, or receiving professional guidance from experts like dietitians or nutritionists, while the inexp. group included those with no prior experience.

The procedure for the needs-finding study consisted of two main stages.
\begin{enumerate}
    \item \textbf{Online Survey}: First, participants were provided with an online questionnaire (Appendix \ref{appx:needs_survey}). This survey was used to gather quantitative data on their demographic information, backgrounds, prior experience with dietary monitoring tools, and eating environments.
    \item \textbf{Semi-Structured Interview:} After completing the survey, each participant attended a one-on-one semi-structured interview session (conducted either face-to-face or remotely, Appendix \ref{appx:needs_interview}), which lasted about 20 minutes per participant. This qualitative session was designed to explore their vision for helpful features, the obstacles they've encountered in dietary tracking, and their specific expectations and concerns regarding future eyewear-based dietary systems.
\end{enumerate}
After completing the interview, each participant was compensated about \$2.5. For the subsequent qualitative analysis of the interview data, the lead author conducted thematic coding and iteratively discussed with other authors to finalize the codes.

\subsection{Key Findings and Design Requirements}
We summarize the following design requirements based on the key findings:

\textbf{R1. Supporting Automated Accurate Diet Identification and Logging.} 
Participants (30 out of 45) confirmed the need for automated detection and logging of diet consumption to reduce manual effort and improve convenience~\cite{bedri2020fitbyte,shin2022mydj}, as the complexity and time demands were the main reasons for both exp. and inexp. participants not or reduce monitoring their diets.
Accurate diet identification is crucial for meaningful dietary analysis.
Thus, the system should automatically and accurately identify and log both the type and amount of food and drink consumed.

\textbf{R2. Facilitating Nutritional Analysis based on Reliable Sources.} 
Participants expected professional nutritional information, including detailed nutrient breakdowns, consistent with prior research emphasizing the importance of nutritional analysis \cite{lo2024dietary,szymanski2024integrating}.
However, nine out of 23 exp. participants emphasized the inaccurate and incomplete databases in existing tools. The system should integrate reliable food databases to support nutrition estimates covering a wide variety of items. 
Twenty-one out of 45 participants mentioned it would be helpful to track nutrition consumption.
This would help users manage their nutrition consumption to maintain health or achieve specific goals.

\textbf{R3. Providing Personalized Dietary Suggestions with Domain-Specific Knowledge.}
33 out of 45 participants were highly interested in reliable and personalized dietary recommendations tailored to their profiles (e.g., age and weight), goals (e.g., muscle development, weight management, general wellness), and specific dietary needs (e.g., diabetes, allergies, vegetarianism), which is consistent with prior work \cite{szymanski2024integrating}. Our findings emphasized the need for seamlessly integrating domain knowledge cost-effectively.
The system should provide dietary suggestions based on individual dietary behaviors and domain-specific knowledge. Participants desired guidance on various topics, including nutritional balance, weight management, customized meal plans, ingredient substitutions, portion control, and dietary assessments.

\textbf{R4. Protecting Privacy.} 
Privacy concerns were raised by 24 participants towards the eyewear-based dietary monitoring, especially regarding the potential of recording their phone screen and conversations with others during meals. Therefore, the system should try to reduce the capture of sensitive information during meal sessions and incorporate a mechanism to protect privacy.

\section{System Design and Implementation}
\label{sec:system}
Based on the design requirements, we developed \shortname, as shown in \figurename{~\ref{fig:pip}}. 
First, we propose a multimodal sensing approach using camera, IMU, and audio sensors on eyeglasses to detect ingestion behaviors (Sec.~\ref{sub:system:episode_detection}) and capture meal images (Sec.~\ref{sub:system:image_capture}) (\textbf{R1}, \textbf{R4}).
The processed images and user profiles were analyzed using GPT-4V to identify the type and amount of food consumed (Sec.~\ref{sub:system:dietary_analysis}, \textbf{R1}).
Unlike previous studies relying solely on LLM or requiring expert input, we further grounded the LLM with a reliable nutrition library through the RAG module to obtain nutritional analysis and personalized dietary suggestions (Sec.~\ref{sub:system:dietary_analysis}, \textbf{R2}, \textbf{R3}).
Finally, we developed a novel user interface, distinct from previous studies, that allows users to interactively review and correct data, add details, explore results, and obtain more information (Sec.~\ref{sub:system:interface}, \textbf{R1}-\textbf{R4}). All prompts are available in Appendix \ref{appx1:prompts}.

    \begin{figure}[!t]
          \centering
          \includegraphics[width=.8\textwidth]{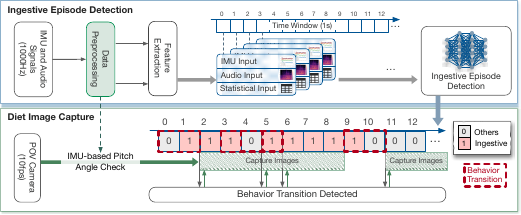}
          \caption{An illustration of the multimodal sensing framework for ingestive episode detection and diet image capture. Image capture is triggered for three seconds after both an IMU-based pitch angle check and a behavior transition (e.g., from non-ingestive '0' to ingestive '1', and vice-versa).}
          \label{fig:image-cap}
           % \vspace{-0.5cm}
      \end{figure}
      
\subsection{Ingestive Episode Detection}
\label{sub:system:episode_detection}
Accurate detection of ingestive episodes is the first step of 
an automated diet analysis system.
\shortname leveraged IMU and audio signals and built a multimodal machine learning model to achieve the goal (\figurename{~\ref{fig:image-cap}}).
\subsubsection{Data Preprocessing and Feature Extraction}
The IMU data used included tri-axial accelerometer and gyroscope data, sampled at 800Hz and 1000Hz from two IMUs on each side of the Aria Glasses. 
The 800Hz IMU data was upsampled to 1000Hz. Audio signals, captured at 48kHz, were down-sampled to 1000Hz, at which rate the speech data is largely unintelligible, to preserve sensitive content \cite{mollyn2022samosa} (\textbf{R4}).

IMU and audio signals were normalized and segmented into one-second non-overlapping windows.
The IMU data were normalized and concatenated into a 12-channel input (two IMUs, each with a tri-axial accelerometer and a tri-axial gyroscope). The audio signals were processed by Short-Time Fourier Transform (STFT) into Mel-spectrograms. We further extracted statistical features for each window. For IMU, the mean, standard deviation, min, max, kurtosis, skewness, and frequency features were extracted for the two tri-axial acceleration data and two gyroscope data. To obtain the frequency features, we first computed the Fast Fourier Transform (FFT) of the IMU data and then computed the mean and peak of the magnitude of the frequency domain. For audio signals, we computed the mean, standard deviation of their Mel-spectrograms, Zero Crossing Rate, Spectral Centroid, Spectral Bandwidth, Spectral Rolloff, Chroma STFT, and Root Mean Square Error. The statistical features of IMU and audio data for each time window were concatenated as the statistical feature.

\subsubsection{Ingestive Episode Detection}

\begin{figure}[!t]
          \centering
          \includegraphics[width=.9\textwidth]{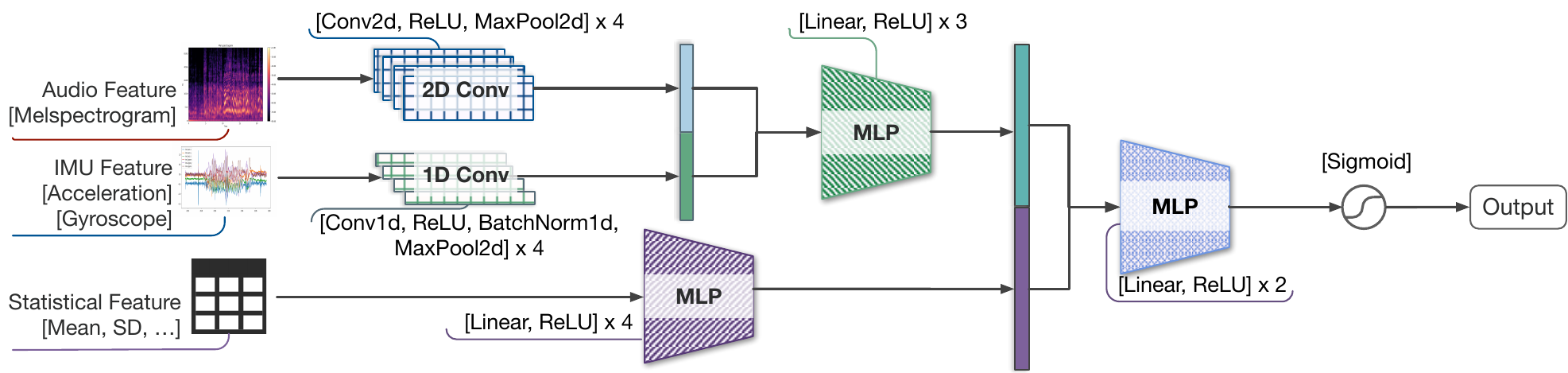}
          \caption{The multimodal learning model for classifying ingestive episodes and others.}
          \label{fig:nn}
           % \vspace{-0.5cm}
      \end{figure}

As illustrated in \figurename{~\ref{fig:nn}}, the model consists of three parallel input branches that are processed in two main stages. First, the Audio Feature (Melspectrogram) is processed by a 2D Convolutional Network (a stack of [Conv2d, ReLU, MaxPool2d] repeated 4 times), and the IMU Feature (Acceleration, Gyroscope) is processed by a 1D Convolutional Network ([Conv1d, ReLU, BatchNorm1d, MaxPool2d] repeated 4 times). 
The outputs of these two convolutional branches are then concatenated and fed into a shared MLP ([Linear, ReLU] repeated 3 times). 
In parallel, the Statistical Feature is processed by its own MLP ([Linear, ReLU] repeated 4 times). 
Finally, the outputs of the Audio and IMU MLP branch and the Statistical MLP branch are concatenated and fed into a final MLP ([Linear, ReLU] repeated 2 times).
The output layer uses a Sigmoid activation function to produce a probability score between 0 and 1, classifying the window as either non-ingestive (0) or ingestive (1).

Specifically, instead of distinguishing between eating and drinking, we categorized both under the broader term ``ingestive behaviors.'' This decision accounts for practical challenges, such as the overlap in actions like consuming noodle soup (involving both eating and drinking) or swallowing soft foods without chewing. This simplification aligns with the study's goals and improves usability in real-world applications.

\subsection{Diet Image Capture and Segment}
\label{sub:system:image_capture}
We proposed to capture the meal images considering the pitch angle to increase the completeness of meals captured and segment the food, drink, and tableware in the images to reduce privacy concerns (\textbf{R1}, \textbf{R4}).
\subsubsection{IMU-based Pitch Angle Check}
Simultaneously with ingestive episode detection, we employ a pitch angle check as a practical filter to ensure the egocentric camera is physically pointed at the meal during image capture. The pitch angle was calculated from tri-axial acceleration data ($a_x, a_y, a_z$) as $pitch = arctan(a_x/\sqrt{a_y^2+a_z^2})$, where a value exceeding a threshold ($pitch_{th}=5$) indicated a state of being tilted forward.
The pitch angle check helps increase the comprehensiveness of the meal capture while also avoiding the capture of the surrounding environment, which benefits privacy.
Our image capture is then triggered by a dual-condition logic as illustrated in \figurename{~\ref{fig:image-cap}}: images are captured at 10fps for three seconds only when (1) a behavior transition is detected (i.e., from non-ingestive to ingestive, or vice-versa), and (2) the user's head is in this downward-tilted state.
This approach allowed the system to capture a series of images throughout the meal, including dishes added midway.

Although users may tilt their heads back while drinking, capturing beverages from this angle in egocentric images is challenging. By focusing on images captured when users tilted their heads downward, beverages and other meal components were recorded more effectively.

\subsubsection{Food, Drink, Tableware Segmentation}
\begin{figure}[!t]
          \centering
          \includegraphics[width=\textwidth]{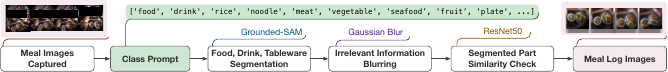}
          \caption{An illustration of diet image capture and segmentation.}
          \label{fig:image-segment}
           % \vspace{-0.5cm}
      \end{figure}

We further segmented the food, drinks, and tableware within the images.
To exclude the food from nearby diners and include the user's meal more completely in the egocentric images, we set the bottom one-third region as the primary area of interest (detailed in Appendix \ref{appx:image_seg}). For segmentation, we employed Grounded-Segment-Anything (Grounded-SAM) \cite{ren2024grounded}, which enables precise identification based on textual descriptions. This approach allows the segmentation of a wide range of foods and drinks using class prompts and can easily accommodate new food types by adding new prompts. 
As illustrated in \figurename{~\ref{fig:image-segment}}, we first utilized the pre-trained GroundingDINO model to detect objects based on the class prompt, then applied the pre-trained SAM model to convert detections into masks. Additionally, the class prompt (detailed in Appendix \ref{appx1:class_prompts}) included not only food and drink categories but also tableware, as these are often the only remaining objects post-meal.

\subsubsection{Irrelevant Information Blurring and Similarity Check}
After segmentation, we blurred the images except for the segmented part with Gaussian Blur \cite{jiang2023dartblur} to reduce privacy concerns and also highlight the area of interest (i.e., the meal of the user).
We then screened these images by comparing the Cosine Similarities of the features extracted from segmented parts using ResNet50 \cite{rajpal2021using} to remove redundancy. If the similarity between the segmented parts was larger than 0.75, we removed one of the images. The blurred part of the images would be further automatically cropped (to the bounding box of the segmented meal region) to highlight the meal items and enhance privacy protection. In this way, we finally obtained a series of privacy-preserving meal images.

\subsection{Dietary Analysis}
\label{sub:system:dietary_analysis}
Based on the meal images captured across the meal session, we used a vision language model to identify various food and drink items and estimate the consumed amount (\textbf{R1}). Furthermore, we built a nutrition library and proposed to augment the LLM with a RAG module for nutritional analysis and dietary suggestions (\textbf{R2}, \textbf{R3}).
\subsubsection{Diet Identification}
\begin{figure}[!t]
          \centering
          \includegraphics[width=\textwidth]{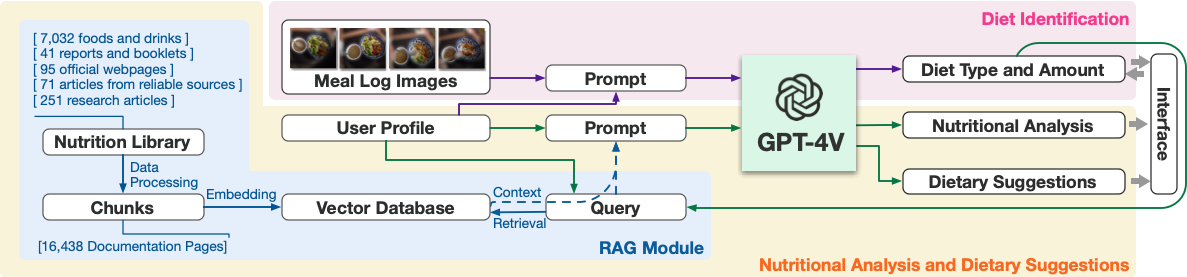}
          \caption{An illustration of RAG-grounded GPT-4V for diet identification, nutritional analysis, and dietary suggestion.}
          \label{fig:llm}
           % \vspace{-0.5cm}
      \end{figure}

LLMs, trained on diverse datasets, excel at identifying various cuisines and ingredients in images and estimating food quantities through visual cues in a zero-shot manner, offering detailed nutritional information. In this study, we select GPT-4V among all LLMs due to its superior performance in diet recognition and estimation in open-world settings \cite{lo2024dietary} (prompt available in Appendix \ref{appx1:diet_identify}). As shown in \figurename{~\ref{fig:llm}} (pink part), the meal log images and user profile (including gender, age, height, and weight) were fed into the GPT-4V to obtain the types and amount of each food and drink consumed by the user in this meal session.

\subsubsection{RAG Module Construction}
While LLMs excel in diet identification, they may generate seemingly plausible but inaccurate results, lack health guidelines, and miss domain-specific knowledge.
To address this, we integrated RAG to enhance the generative capability of LLM by incorporating knowledge from an extensive nutrition library, thereby providing a more reliable, contextually relevant, and factual response \cite{lewis2020retrieval}.

To build the RAG module, we have constructed a comprehensive nutrition library consisting of documents from a diverse collection of reliable resources.
The inclusion criteria included (1) the government-approved food nutrient database, (2) official reports and booklets on food consumption and dietary suggestions for the general public and specific groups, issued by the government or the World Health Organization (WHO), (3) online resources from official government websites, containing recipes, scientific articles about food and nutrition, and healthy diet suggestions, (4) articles published by highly trustworthy organizations (e.g., National Institutes of Health), (5) peer-reviewed research articles from prestigious journals with 28 different needs (e.g., weight management, muscle development, and blood glucose stabilization). These inclusion criteria were designed to ensure the corpus was composed of high-quality, expert-vetted sources to maximize accuracy and minimize the risk of conflicting information. Detailed information is available in Appendix \ref{appx2:nutrition}.

We then built the RAG module using the LangChain framework \cite{langchain2024}. The documents in the nutrition library were first split into chunks with LangChain's \texttt{RecursiveCharacterTextSplitter}, resulting in 16,438 pages, as shown in \figurename{~\ref{fig:llm}} (blue part). Then OpenAI's embedding model \texttt{text-embedding-ada-002} created embeddings stored with Facebook AI Similarity Search (Faiss) vector database. Based on these embeddings, the most relevant chunks will be searched based on the query to provide context for answering the query, enhancing the relevance and accuracy of the responses by incorporating the contexts. By invoking the retrieval chain created by LangChain, it would not only return the answer based on the prompt and query but also the sources of the answer from the relevant context, making the sources of dietary suggestions more transparent.

\subsubsection{Nutritional Analysis and Dietary Suggestion}

With the RAG-grounded LLM, we first use a crafted prompt (available in Appendix \ref{appx1:nutritional_analysis}) to derive nutritional analysis based on the description and amount of food and drinks, including the nutrient and corresponding values. The total nutrition consumed per meal was calculated from the estimated portions. The system then checked if each nutrient intake was too high, too low, or within a reasonable range, offering suggested reference values.

We further fed the meal logs, nutrition analysis, and user profile into the RAG-grounded LLM to generate general dietary suggestions and personalized suggestions tailored to users' dietary goals, with references for these suggestions. User profiles could include general demographic information (i.e., gender, age, height, weight) and personalized dietary goals, such as muscle development, weight loss, weight control, allergens, and vegetarianism. The meal logs contained the description of meals, including the consumed food and drink, and the meal timing. The system used this comprehensive information to generate dietary suggestions using the RAG-grounded LLM. As shown in \figurename{~\ref{fig:llm}} (yellow part), a query containing the information of the user profile, meal logs, nutritional analysis, and the dietary goal is first used to retrieve the relevant contexts from the nutrition library. The contexts and the query will be used to build the prompt for producing tailored dietary suggestions leveraging the contextual and reasoning capabilities of the LLM.
For those without specific goals, the system provided balanced diet suggestions based on their profile and dietary behavior.

    \begin{figure}[!t]
          \centering
          \includegraphics[width=\textwidth]{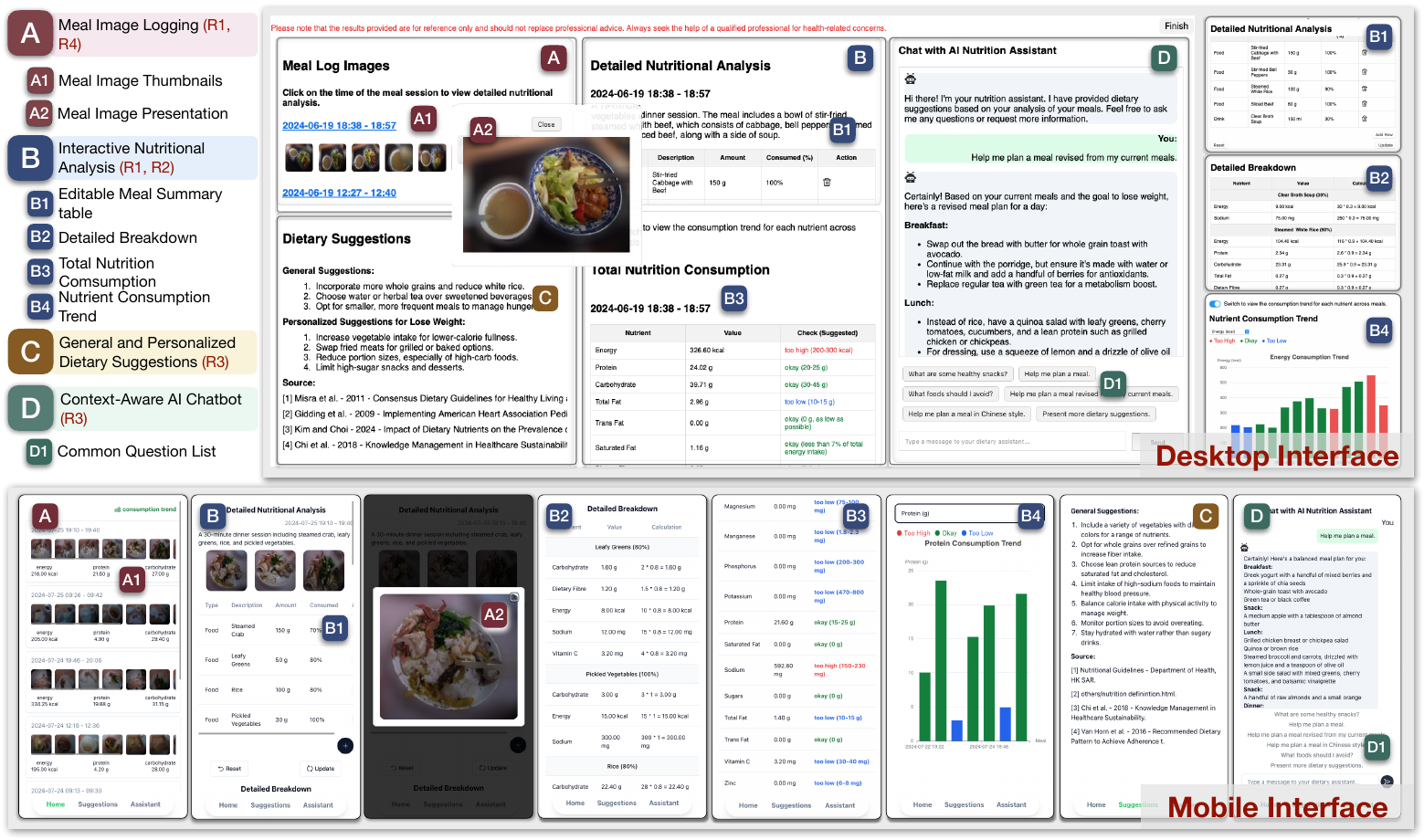}
          \caption{Interface Overview. There are desktop and mobile versions. The interface includes four main parts: (A) Meal Image Logging, (B) Interactive Nutritional Analysis, (C) General and Personalized Dietary Suggestions, and (D) Context-Aware AI Chatbot.}
          \label{fig:interface}
           % \vspace{-0.5cm}
      \end{figure}

\subsection{Interface Design}
\label{sub:system:interface}
To support users exploring the results, seeking further details, such as explanations for specific nutrients, alternative food options, or personalized meal plans, and providing contextual information, we developed an interactive desktop and mobile interface (\figurename{~\ref{fig:interface}}) that not only displays results and suggestions but also allows users to correct errors, add missing details, and access more information. 
The interface consists of four major components: (1) Meal Image Logging (\figurename{~\ref{fig:interface}-A}), (2) Interactive Nutritional Analysis (\figurename{\ref{fig:interface}-B}), (3) Dietary Suggestion (\figurename{~\ref{fig:interface}-C}), and (4) Context-Aware AI Chatbot (\figurename{~\ref{fig:interface}-D}). 
Chronologically displayed meal log thumbnails allowed users to review their dietary patterns and access privacy-protected meal images (\figurename{~\ref{fig:interface}-A}, \textbf{R1}, \textbf{R4}), while clicking on a meal session provided detailed nutritional analysis and total nutrient consumption (\figurename{\ref{fig:interface}-B}, \textbf{R1}, \textbf{R2}). The interactive analysis enabled users to edit meal summaries and dynamically update nutritional insights and suggestions. Personalized dietary suggestions (\figurename{\ref{fig:interface}-C}, \textbf{R3}), limited to seven items for clarity, offered general guidance for a balanced diet and tailored suggestions based on user goals, with transparent source references to enhance trust. The AI chatbot (\figurename{\ref{fig:interface}-D}, \textbf{R3}) provided context-aware guidance, leveraging user profiles, meal logs, and prior interactions to deliver accurate, personalized responses, further streamlining the experience with preloaded common questions. This combination of features supported reflective, informed, and personalized dietary management. More details are available in the Appendix \ref{appx:system_interface}.

\section{Study I: Short-Term Study for Evaluating \shortname from the four design requirements}

We evaluated \shortname following the four design requirements.
Our comprehensive evaluation involves multiple stakeholders, including (1) Study I.1: short-term real-world user study to explore users' perceptions towards \shortname (\textbf{R1}-\textbf{R4}), (2) Study I.2: quantitative evaluation of \shortname's performance on diet identification and logging (\textbf{R1}), (3) Study I.3: quantitative evaluation with domain experts of \shortname's performance on nutritional analysis (\textbf{R2}), and (4) Study I.4: comparative analysis for the RAG module (\textbf{R2}, \textbf{R3}).
This study has obtained IRB approval.

\subsection{Study I.1: Short-Term User Study to Explore Users' Perceptions towards \shortname (\textbf{R1}-\textbf{R4})}
We implemented the proposed system and conducted a user study, recruiting participants to collect data using Aria Glasses in their daily routine and explore the dietary monitoring analysis results via the user interface. Through this study, we aimed to evaluate the system's performance on the four design requirements qualitatively and quantitatively.

\subsubsection{Participants and Data Collection} 

\label{sec:par}
We recruited 33 participants (P1-P33), including 21 self-declared males (63.6\%) and 12 self-declared females (36.4\%), aged 23 to 73 ($M=30.73$, $SD=10.51$). 
The group consisted of twenty-one university students, nine senior researchers, one accountant, one retired person, and one engineer.
Twenty-one (63.6\%) participants self-reported exp. with dietary monitoring and analysis, while the remaining participants (36.4\%) were inexp. More details are available in Appendix \ref{appx:participants}.

Participants used Aria Glasses before, during, and after their meal sessions, capturing the ground truth video recordings, audio, and IMU data. Due to the power limitation of Aria Glasses, we did not ask the participants to wear the Aria Glasses all day, but for periods that covered the meal sessions, which also included various behaviors like walking, sitting, talking, using mobile phones, etc. Participants reviewed their recordings and labeled the start and end of each ingestive episode, verified by the authors. Participants earned an average wage of about \$8 per hour of recording and would get about \$6.42 after finishing the user study.

\subsubsection{Data Processing and Ingestive Episode Detection Performance}
Recordings were segmented into one-second windows, and if ingestive behavior occurred for more than 50\% of the window, it was labeled as ingestive; otherwise, it was labeled as others (non-eating or drinking). In total, we collected 38.23 hours of recordings, covering 144 meal sessions ($M=15.9$ minutes, Min$=3.3$, Max$=66.2$, $SD=15.9$). Each participant contributed between two (one day without breakfast) and nine sessions ($M=4.4$, $SD$ = 1.74, spanning one to three days). Of the total, 24.78 hours were ingestive episodes, and 13.46 hours were other behaviors.
The performance of ingestive episode detection was evaluated using the leave-one-user-out (LOUO) approach. Our system achieved an F1-score of 0.925, with precision at 0.939 and recall at 0.912. The LOUO method evaluates the model's generalization capacity by iteratively training it on data from all users except one, testing on the excluded user’s data, and repeating this process for each user, thereby offering a comprehensive, user-specific assessment of model performance.
Despite some time windows being misclassified, the system successfully captured the meal images across the entire meal sessions. Additionally, during image segmentation, images without target items were excluded, further enhancing system robustness and ensuring successful image capture for all 144 meal sessions.

\subsubsection{Study Settings} 
After obtaining participants' consent, we introduced the system and provided a brief tutorial on the user interface. Participants were then asked to explore their dietary monitoring results, review meal identifications for accuracy, examine dietary suggestions and nutritional analysis, and use the chatbot. We recorded their screens and tracked interactions using Google Analytics \cite{ledford2011google}.
Participants then completed a survey, including the System Usability Scale (SUS) \cite{lewis2018system}, Net Promoter Score (NPS) \cite{score2018net}, User Experience Questionnaire (UEQ) \cite{schrepp2014applying}, and questions corresponding to the four design requirements.
Finally, semi-structured interviews were conducted for open-ended feedback on their experience. Survey scales were converted to a 5-point Likert scale (1 = strongly disagree, 5 = strongly agree) to make it consistent. Detailed scale items and interview scripts are in Appendix \ref{appx:scale}. The lead author first performed a deductive thematic analysis \cite{bowman2023using} using predefined themes derived from the system's design requirements for the interviews and textual data from participants, and iteratively discussed with other authors to validate the codes and address any disagreements.

\subsubsection{User Study Results}
\label{sec:study_I_results}
We summarized the participants' perceptions towards \shortname and evaluated the system's overall usability, user satisfaction, experience, and users' interaction behaviors with the system.

\paragraph{Perception on Diet Identification and Logging (\textbf{R1})}
% Twelve participants (36.4\%) strongly agreed and twenty (60.6\%) somewhat agreed that the system accurately identified and logged their meals.
Most of the participants (12 strong, 20 somewhat; average rating 4.3/5) agreed that the system accurately identified and logged their meals.
A few challenges were noted, including difficulty identifying mixed dishes (P8, inexp.), visually similar items like sugar-free vs. regular Coke (P7, P10, P21, exp.), estimating portions during shared meals (P29, exp., P33, inexp.), and identifying cooking methods (P24, exp.). Despite these issues, participants appreciated the system's accuracy, with exp. participants valuing its automation over manual tools they had used and its ability to identify culturally specific or less common items. Both groups emphasized its effectiveness in handling a wide range of foods.

\paragraph{Perception on Nutritional Analysis (\textbf{R2})}
Most participants (90.9\%) somewhat agreed or strongly agreed with the system's nutritional analysis accuracy, citing alignment with their knowledge (e.g., seafood as high in protein, P23, exp.) and the detailed nutrient breakdown that built trust (P9, inexp.). Both groups valued actionable insights, such as identifying nutrient imbalances. 
For example, P32 (exp.) adjusted meals based on the analysis, while inexp. participants found the system educational, learning about unexpected nutritional facts (e.g., high sodium in soup, P16, inexp.).
Exp. participants often scrutinized the analysis more deeply, with some seeking clarifications via the chatbot (e.g., P2 resolved confusion about high energy and low fat).
In contrast, inexp. participants tended to rely on the system's explanations and valued its ability to reveal new insights. These findings suggest that the system may help reinforce knowledge for exp. users while serving as an educational tool for newcomers.

\paragraph{Perception on Personalized Dietary Suggestion (\textbf{R3})}
Fourteen participants (42.4\%) strongly agreed, and eight (24.2\%) somewhat agreed that the dietary suggestions aligned with their goals, with exp. participants often highlighting consistency with their existing knowledge (P31). Suggestions were deemed easy to follow by 24 participants (72.7\%), though some noted challenges in following the suggestions due to limited food options or preferences, such as P28 (exp.), who cited university canteen constraints, and P18 (exp.), who disliked certain healthy recommendations.
Trust in the suggestions was enhanced by credible sources provided, with 15 (45.5\%) strongly and 13 (39.4\%) somewhat trusting the recommendations. Exp. participants leveraged the suggestions to refine and validate their dietary practices, while inexp. participants viewed them as practical guidance for incremental improvements. Despite differing levels of reliance, both groups acknowledged the system's potential to fill knowledge gaps and facilitate healthier, goal-oriented choices.

\paragraph{Perception on Privacy Protection (\textbf{R4})}
Twelve participants (36.4\%) strongly agreed and 18 (54.5\%) somewhat agreed that their privacy was secure, while three (9.1\%) were neutral or disagreed. Similarly, 14 (42.4\%) strongly and 16 (48.5\%) somewhat agreed that data usage was transparent.
Most participants trusted the system's privacy protection due to clear communication and technical safeguards like data anonymization and secure storage. Exp. participants particularly appreciated the transparency of data handling (P15, exp.). Both groups valued visual privacy measures like background blurring, which ``\textit{made everything except the food unrecognizable}'' (P9, inexp.). However, some concerns remained about external influences: ``\textit{While I wasn't worried, my friend sitting next to me was}'' (P27, inexp.). ``\textit{Even though the speech was unrecognizable, I was still reserved when discussing sensitive topics}'' (P26, exp.).
In general, exp. participants showed greater confidence and understanding of privacy protocols, while inexp. participants required more reassurance. These findings highlight the importance of improving privacy communication to address privacy concerns and enhance trust.

\paragraph{General Usability, Satisfaction, and Experience} Our evaluation revealed broadly positive usability, satisfaction, and user experience across both exp. and inexp. participants. Exp. participants consistently rated the system higher in terms of ease of use, automation, and efficiency, while inexp. participants highlighted challenges in adapting to workflows and interacting with the system. Satisfaction scores averaged 8.3 (NPS), with high utility acknowledged by most participants. However, the physical discomfort from prolonged wearing of glasses (such as heat buildup around the glasses' legs, sweat accumulation, and slippage due to the limited glasses sizes provided) and workflow complexity emerged as areas for improvement, particularly for less tech-savvy users. Details on these assessments, including SUS, NPS, and UEQ metrics, are provided in Appendix \ref{appx:general}.

\paragraph{User Interactions and Engagement}
Participants spent an average of 10.10 minutes per session (2.59 minutes per meal).
Reviewing dietary suggestions averaged 2.82 minutes. Inspecting nutritional analysis averaged 5.05 minutes, indicating participants prioritized understanding their diet over simply viewing suggestions.
Manual corrections were minimal, with 27 changes across 144 meal sessions, mostly involving beverages (e.g., from ``\textit{Coke}'' to ``\textit{Coke Zero}''). Users also corrected misidentified items or cooking methods (e.g., ``\textit{grilled pork}'' to ``\textit{grilled duck}'' or ``\textit{steamed}'' to ``\textit{fried}''). Despite features designed to reduce errors of including nearby meals, occasional deletions of extra items indicate room for improvement.
Participants also engaged actively with the chatbot, logging 125 conversation pairs (average of 3.79 per participant), primarily for meal planning with constraints such as ``\textit{in Chinese style}.'' 
Iterative interactions, like P6 refining meal plans to exclude breakfast or avoid certain foods (full chat history is available in Appendix \ref{appx:chat}), demonstrated the chatbot's adaptability in addressing preferences and fostering personalized dietary management.
Overall, the system's core features were actively utilized, with strong engagement enhancing the personalized user experience and demonstrating the system's value in dietary monitoring and analysis.

\paragraph{Comparison with Other Dietary Tracking Technologies}
Participants highlighted that the system's hands-free, automatic tracking made it more convenient than many mobile apps requiring manual logging.
Additionally, participants valued the interactive features of our system, which allowed them to follow up on initial analysis. Unlike traditional apps with static data entry, our system allowed users to ask for more information. This dynamic, engaging experience enabled users to actively adjust their dietary behavior based on continuous feedback, enhancing their understanding of nutrition and empowering better food choices.

\subsection{Study I.2: Quantitative Evaluation with Crowd Workers on Diet Identification and Logging Performance (\textbf{R1})}
With the successfully captured meal images for all 144 sessions, we assessed diet identification accuracy by comparing the ground truth food and drink items with the text descriptions inferred by the LLM from the processed images.
However, this agreement can be complex and ambiguous due to the intricate nature of the dishes. 
For instance, a bowl of noodles with soup, carrots, and beef could be viewed as one item (the whole dish), three items (noodle soup, carrots, beef), or four distinct components (noodles, soup, carrots, beef). Text descriptions may only mention ``\textit{noodle soup and beef},'' leading to varying accuracy scores: 0/1 (missed the whole dish), 2/3 (correctly identified two out of three items), or 3/4 (identified three out of four items).
Besides, some dishes, like Malatang or Hot Pot, are especially challenging to identify every ingredient, even for humans. Therefore, we set the ground truth based on items visually identifiable by humans. We used a crowdsourcing approach to quantitatively evaluate the diet identification and logging performance.

\subsubsection{Evaluation Settings}
We used a crowdsourcing approach, presenting processed meal images (extracted from the video but cropped for privacy) alongside the LLM-generated text descriptions to crowd workers recruited from AWS MTurk \cite{cheung2017amazon} who met specific qualification criteria i.e., Human Intelligence Task (HIT) Approval Rate $\geq 90\%$ and Approved HITs $\geq 500$. Crowd workers worked independently. Specifically, due to the complexity of dishes and the high burden of measuring exact ingredient quantities, we employed an indirect evaluation of diet item quantity estimation by evaluating the nutrient estimation performance, since nutrient estimation involves quantity inference, as detailed in Sec.~\ref{subsec:eva_nutrition}.

\subsubsection{Evaluation Metrics}

The crowd workers were tasked with verifying the alignment between the images and the generated text by counting the following:

\begin{itemize}
    \item Correct Identified Items ($C$): Items in the ground truth images correctly identified in the text description.
    \item Image Items ($I$): Items presented in the meal log images.
    \item Text Items ($T$): Items mentioned in the text description. 
\end{itemize}

We filtered out those unqualified answers with the following rules:

\begin{itemize}
    \item The number of correct items ($C$) should not exceed that of items in images ($I$) or text ($T$), i.e., $C\leq I$ and $C\leq T$.
    \item The numbers of items in images ($I$) or text ($T$) should be greater than zero, i.e., $I>0$ and $T>0$.
    \item Task completion time must be at least 60 seconds. 
    \item The number of text items must align with the number of items after splitting the text using commas.
\end{itemize}

Submissions failing these quality checks were automatically re-released to new workers to maintain the target sample size. Each meal was labeled by 10 qualified crowd workers. Workers could get \$0.1 for each qualified label. We then calculated precision, recall, and F1-score.

\subsubsection{Results}
Based on the crowd worker labels, the proposed system achieved an average F1-score of 0.972, with a precision of 0.957 and a recall of 0.989, demonstrating the effectiveness of \shortname in diet identification. 
The meal that achieved the lowest precision (0.643) included a bowl of noodles with various ingredients, such as blood cubes, bean sprouts, sliced meat, green onions, and tomatoes, making it difficult for LLM to correctly identify the ingredients from the processed meal log images. 
The meal with the lowest recall (0.853) contained a bowl of rice with grilled fish, lemonade (with small pieces of onion and Seaweed on the top), a bowl of clear soup, and a cup of milk tea. \shortname failed to describe those small ingredients. 
Overall, the system performed well, particularly on Western-style dishes with clearly separated components. Highly complex meals like Malatang and Hot Pot did not yield the lowest F1-scores; this might be because their ingredients are very difficult to describe even for the crowd workers.

\subsection{Study I.3: Quantitative Evaluation with Experts on Nutritional Analysis Performance (\textbf{R2})}
\label{subsec:eva_nutrition}
We held a panel with Food and Nutrition Science experts who highlighted the difficulty of determining nutrient values due to factors like ingredient freshness, cooking methods, portion sizes, storage conditions, and ingredient interactions during cooking \cite{willett2012nutritional}, etc.
Despite this variability, experts can estimate nutrients using food composition databases, standardized recipes, and their experience with similar meals. 
\subsubsection{Evaluation Settings}
Recognizing the variability in nutrient estimation, we recruited ten food and nutrition experts, including six nutrition interns, three researchers, and one Accredited Practicing Dietitian. They independently analyzed meal data (the ground truth meal images across the eating process), referencing the database provided by the Centre of Food Safety of the Government of the Hong Kong SAR database, with each meal randomly assigned to five experts for estimation. Each expert received about \$25 for compensation.

\subsubsection{Evaluation Metrics} We measured expert agreement using the Intraclass Correlation Coefficient (ICC), showing high reliability of the average estimation from experts with ICC(2,k) > 0.9 (more details in Appendix \ref{appx:agreement}). 
Therefore, to evaluate \shortname's performance, we compared its estimation to the expert averages using Mean Absolute Percentage Error (MAPE) due to varying nutrient units and scales, calculated as: ${MAPE = \frac{1}{N}\sum_{n=1}^{N}|\frac{\overline{E}_n-S_n}{\overline{E}_n}|\times 100\%}$,
where $N$ is the number of meals. $\overline{E}_n$ and $S_n$ are the average estimations of experts and the system, respectively.

    \begin{figure}[!t]
          \centering
          \includegraphics[width=.9\textwidth]{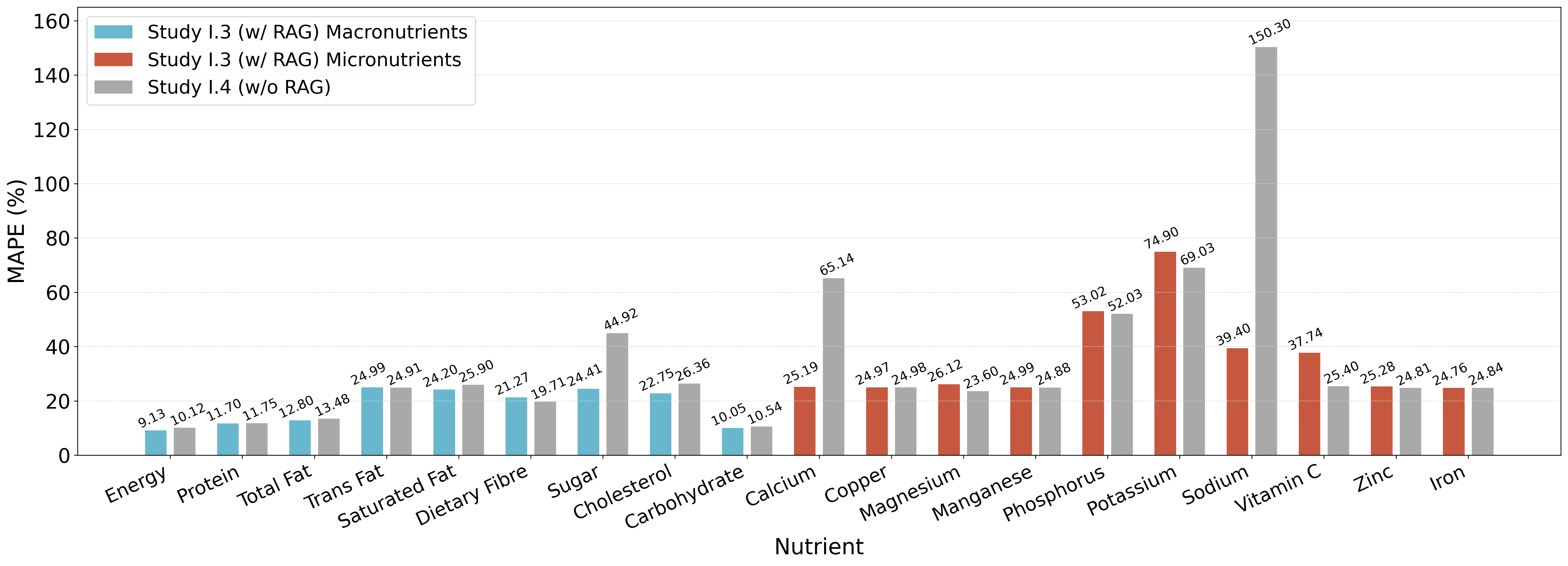}
          \caption{The results of MAPE (\%) in Studies I.3 and I.4. Lower values indicate better performance.}
          \label{fig:mape}
           % \vspace{-0.5cm}
      \end{figure}

\subsubsection{Results}
As shown in \figurename{~\ref{fig:mape}}, the system demonstrated high accuracy for key macronutrients, including energy (9.13\%), protein (11.70\%), total fat (12.80\%), and carbohydrates (10.05\%). These macronutrients are major food components and are typically well-documented in databases, making them easier for \shortname to estimate accurately. Given the challenges in nutrient estimation, such as variations in food composition databases, portion size estimation errors, and the difficulty of accurately assessing mixed dishes, a MAPE around 10\% is considered a strong result \cite{kipnis2002bias, thames2021nutrition5k}.

Moderate accuracy was achieved for nutrients like calcium (25.19\%), iron (24.76\%), cholesterol (22.75\%), dietary fiber (21.27\%), and zinc (25.28\%), possibly due to their greater variability across food sources and bioavailability, complicating precise estimation. The system exhibited poor alignment with experts' estimation of phosphorus (53.02\%) and potassium (74.90\%). These nutrients are known to exhibit significant variability across different foods due to a variety of factors, such as plant and animal growth conditions, differences in food type, preparation, and source \cite{world2004vitamin}. This variability, combined with the effects of processing and cooking methods, makes accurate estimation particularly challenging.

Overall, the system highlighted its strengths in estimating macronutrients with an average MAPE of 17.92\% but faced difficulties with more variable micronutrients (MAPE=35.64\%).

\subsubsection{Post-Analysis Discussion with Experts}
We held a panel discussion with the experts, moderated by the lead researchers, to interpret the quantitative results. The discussion was structured around specific guiding questions focusing on two themes: (1) reasons for low performance in specific nutrients and (2) key nutrients for dietary analysis and suggestions.

(1) Experts observed a lack of detailed ingredient information and documentation, especially for compound dishes like Chinese-style food, where oils, seasonings, and cooking methods significantly affect nutrient content. 
For example, the official nutrition database used for estimations often failed to include many compound dishes, requiring experts to estimate ingredients separately without accounting for preparation methods, such as temperature and seasoning.
This lack of detail led to inaccuracies in estimating micronutrients like potassium and phosphorus, as the types of vegetables or seafood can significantly influence these calculations but were difficult to distinguish from the images.

(2) The experts valued the system's overall performance for daily dietary tracking, since the important nutrients, such as the macronutrients, calcium, and iron, have a relatively lower MAPE. While all nutrients are important, they noted that potassium and phosphorus, which showed poor performance, are typically less emphasized in general dietary analysis. 
Experts suggested improving the accuracy of magnesium, vitamin C, and sodium estimates and highlighted potential biases due to dish complexity and portion size estimation. 
However, higher accuracy was expected in clinical contexts where more precise nutrient intake is essential. In this context, the important nutrients depend on the purpose of conducting a nutritional analysis. For example, potassium and phosphorus are critical to monitor in patients with chronic kidney disease, who often must strictly limit their intake of these minerals.

\subsection{Study I.4: Comparative Analysis for \shortname with and without RAG (\textbf{R2}, \textbf{R3})}
To demonstrate the effectiveness of the RAG module, we conducted a comparative analysis to explore the \shortname's performance on nutritional analysis and personalized dietary suggestions with (w/) and without (w/o) RAG, including (1) comparing the nutritional analysis results of \shortname w/ RAG (Study I.3) and w/o RAG, and (2) comparing the quality of dietary suggestions w/ and w/o RAG.

\subsubsection{Comparing the nutritional analysis results of \shortname w/ RAG (Study I.3) and w/o RAG}
We compared the MAPE of nutrition estimation w/o RAG with the results in Study I.3. As shown in \figurename{~\ref{fig:mape}}, the system w/ RAG presented lower or close MAPE values in most of the nutrients compared to the system w/o RAG. The system w/o RAG resulted in significantly higher errors in sugar, calcium, and sodium. This indicates that RAG enhances the system's predictive accuracy, especially for some nutrients that could be hidden in sauces, condiments, and bone broth. Since most of the meals collected were Chinese-style, which tend to have higher sodium content due to the frequent use of sodium-rich condiments and seasonings, the lack of the RAG module likely exacerbated the system's inability to account for the variability in sodium levels.
The results demonstrated that while the LLM w/o RAG could achieve satisfactory performance on some nutrients, the incorporation of RAG further enhanced the system's nutritional analysis capacity, particularly in handling the complexities and variabilities of diverse and culturally sensitive diets.

\subsubsection{Comparing dietary suggestion quality w/ and w/o RAG}

With the data of 33 participants collected in Study I.1, we obtained two versions of dietary suggestions w/ and w/o the RAG module and compared the quality of dietary suggestions from two perspectives: (1) widely-used metrics for generative AI and (2) expert evaluation.

\textbf{Metrics for Generative AI.} AI-assisted metrics that are widely used for evaluating the AI-generated responses \cite{li2024leveraging} were used to evaluate the quality of dietary suggestions. GPT-4 was used as a grader to rate the responses with a score between 1 and 5, where 1 is the lowest quality and 5 is the highest quality. To compare the responses w/ and w/o RAG, the following metrics from Azure AI Evaluation SDK \cite{Microsoft_AzureAIEvalSDK_2025} were used:

\begin{itemize}
    \item \textit{Relevance-Q} measures the accuracy, completeness, and direct relevance of the response given the query.
    \item \textit{Coherence} measures the logical and orderly presentation of ideas in a response.
    \item \textit{Fluency} measures the effectiveness and clarity of written communication.
\end{itemize}
Since suggestions w/ and w/o RAG were evaluated independently, we used G$^*$Power \cite{faul2009statistical} to determine the sample size by setting the effect size $d=0.5$ (moderate effect), the significant threshold $\alpha=0.05$, and a statistical power $(1-\beta)=0.8$, resulting in a sample size of 128. Each document was graded twice independently, resulting in 66 grades per group (132 total). We then conducted the Mann-Whitney U Test with a significance level of 0.05 to compare the \textit{Relavance-Q}, \textit{Coherence}, and \textit{Fluency} of dietary suggestions generated w/ and w/o RAG.

\textbf{Expert Evaluation.} We employed a pairwise comparison method using 33 dietary suggestions pairs, each consisting of one version w/ RAG and one w/o. Three experts from Study I.3 were recruited with about \$12.9 compensation. They were independently tasked with comparing the paired documents across the following dimensions:

\begin{itemize}
    \item \textit{Relevance-G} measures if the suggestions are tailored to the dietary goals and profile of the participant.
    \item \textit{Actionability} measures if the recommendations are practical and easy to implement for users.
    \item \textit{Accuracy} measures if the suggestions are factually correct and nutritionally valid.
\end{itemize}
To minimize bias, we hid the sources for the RAG version, ensuring consistent output formats. The documents in pairs were randomly labeled as ``A'' and ``B'', with the order of presentation varying randomly for each expert. 
Each pair was evaluated independently by the experts.
The score scale was also between 1 and 5, with 1 being the lowest quality. After obtaining the scores, we conducted statistical analysis and discussed with experts to justify the results.

\textbf{Results.}
The Mann-Whitney U Test revealed that w/-RAG suggestions ($M=3.80$, $SD=0.92$) had significantly higher \textit{Relevance-Q} ($U=2655.5$, $p=0.01$) than w/o-RAG ones ($M=3.39$, $SD=0.81$).
W/-RAG suggestions ($M=4.65$, $SD=0.62$) also had significantly higher \textit{Coherence} ($U=2555.0$, $p=0.04$) than w/o-RAG ones ($M=4.38$, $SD=0.79$). W/-RAG suggestions ($M=4.13$, $SD=0.34$) also outperformed w/o-RAG suggestions ($M=4.0$, $SD=0.00$) in \textit{Fluency} ($U=1881.0$, $p<0.01$). 
The results showed that while both dietary suggestions w/ and w/o RAG achieved high grades in these metrics, the dietary suggestions w/ RAG had significantly higher relevance to the query, coherence, and fluency.

For the expert evaluation, we first assessed inter-rater agreement among the three experts using ICC(2,k), yielding a value of 0.893, indicating good consistency. Based on this, we averaged the experts' ratings and conducted the Wilcoxon Signed-Rank Test at a significance level of 0.05. Calculated using G$^{*}$Power, the achieved power was 0.795, with an effect size of $d=0.5$ and a sample size of 33 pairs.
The results showed that while the mean \textit{Relevance-G} of w/-RAG suggestions ($M=4.23$, $SD=0.70$) was slightly higher than w/o-RAG ones ($M=4.21$, $SD=0.81$), the difference was not significant ($W=37.0$, $p=0.87$). Experts noted that although RAG suggestions were tailored, some contained overly complex or nuanced recommendations, potentially reducing their perceived relevance, likely due to the inclusion of many academic articles in the nutrition library.
For \textit{Actionability}, with-RAG suggestions ($M=4.25$, $SD=0.37$) showed a slight but not significant improvement over without-RAG ones ($M=4.09$, $SD=0.31$, $W=26.5$, $p=0.06$). Experts observed that RAG suggestions often included practical advice (e.g., meal preparation strategies) but were sometimes challenging to implement.
Finally, \textit{Accuracy} was significantly higher ($W=32.0$, $p=0.01$) for with-RAG suggestions ($M=4.78$, $SD=0.27$) compared to without-RAG suggestions ($M=4.62$, $SD=0.36$). Experts emphasized that RAG-based suggestions were factually accurate and nutritionally valid, while without-RAG suggestions, though accurate, lacked the context-specific precision offered by RAG.

In summary, this comparative analysis demonstrates the specific and critical advantages of integrating the RAG module. While the baseline LLM (w/o RAG) could generate plausible responses, the RAG-grounded system showed significantly better performance in areas essential for a health application. Specifically, for nutritional analysis, RAG provided more accurate estimations for key nutrients like sugar, calcium, and sodium. For dietary suggestions, the RAG-enhanced version received significantly higher expert ratings for Accuracy. Furthermore, unlike the baseline, the RAG module provides references for its claims, ensuring that the system's advice is traceable and verifiable, which is a critical enhancement for preventing misinformation in health-sensitive applications.

\section{Study II: Four-Week Longitudinal Study}
To evaluate the long-term effect of \shortname on users' dietary patterns, we recruited 16 participants from Study I in Hong Kong for a 4-week longitudinal study with IRB approval, including 4 females and 12 males (P1, P6, P7, P10, P13, P15, P18, P20, P21, exp., P3, P8, P9, P14, P16, P17, P22, inexp.), aged 24 to 33 ($M=28.3$, $SD=2.4$).
Building on the design requirements and findings from Study I, this study aims to explore how exp. and inexp. participants perceive and use \shortname over time, and how it affects their dietary patterns.

\subsection{Study Settings}

Before conducting the longitudinal study, we made system updates based on feedback from Study I:
\begin{itemize}
    \item Added frequently asked questions to the chatbot's common questions list.
    \item Incorporated users' habits (e.g., drinking Coke Zero instead of regular Coke) into prompts to refine diet identification.
    \item Asked the participants to report the number of people sharing a meal, adjusting consumption proportion accordingly (assume equal portions for all).
\end{itemize}

During the study, participants were encouraged to use Aria Glasses covering as many ingestive behaviors (meals and snacks) as possible to gain comprehensive insights into their habits and use the interface to review their dietary monitoring results. 
The interface was available 24/7, except from 12:00 am to 5:59 am for potential maintenance and backup, and was tracked via Google Analytics. 
Meal data older than seven days was archived, with analysis based on the most recent seven days to reflect current dietary patterns. Participants provided weekly feedback through questionnaires and interviews (details in Appendix \ref{appx:scale}).
Participants received about \$17 every week and could review their monitoring and analysis results whenever the interface was available.

We presented the descriptive statistics of the participants' ratings towards the four design requirements, general usability, user satisfaction, user experience, and nutrient consumption, to show the trends across the four weeks. For interviews and textual data, the lead author performed the initial thematic coding and iteratively discussed with other authors to finalize the codes to analyze the impact of \shortname on participants' dietary behaviors and their suggestions for improvement.

\subsection{Results}
\label{sec:study_II_results}
During the study, 821 meal or snack sessions were involved ($M=51.3$, $SD=13.6$).
The variation in session collected was mainly due to individual dietary habits (e.g., some participants skipped meals like breakfast) and practical challenges. Specifically, the Aria Glasses had limited battery life, and some participants had vision correction needs. We were unable to replace the lenses to match participants' prescriptions, which affected their ability to consistently wear the glasses and log every meal, especially in some formal dining occasions involving social or business gatherings. To accommodate these challenges and reduce participant burden, we allowed occasional missed sessions to maintain engagement. As a result, a total of 245 meal sessions (23.0\%, 15.3 per participant) were missed.

Even though some participants had fewer sessions, analyzing these data still revealed consistent patterns and unique behaviors, offering insights into dietary practices. This approach captures real-world variability, providing a more comprehensive understanding of different eating habits. Detailed analysis results are discussed below.

 \begin{figure*}[!t]
          \centering
          \subfloat[]{
            \centering
            \includegraphics[width=.25\textwidth]{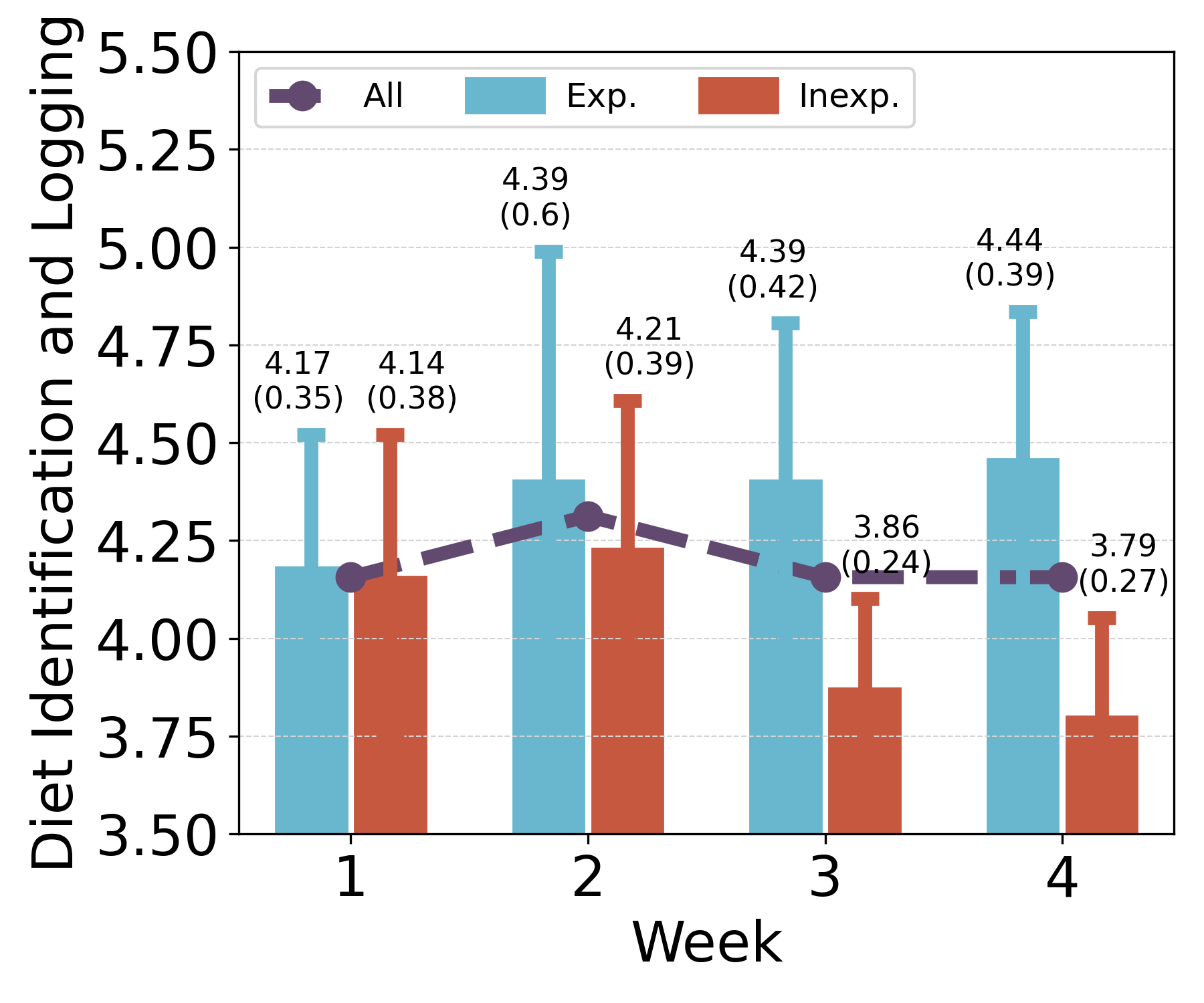}
            \label{fig:r1}
          }
          \subfloat[]{    
                  \centering
                  \includegraphics[width=.25\textwidth]{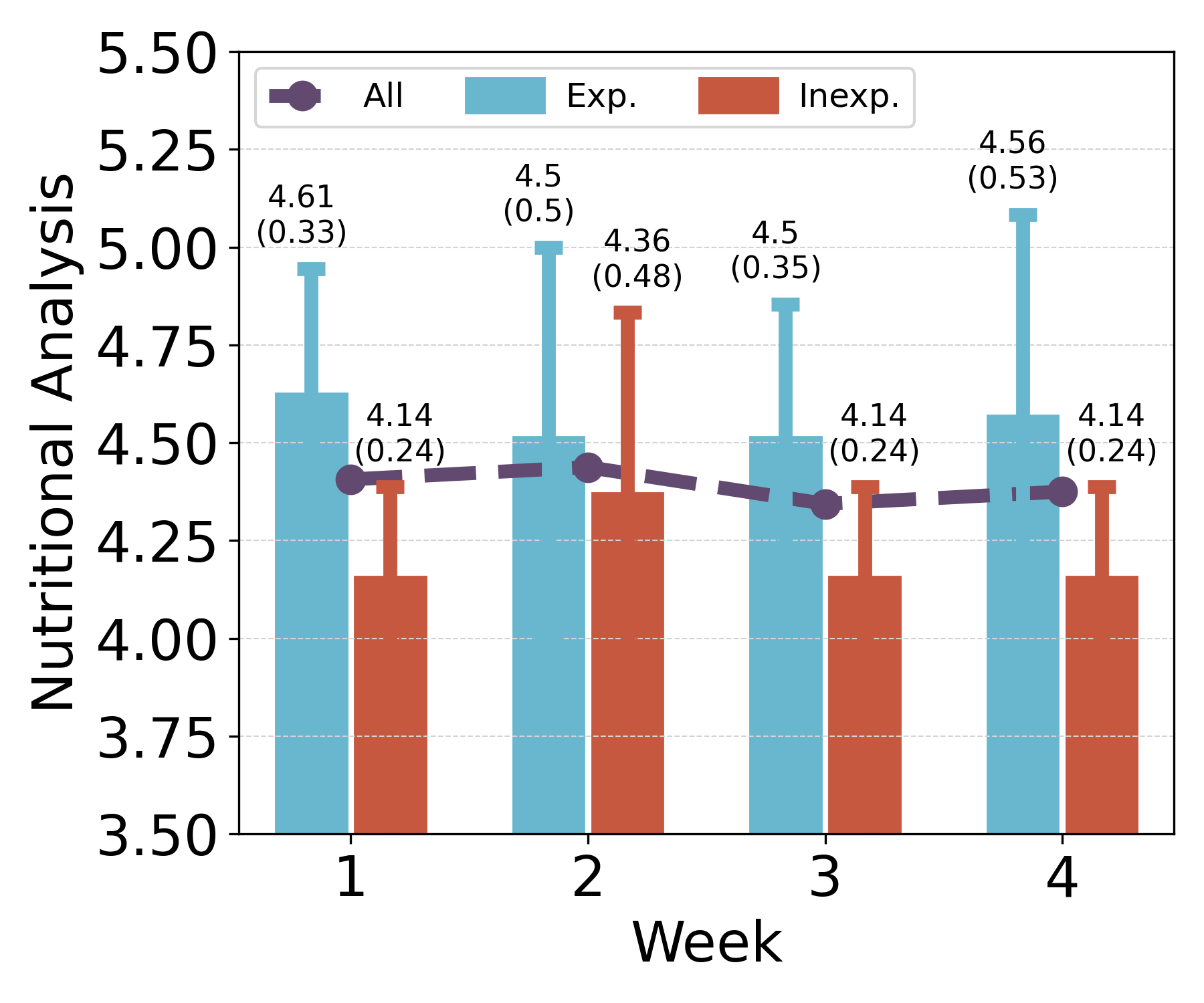}
              \label{fig:r2}
          }
          \subfloat[]{    
                  \centering
                  \includegraphics[width=.25\textwidth]{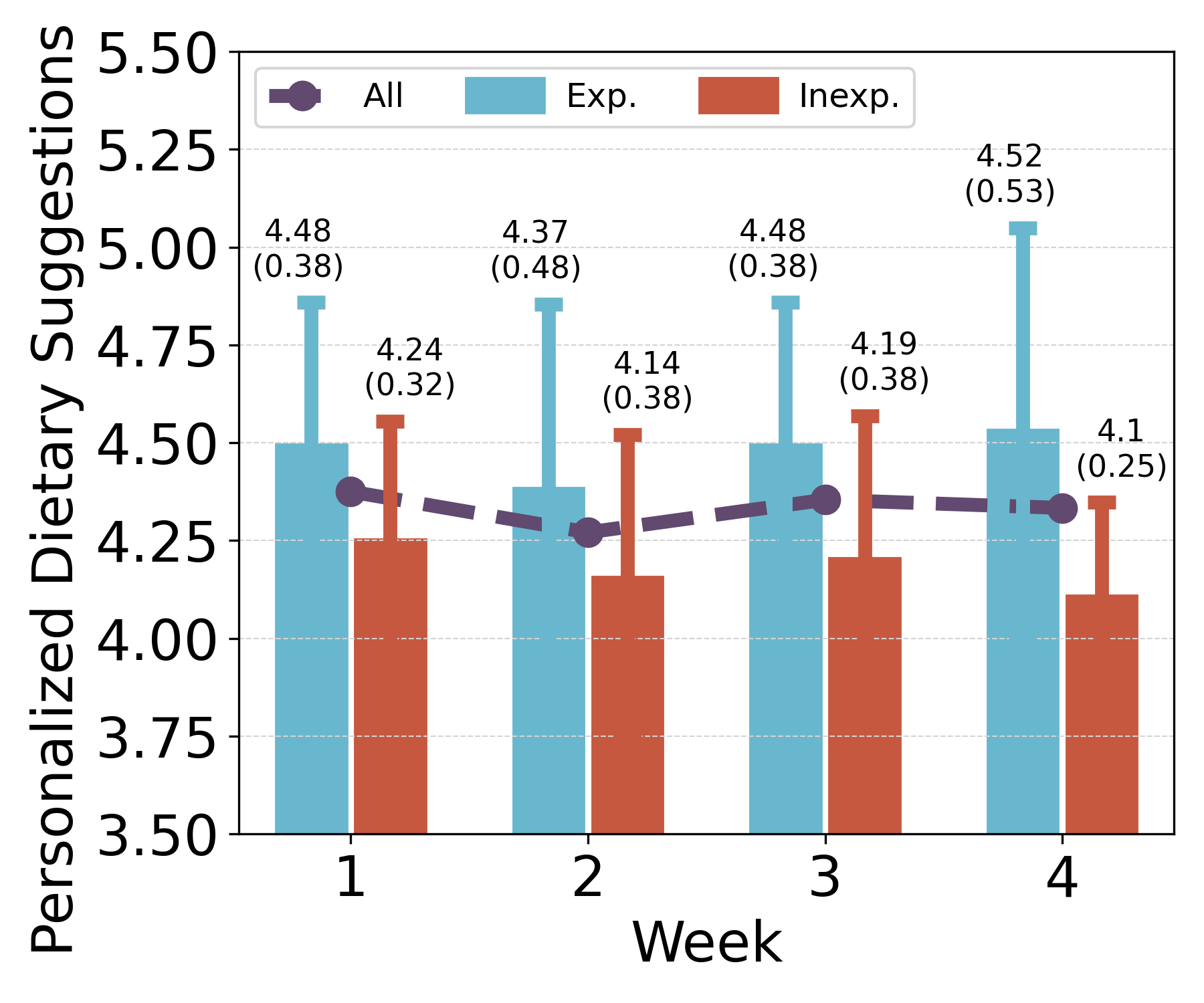}
              \label{fig:r3}
          }
          \subfloat[]{    
                  \centering
                  \includegraphics[width=.25\textwidth]{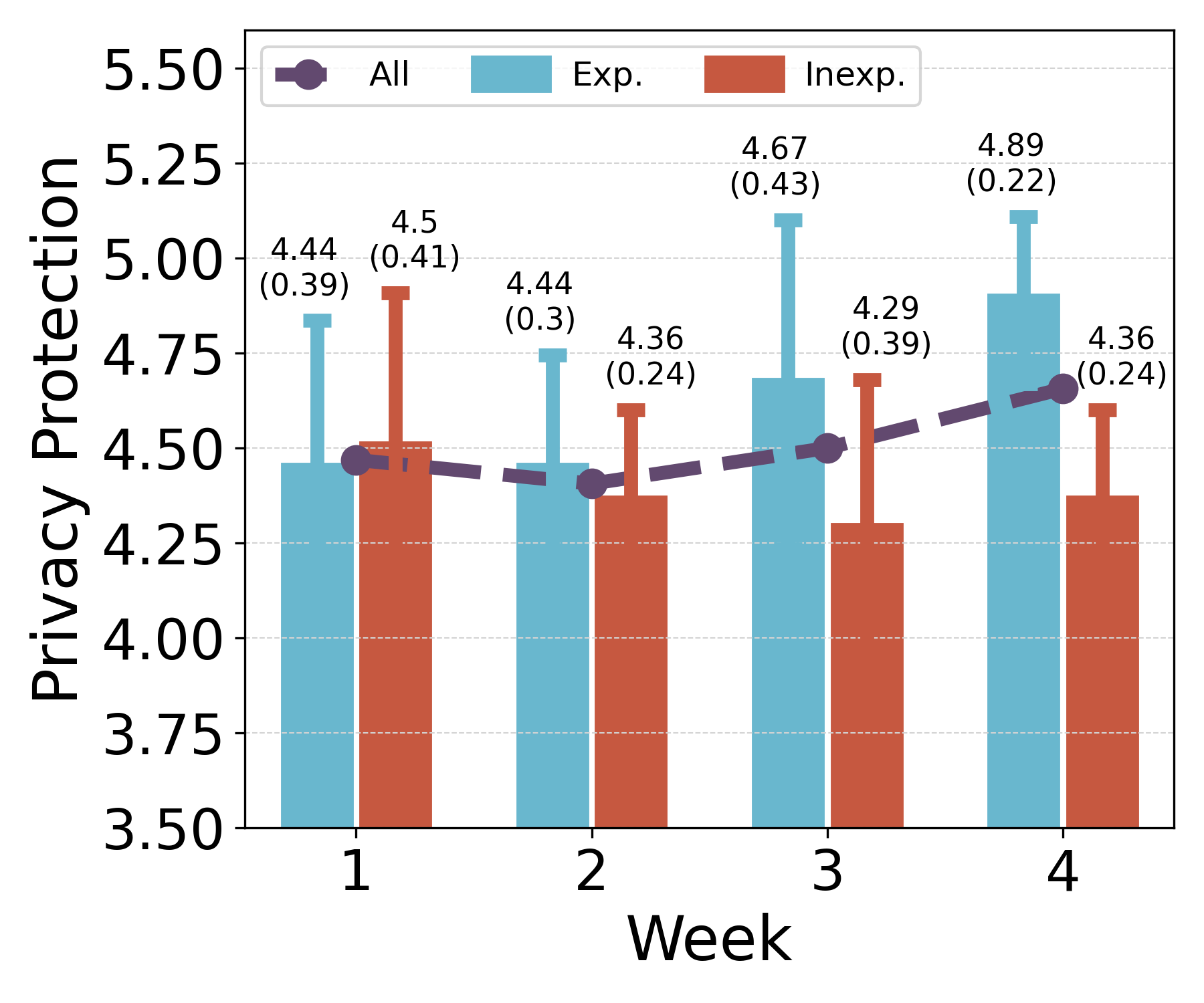}
              \label{fig:r4}
          }
          \caption{The weekly trends of the scores of exp., inexp., and all participants on (a) Diet identification and logging, (b) Nutritional Analysis, (c) Personalized Dietary Suggestions, and (d) Privacy Protection. The values above the bars are the mean (standard deviation).}
          \label{fig:scales-r}
           % \vspace{-0.5cm}
      \end{figure*}

\subsubsection{Long-Term Perception on Diet Identification and Logging, Nutritional Analysis, Personalized Dietary Suggestions, and Privacy Protection}

\figurename{~\ref{fig:scales-r}} shows that generally the ratings of exp. participants were higher than inexp. participants, except for the privacy protection in the first week. While exp. participants demonstrated steady or increasing satisfaction, particularly for privacy protection and personalized suggestions, inexp. participants often struggled, with ratings plateauing or declining over time, as seen in diet identification and logging. Privacy protection scores remained consistently high and even increased, reflecting growing user trust in the system's privacy safeguards. The ratings of the diet identification and logging and nutritional analysis decreased, potentially due to the increasing variety of diets. 
Personalized dietary suggestions showed a notable gap, with exp. participants finding them more effective, suggesting that exp. participants are likely more skilled in leveraging the system's features to tailor them effectively to their specific needs and preferences.

\subsubsection{Usefulness}
There were nuanced differences in how exp. and inexp. participants engaged with the system.
In the first week, participants highlighted the accuracy of diet identification, with its success in identifying less common or culturally specific dishes, such as soba noodles (P21) and Hainanese chicken (P15).
Throughout the study, nutritional analysis and dietary suggestions were consistently valued for providing insights into nutrient consumption and deficiencies (P15, P18). 
Exp. participants appreciated source-backed suggestions and the ability to track nutrition trends (P6), while both groups valued meal timing feedback, like reminders to eat regularly and slow down (P17, P15, P20).
The AI chatbot was helpful in planning meals and offering actionable nutritional knowledge (P1, P7, P15, exp.). The chatbot's professional tone and cited sources made it feel like ``\textit{chatting with a real nutritionist}'' (P16), suggesting its effectiveness in bridging gaps in users' nutritional knowledge.
Meal log images were useful to both groups for meal recall, with exp. participants tending to use them to verify and refine diet summaries. These differences highlight exp. participants' active interrogation of the system and the supportive reliance of inexp. users.

\subsubsection{Long-Term General Usability, User Satisfaction, and User Experience}
Usability improved throughout the study, with SUS scores rising from 76.09 in Study I to 85.00 by the end of Study II, reflecting increased perceived usability. Exp. participants consistently rated the system higher, while inexp. ones showed steady improvement, highlighting adaptability. NPS scores also rose, indicating sustained satisfaction across both groups. User experience, measured through UEQ dimensions, showed positive trends in attractiveness, efficiency, and dependability, while stimulation and novelty declined slightly, reflecting reduced novelty as users became familiar with the system. These findings emphasize the \shortname's high usability, satisfaction, and user experience, highlighting the need for tailoring onboarding processes and maintaining engagement, particularly for inexp. users. More details are in Appendix \ref{appx:long_term}.

\subsubsection{Long-Term Interaction Behaviors with \shortname}
Participants demonstrated consistent engagement with the system over four weeks, averaging 38.13 usage sessions. Exp. showed slightly higher engagement than inexp. ones. Initial novelty drove higher interactions in Week 1, which gradually stabilized into routine usage by Week 3. The average engagement time per session was 5.95 minutes, with exp. participants spending longer on average, reflecting deeper interaction with the system. The relatively low number of logged updates to diet summaries highlighted the system's accuracy, while sustained chatbot interactions demonstrated its role in enhancing personalization and feedback. More details are available in the Appendix \ref{appx:behavior}.

 \begin{figure*}[!t]
          \centering
          \subfloat[Energy$^{***}$]{
            \centering
            \includegraphics[width=.19\textwidth]{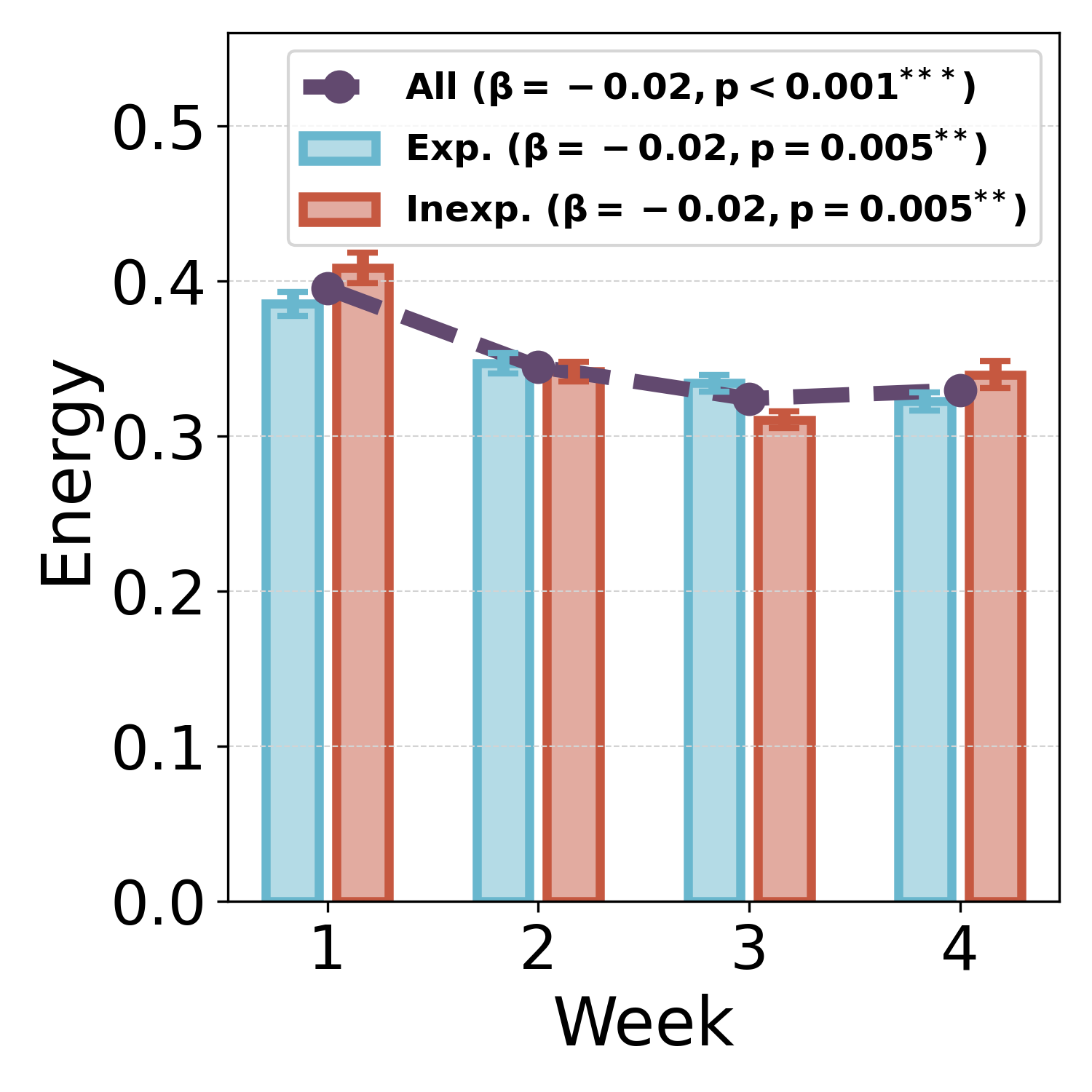}
            \label{fig:energy}
          }
          \subfloat[Protein$^{*}$]{    
                  \centering
                  \includegraphics[width=.19\textwidth]{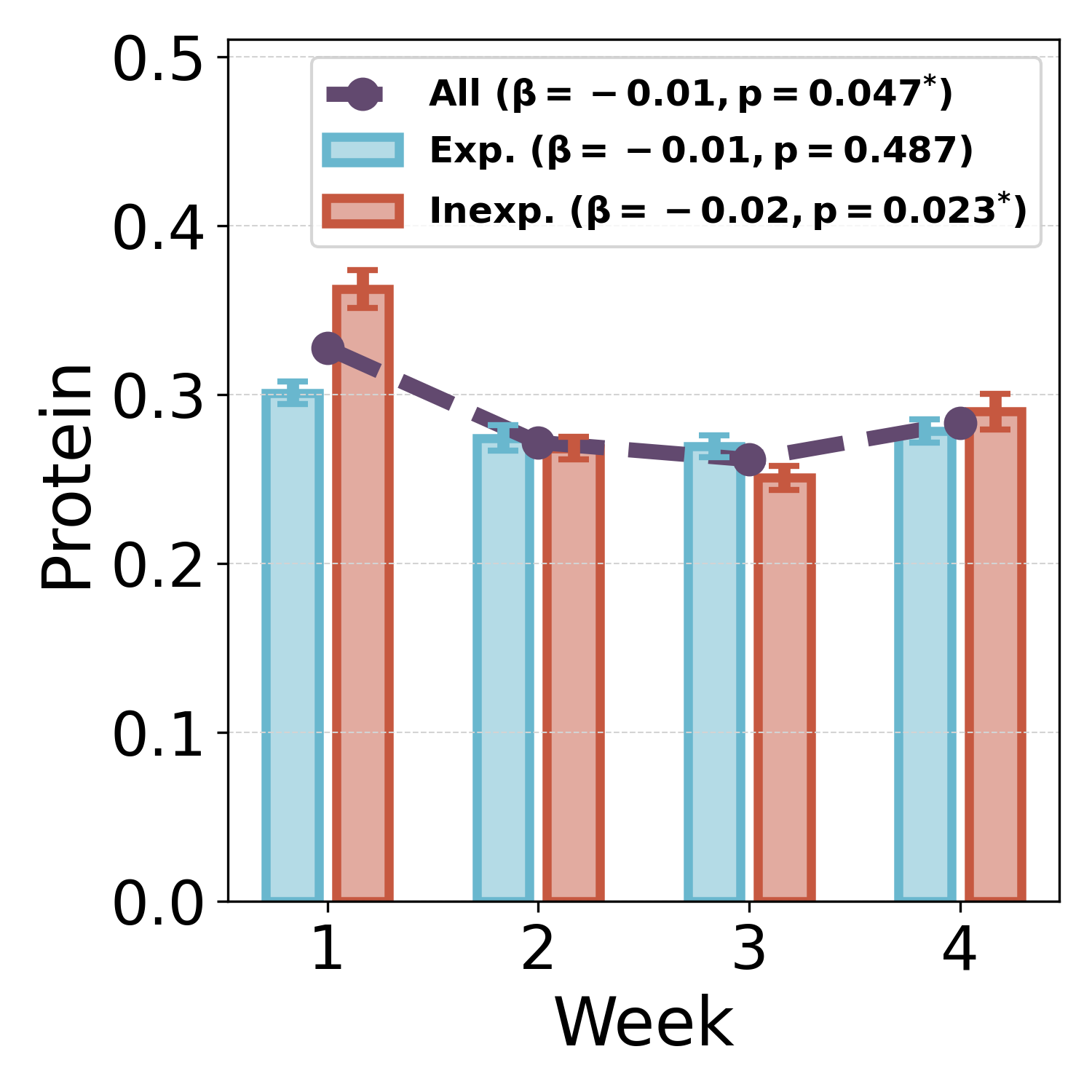}
              \label{fig:protein}
          }
          \subfloat[Total Fat$^{***}$]{    
                  \centering
                  \includegraphics[width=.195\textwidth]{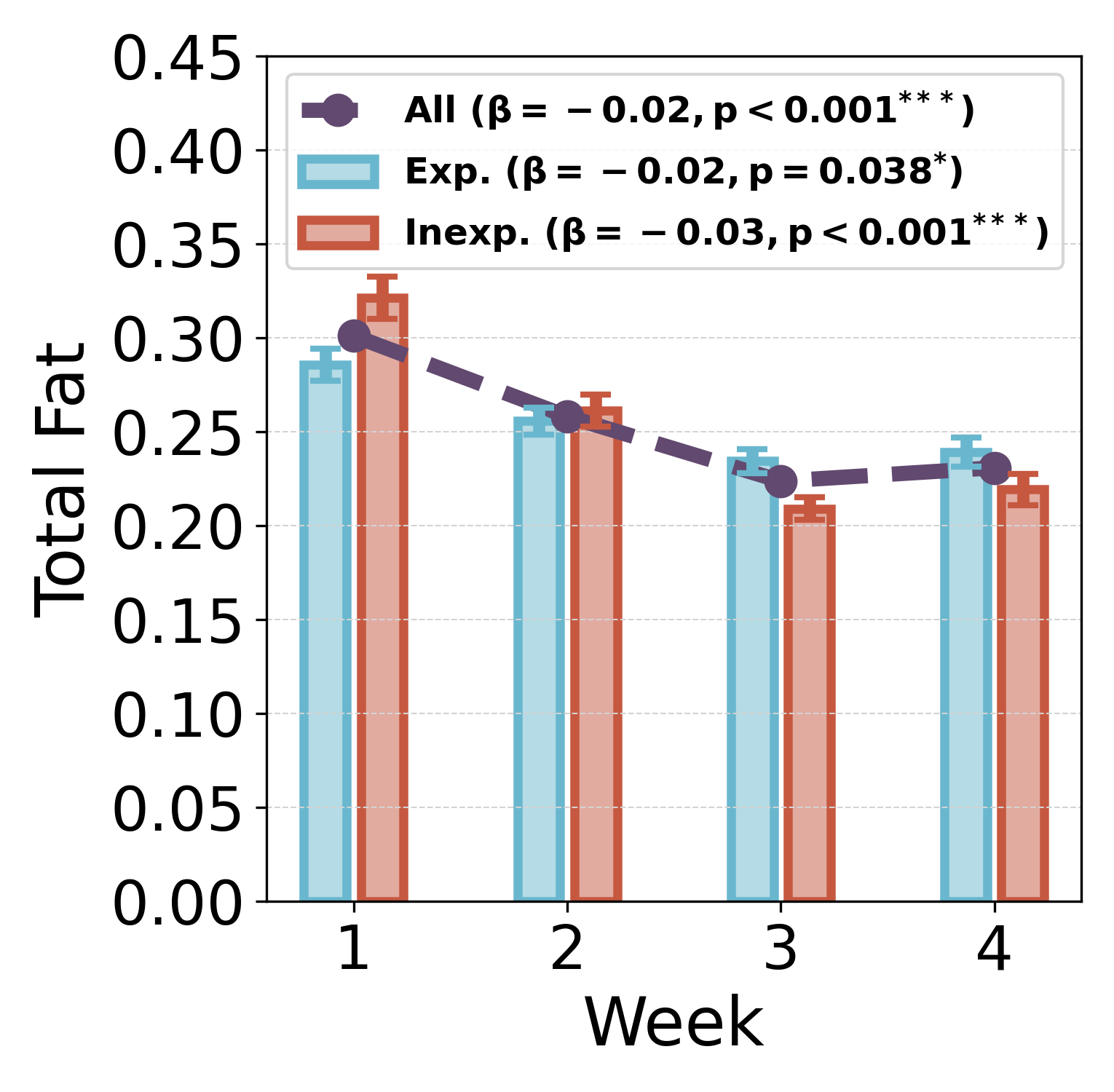}
              \label{fig:total_fat}
          }
          \subfloat[Trans Fat$^{*}$]{    
                  \centering
                  \includegraphics[width=.189\textwidth]{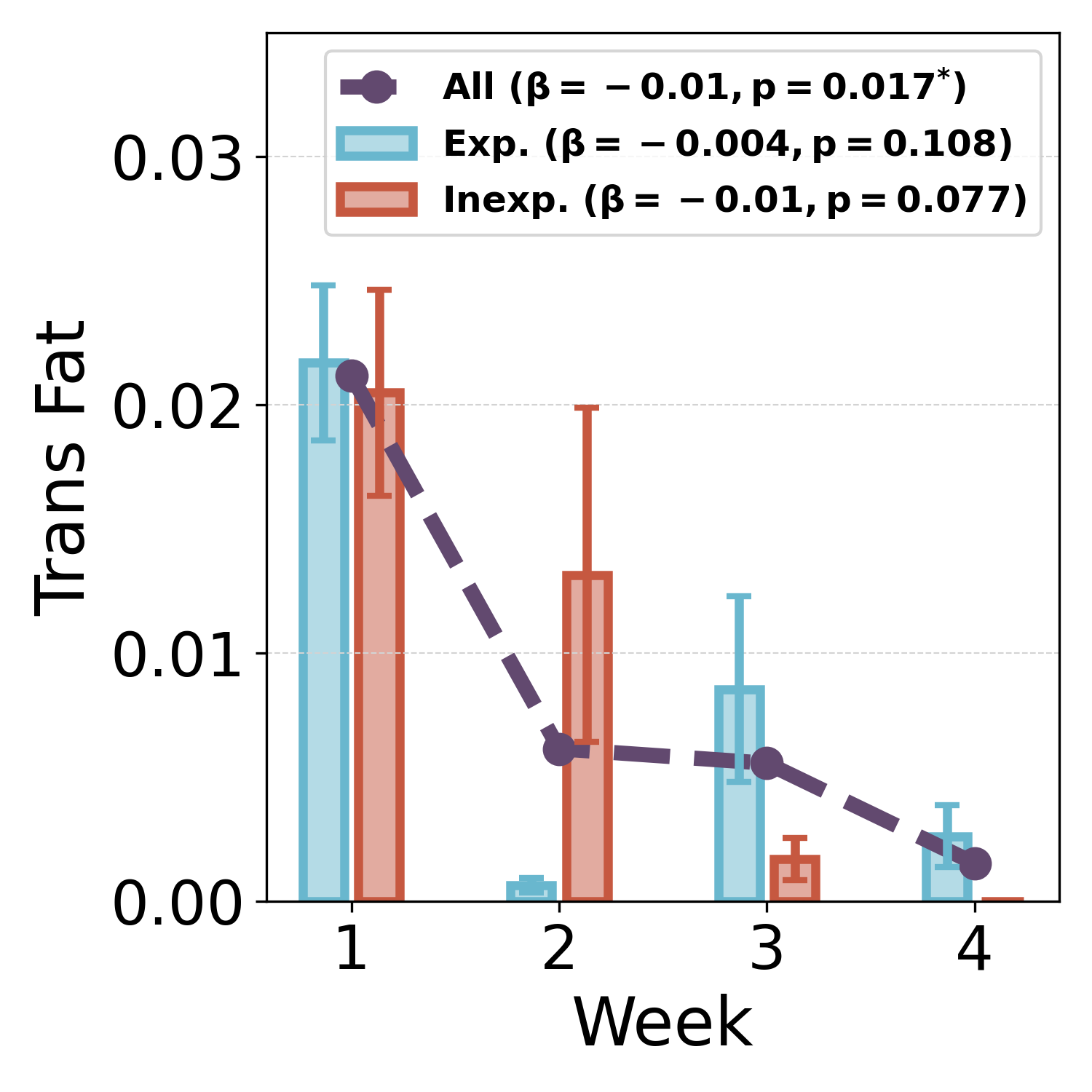}
              \label{fig:trans_fat}
          }
            \subfloat[Saturated Fat$^{***}$]{    
                  \centering
                  \includegraphics[width=.195\textwidth]{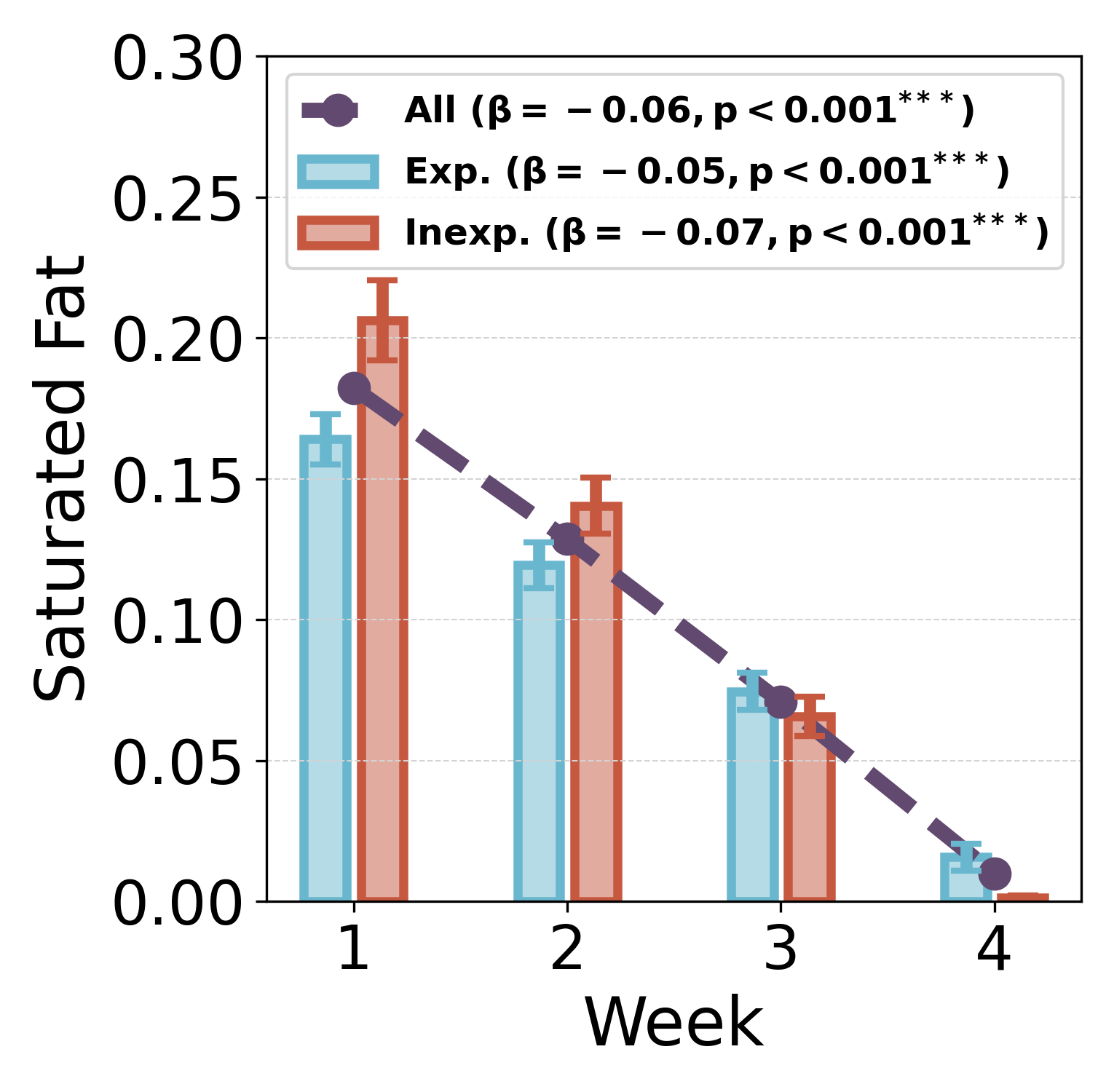}
              \label{fig:saturated_fat}
          }\\
          \subfloat[Dietary Fibre$^{***}$]{    
                  \centering
                  \includegraphics[width=.19\textwidth]{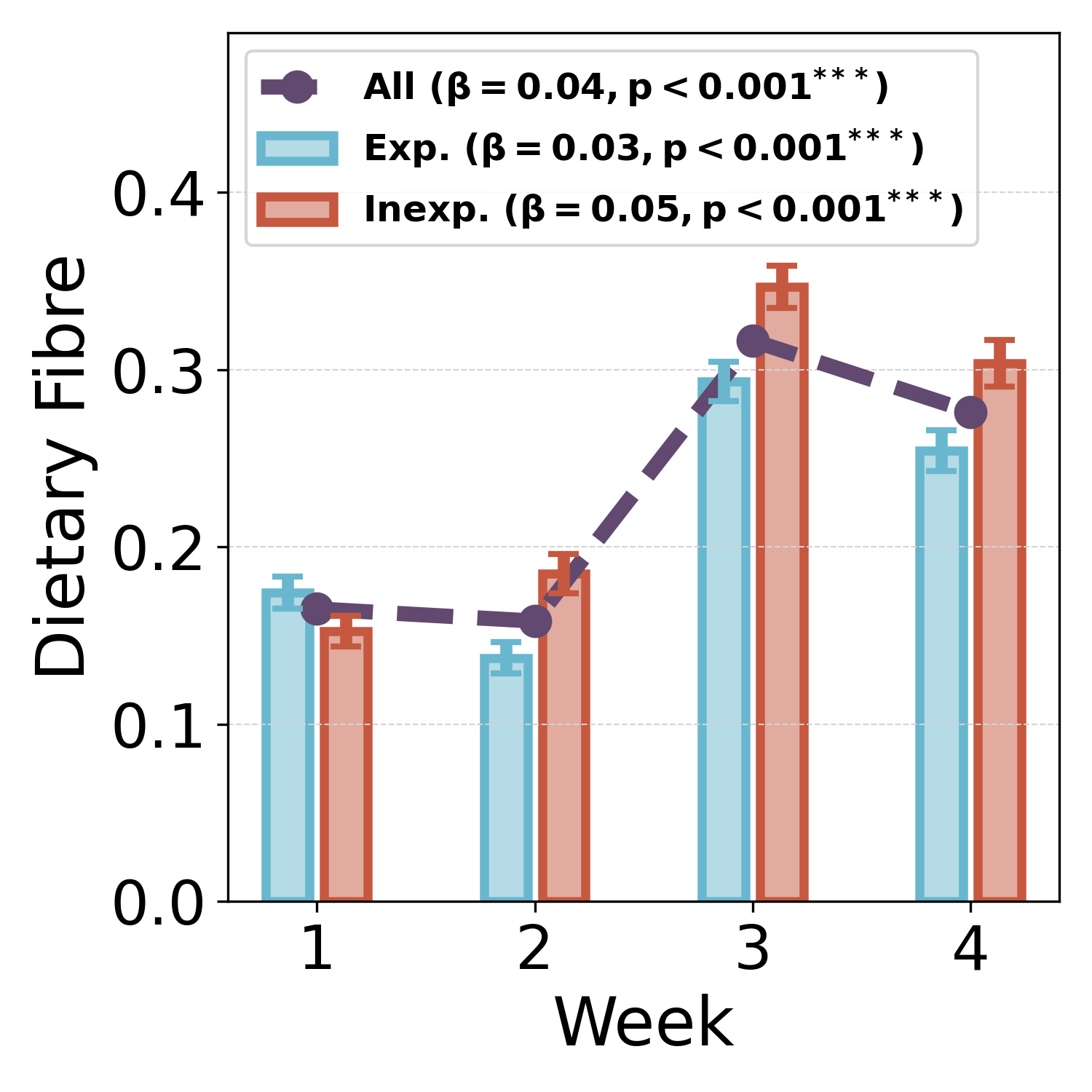}
              \label{fig:dietary_fibre}
          }
        \subfloat[Sugars$^{***}$]{    
                  \centering
                  \includegraphics[width=.19\textwidth]{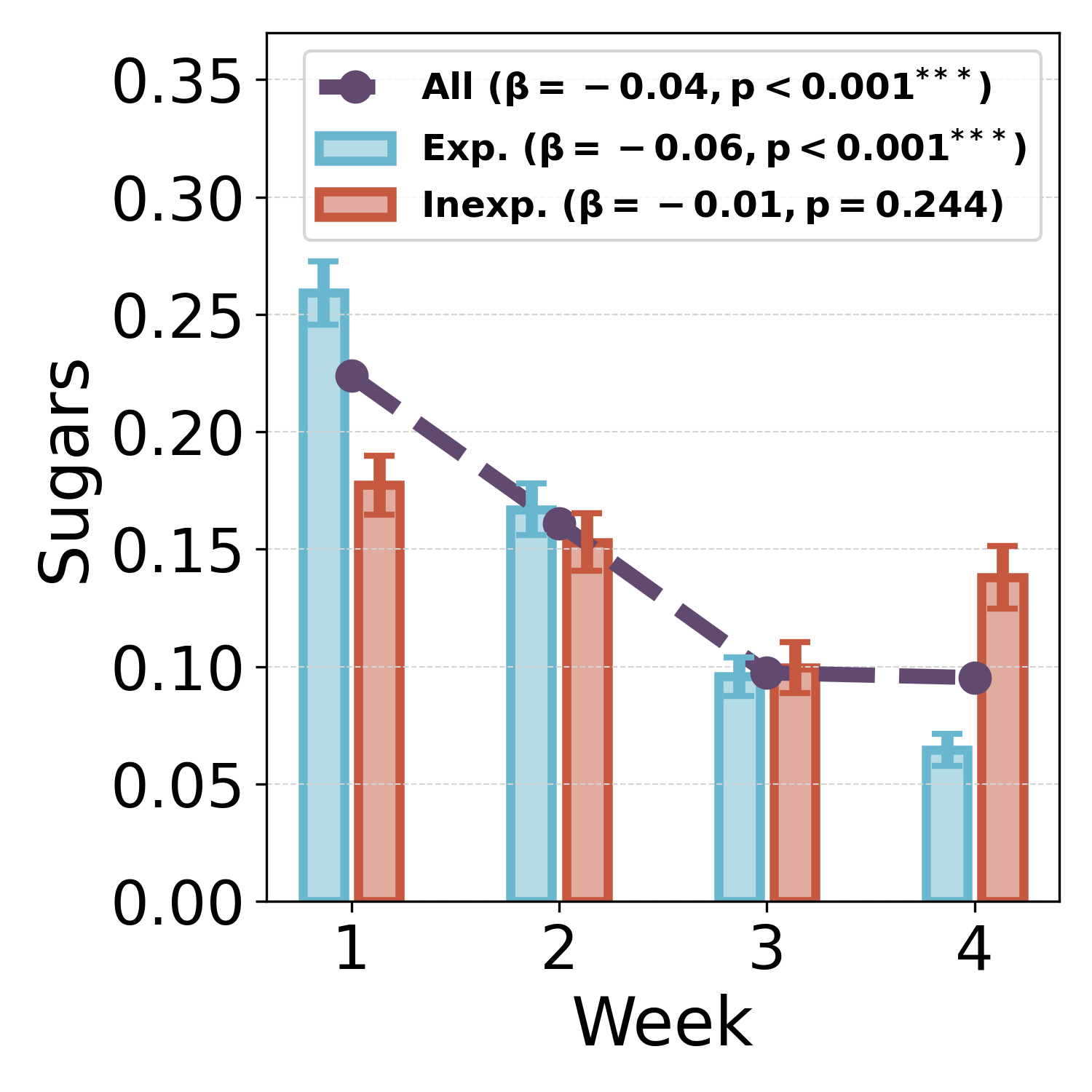}
              \label{fig:sugar}
          }
          \subfloat[Cholesterol$^{***}$]{    
                  \centering
                  \includegraphics[width=.19\textwidth]{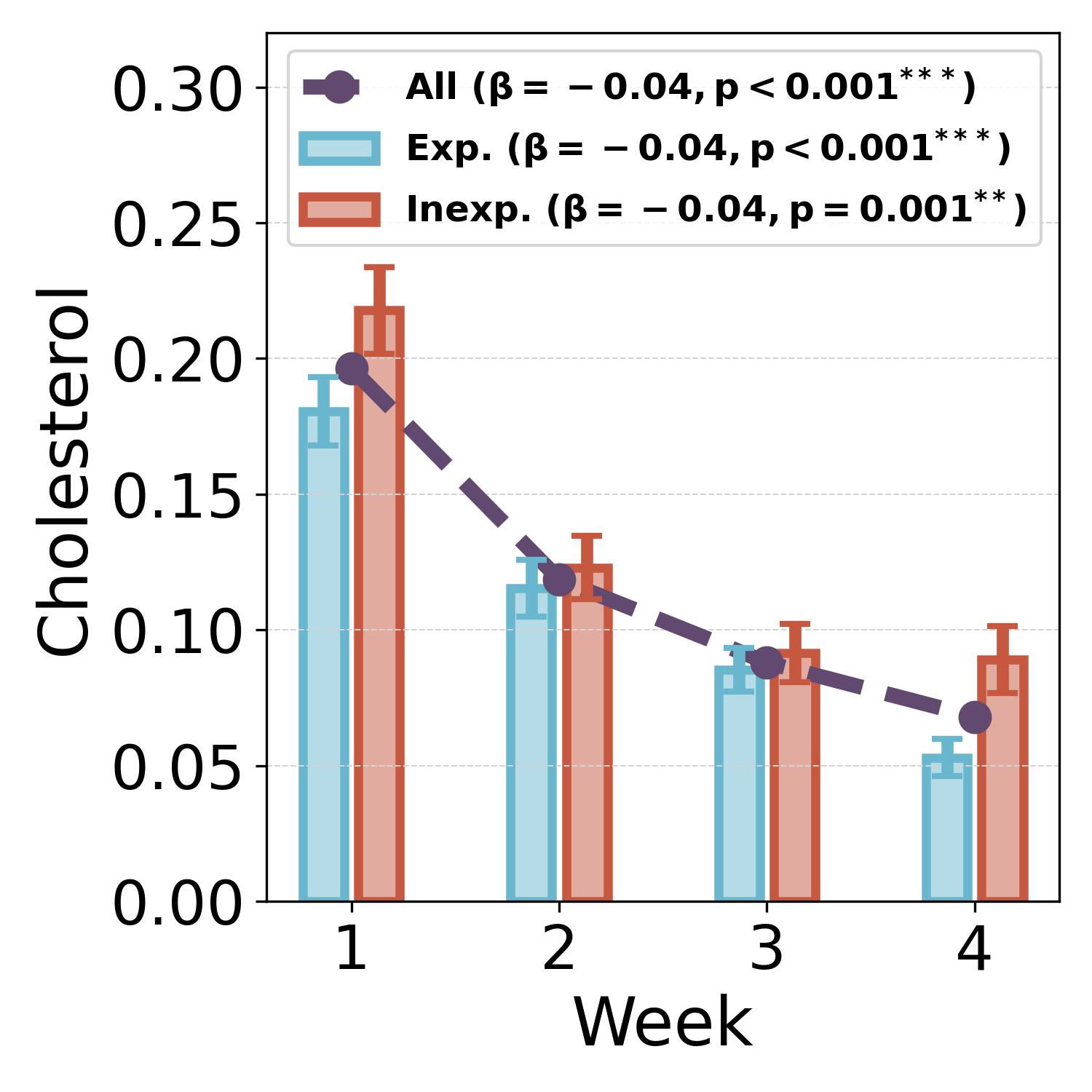}
              \label{fig:cholesterol}
          }
          \subfloat[Carbohydrate$^{***}$]{    
                  \centering
                  \includegraphics[width=.19\textwidth]{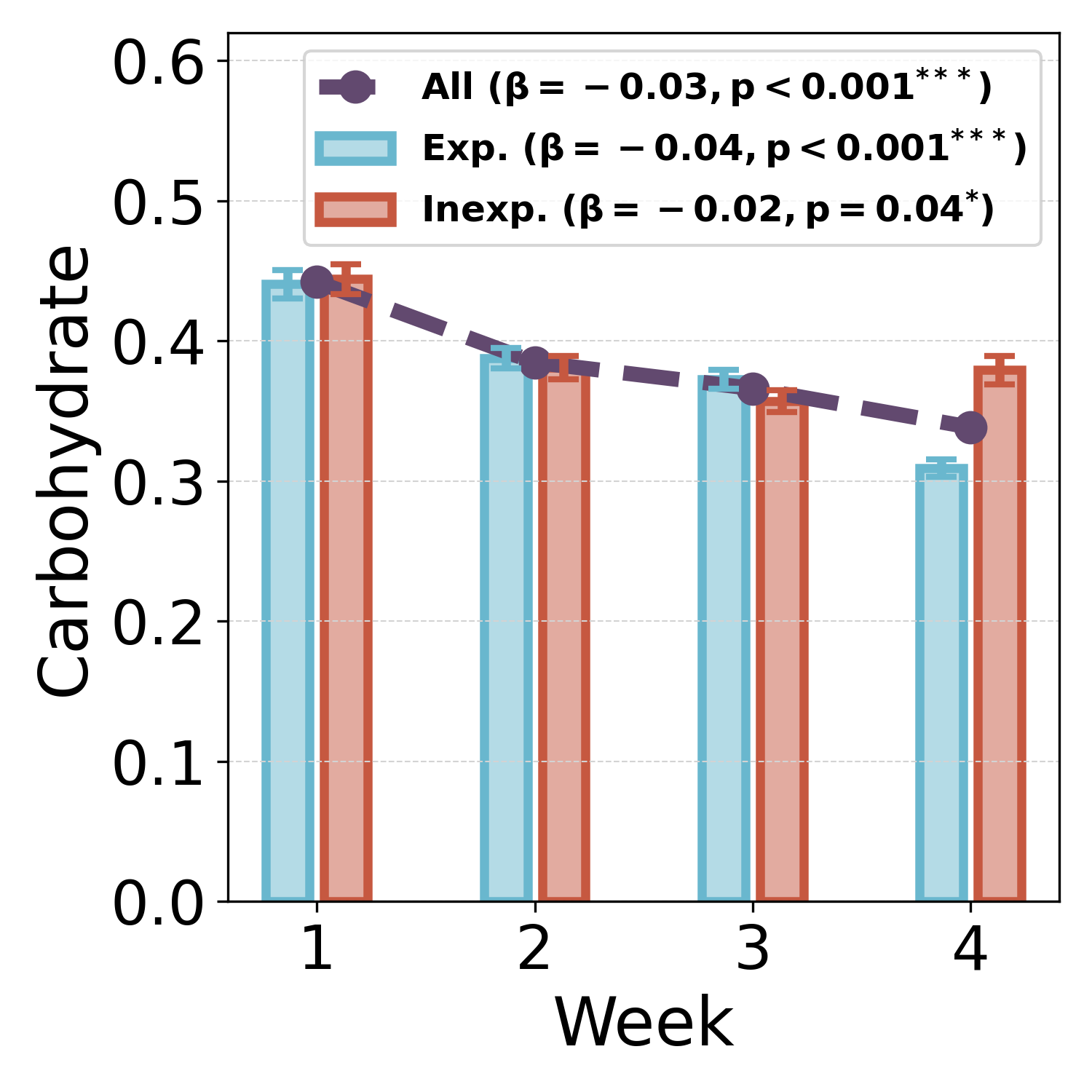}
              \label{fig:carbohydrate}
          }
          \subfloat[Calcium]{    
                  \centering
                  \includegraphics[width=.19\textwidth]{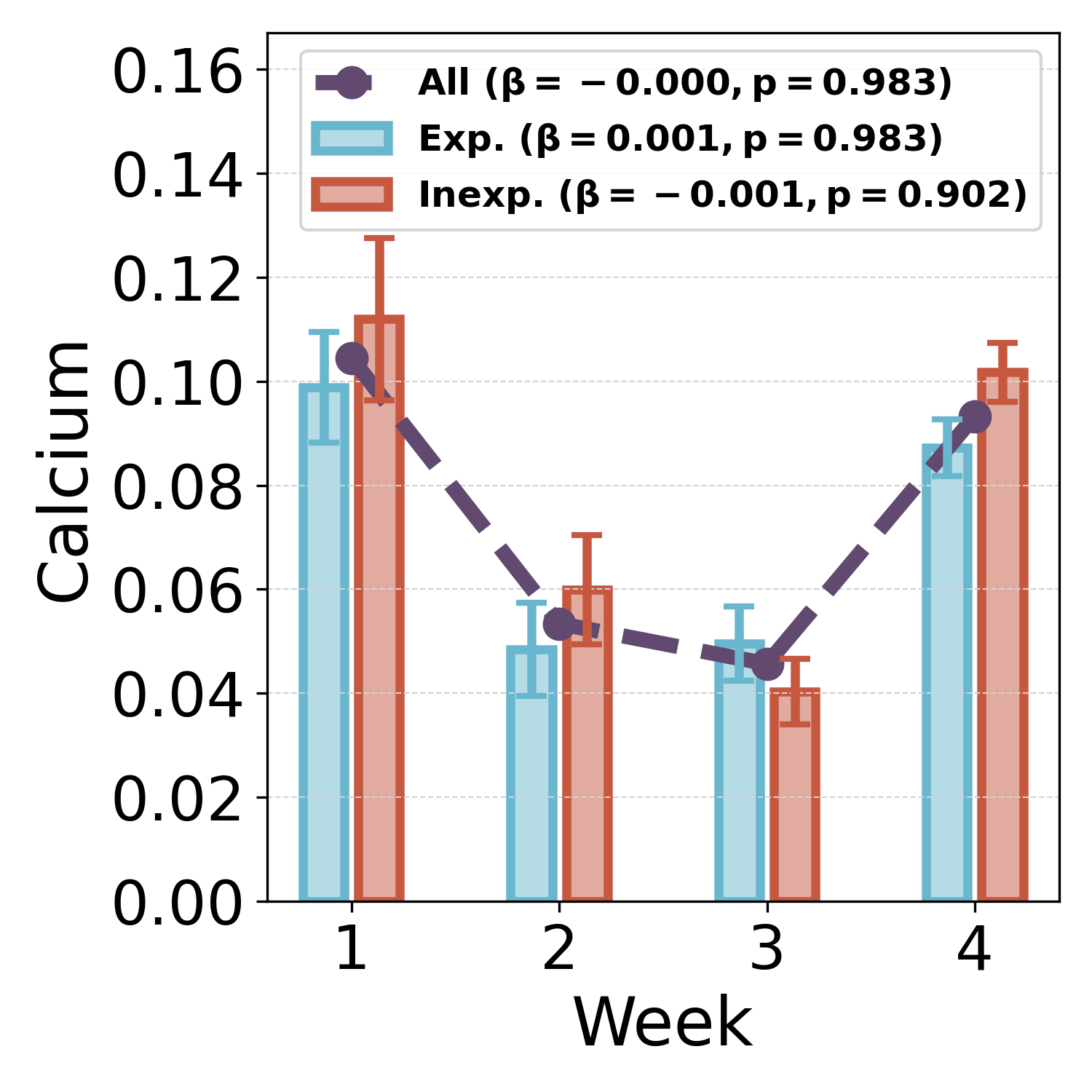}
              \label{fig:calcium}
          }\\
          \subfloat[Copper]{    
                  \centering
                  \includegraphics[width=.19\textwidth]{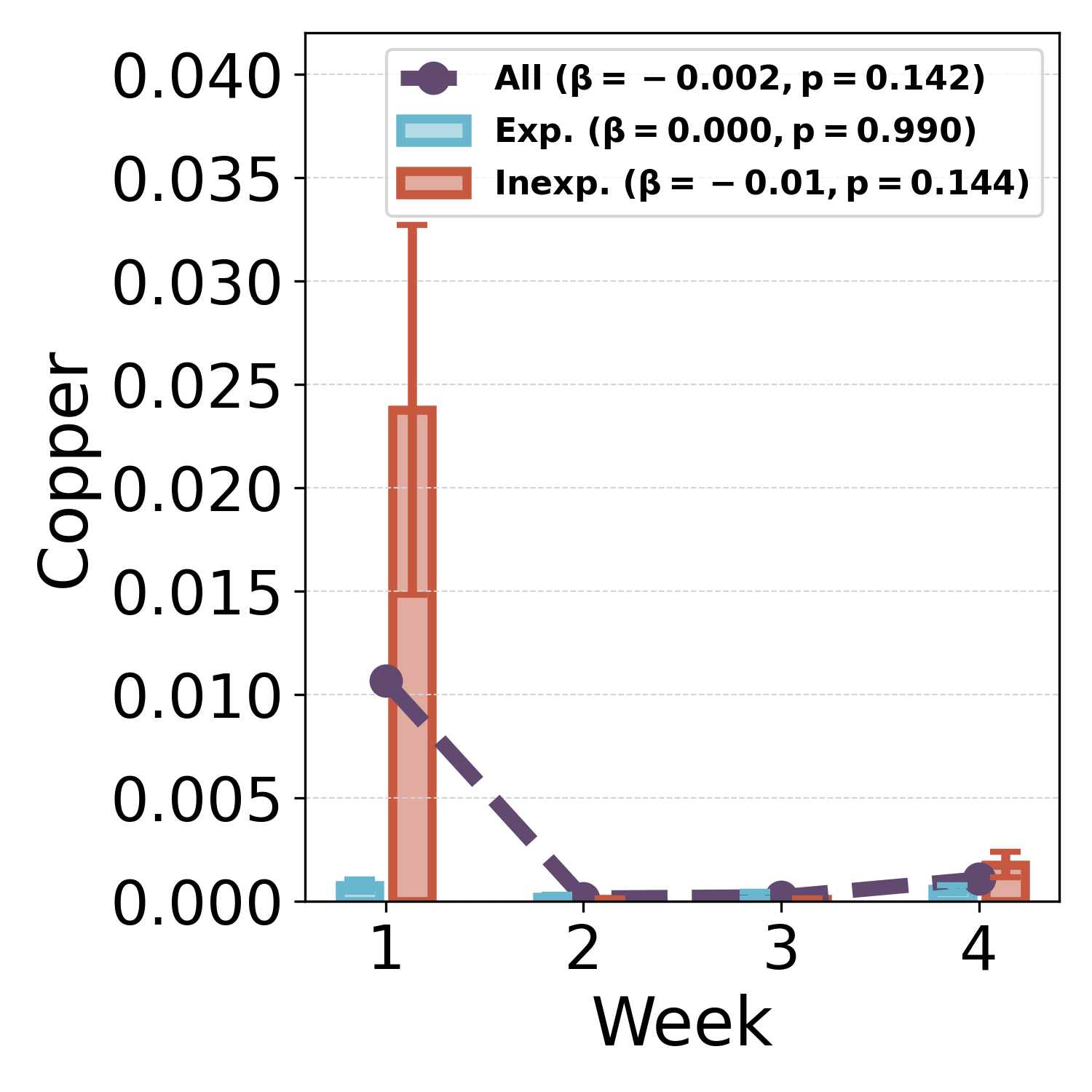}
              \label{fig:copper}
          }
          \subfloat[Magnesium$^{*}$]{    
                  \centering
                  \includegraphics[width=.19\textwidth]{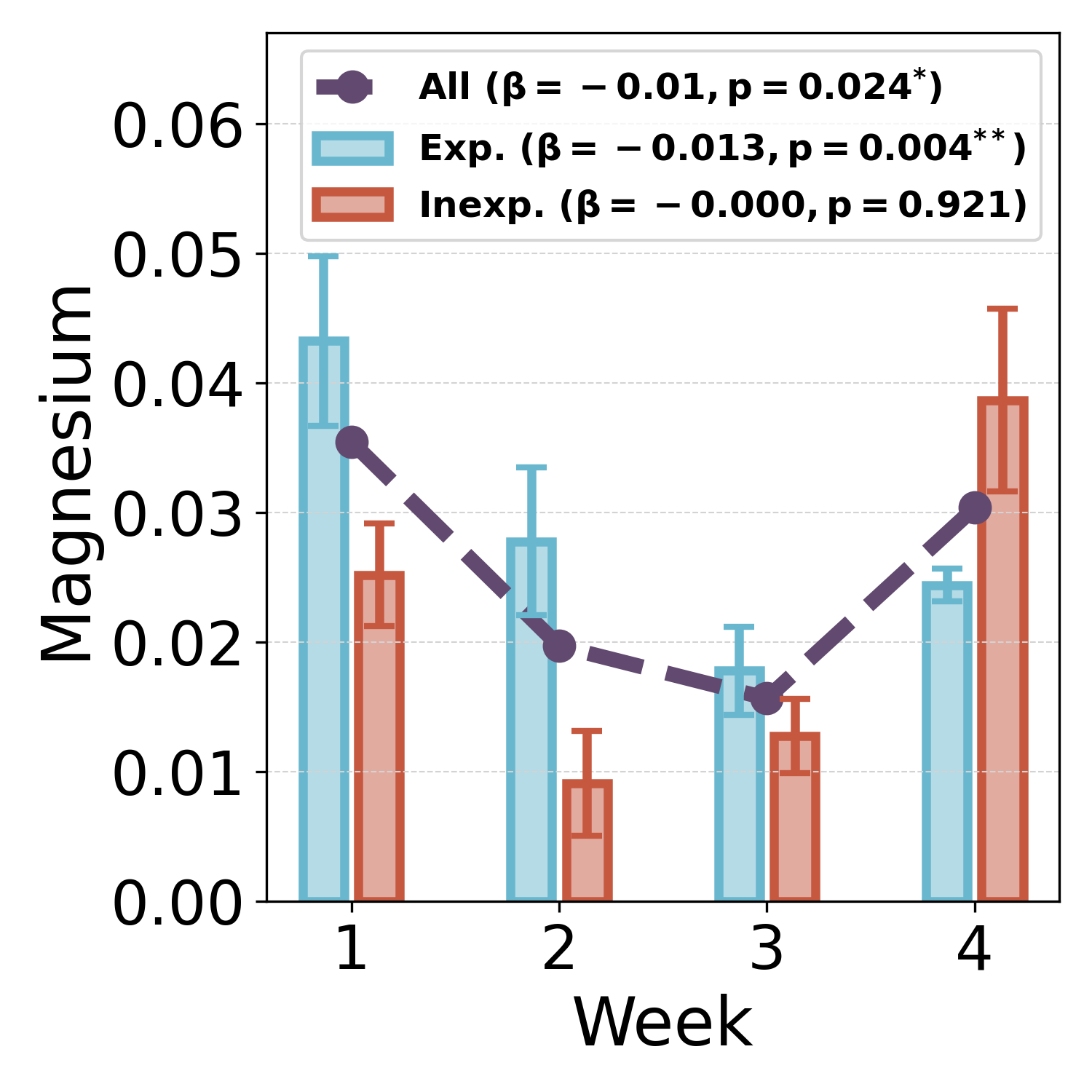}
              \label{fig:magnesium}
          }
          \subfloat[Manganese]{    
                  \centering
                  \includegraphics[width=.19\textwidth]{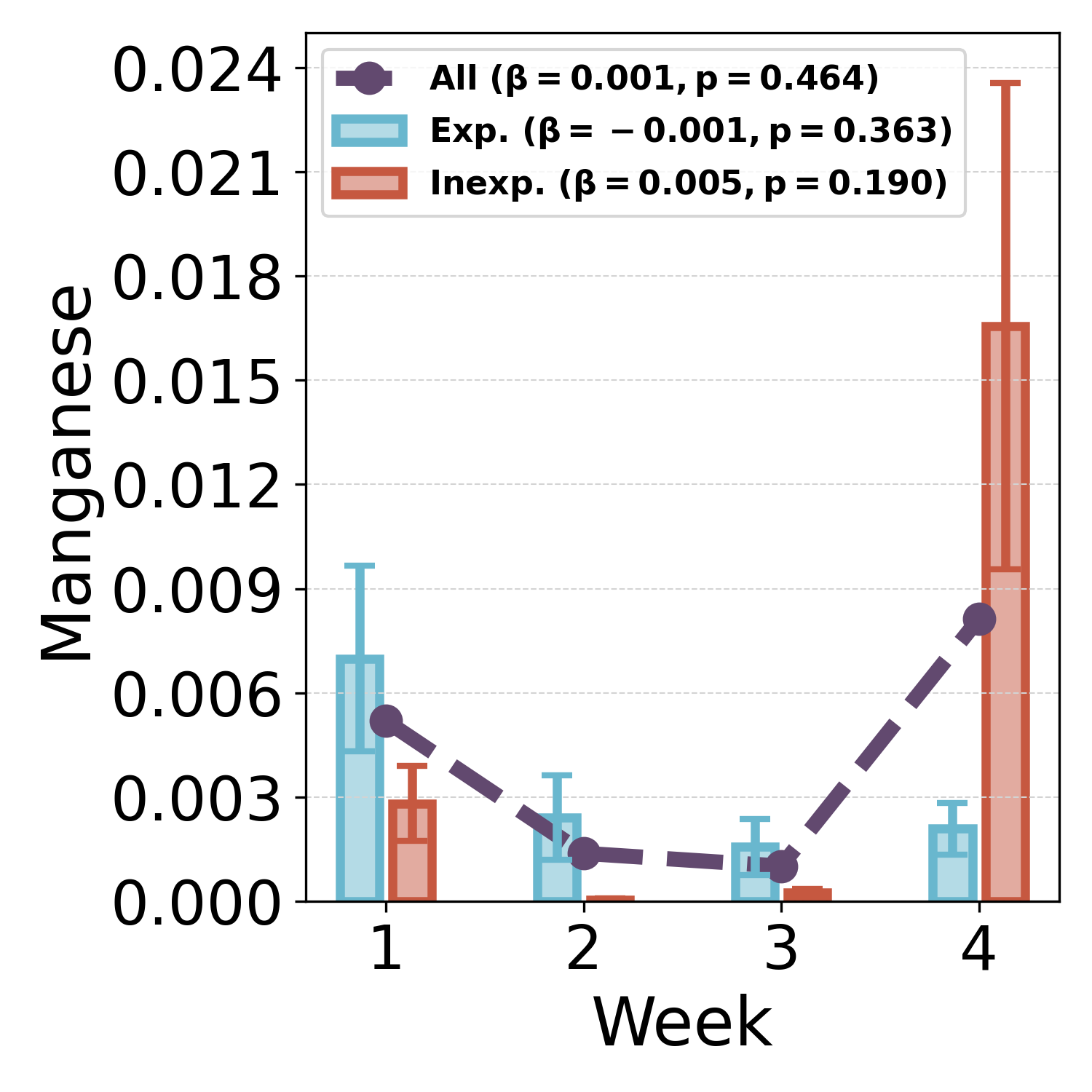}
              \label{fig:manganese}
          }
          \subfloat[Zinc]{    
                  \centering
                  \includegraphics[width=.195\textwidth]{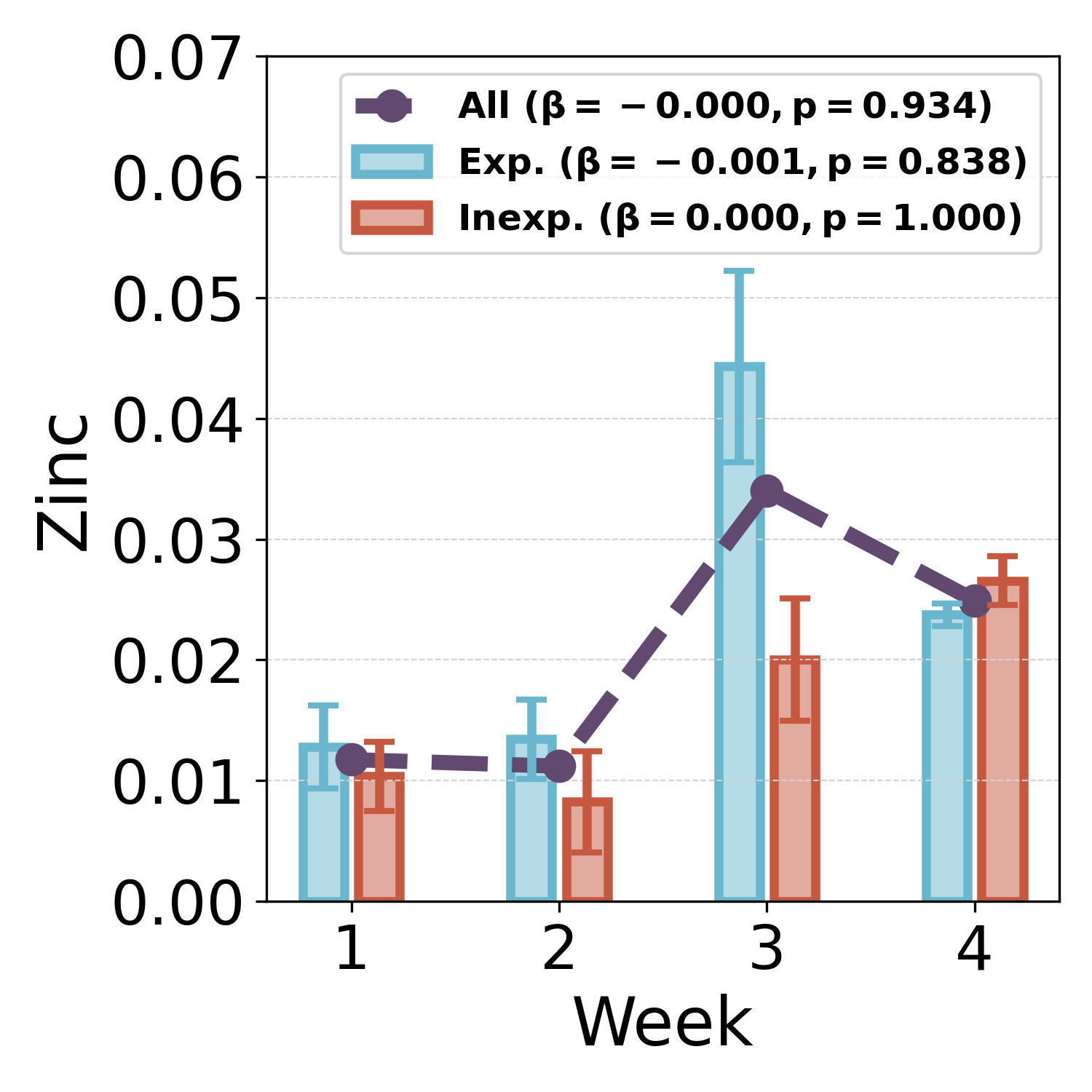}
              \label{fig:zinc}
          }
          \subfloat[Iron$^{***}$]{    
                  \centering
                  \includegraphics[width=.195\textwidth]{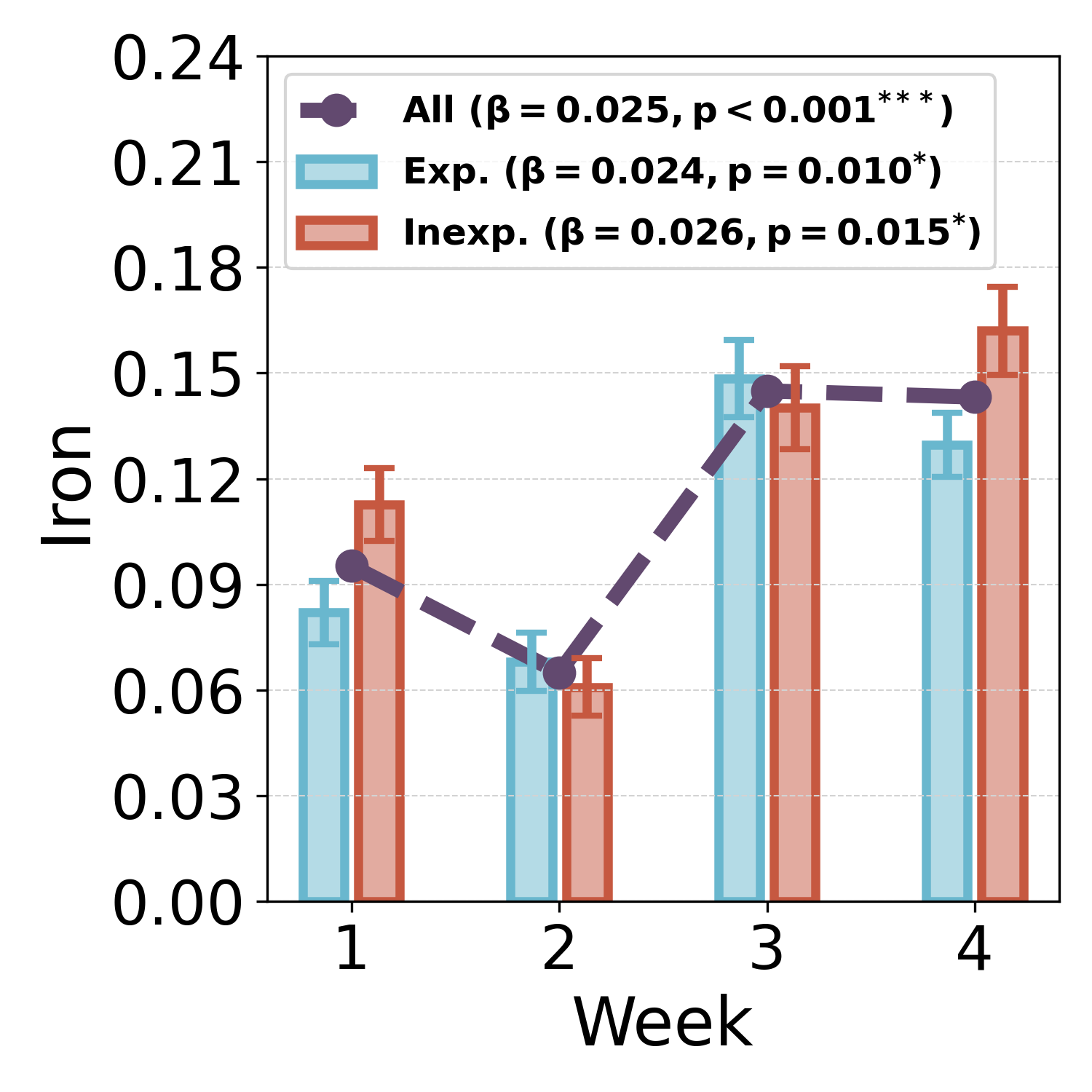}
              \label{fig:iron}
          }
          \caption{The consumption trends of different nutrients in Study II. The error bar shows the standard error. Coefficients ($\beta$) and p-values ($p$) from the mixed linear model are shown in the legends with a significant level of 0.05.}
          \label{fig:nutrient-trend}
           % \vspace{-0.5cm}
      \end{figure*}

\subsubsection{Impact of \shortname on Users' Dietary Patterns}

\

\textbf{Nutrient Consumption Trends.}
To present the nutrition consumption trends across four weeks, we calculated the weekly mean nutrient consumption with min-max normalization. Only nutrients with MAPE < 30\% were included. A Mixed Linear Model assessed the trends.
As shown in \figurename{~\ref{fig:nutrient-trend}}, significant negative trends were observed in energy, total fat, saturated fat, cholesterol, and carbohydrates for both exp. and inexp. groups, suggesting a shift towards a lower-calorie, lower-fat diet. The exp. group also showed a significant reduction in sugar and magnesium, while dietary fiber and iron saw positive trends for both groups, indicating improved consumption of these beneficial nutrients.
Fluctuating trends were noted for trans fat, calcium, copper, manganese, and zinc, potentially reflecting variations in food choices and external factors like meal composition and food availability, while the inexp. group showed more variation in sugar and magnesium.
Overall, these trends indicate a potential shift towards healthier diets, with reductions in unhealthy fats and sugars, and increases in fiber and iron.

\textbf{Increasing Understanding of Current Dietary}. 
Both exp. and inexp. participants reported an increased understanding of their dietary habits. P6 (exp.) noticed the high sodium content in cafeteria soup, which aligned with news she had seen about soups being high in salt and purines. She also observed that eating pizza consistently pushed her calorie consumption above the recommended levels. Inexp. participants (e.g., P22) also benefited from understanding their food choices but tended to rely more on general suggestions without deeper dietary analysis.
Nearly all participants realized their meals lacked vegetables and fruit, which is a common issue in Hong Kong, where this study was conducted.
P7 (exp.) mentioned that using the system made him more mindful of food nutrient content.

\textbf{Raising Awareness of Regular Meal Timing.} 
Over the second to fourth weeks, many participants reported being more conscious of their meal timing. Some participants who used to skip breakfast started to have some breakfast intentionally (P1, P6 exp., P16, inexp). P6 also reported that the system reminded her to eat slowly and enjoy your food without distractions, which was surprising and helpful.

\textbf{Promoting Healthier Dietary Habits.}
While many participants had low fruit and vegetable consumption before, both exp. and inexp. participants reported that \shortname made them intentionally have more vegetables and fruits (P1, P10, P20, exp., P3, P9, P14, P16, P17, P22, inexp.), which also aligns with the nutrient consumption trends in \figurename{~\ref{fig:dietary_fibre}}.
``\textit{Before using this system, I rarely ate vegetables}'' (P22).
Participants also followed system suggestions to achieve dietary goals. For example, P17, aiming to gain weight, increased his consumption of protein-rich foods like tofu and chicken as recommended. 
P9 noted that she began turning down social meals when not hungry, a change from her previous habit of always joining them regardless of hunger.

\textbf{Reducing Consumption of Unhealthy Foods.}
Both exp. and inexp. participants reported reducing their consumption of sugary drinks, high-calorie foods, and other unhealthy items, following the system's recommendations (P6, P10, P13, P15, P20, P21, exp., P14, P16, inexp.), with inexp. participants show greater changes, as demonstrated in Figs. 13(d)(e)(g)(h)(i). 
For instance, P6 reduced carbohydrate intake by switching from Ovaltine to lemonade, and P20 chose low-sugar beverages or avoided drinks altogether.
P21 observed a reduction in his overall calorie intake after cutting back on staple foods like rice and noodles, as presented in the system. P13 also noticed more nutrients falling within the recommended values as he followed the system's dietary advice.

\textbf{No Significant Changes Due to External Constraints.} 
While most participants reported positive changes, P8 (inexp.) noted that it was difficult for him to make significant changes due to limited restaurant options. However, he did make slight adjustments based on the system's suggestions when selecting dishes. 

In summary, after four weeks of using the system, participants reported significant improvements in their dietary awareness and management. 
Exp. participants generally used the system's advanced features to analyze and modify dietary habits in a more targeted way, focusing on specific nutrients and meal planning. In contrast, inexp. participants were more likely to rely on broad suggestions and reported simpler but still effective dietary changes.
Overall, the experience was viewed as beneficial, with users adopting a more mindful and structured approach to their diets and expressing optimism about continued use.

\subsubsection{Suggestions for Improvement}

Based on participant feedback, we identified five key areas for improvement:
\begin{itemize}
    \item \textbf{Nutrients of Interest.} Participants (P1, P6, exp.) found the detailed nutrient analysis overwhelming and suggested a customizable display with automatic prioritization of key nutrients based on individual goals for higher relevance.
    \item \textbf{Incorporation of More Contexts.} Inexp. participants (P8, P9) recommended that the system better accommodate eating environments, such as cafeterias with limited options, by offering meal suggestions based on available menus.
\item \textbf{Increasing Specificity in Dietary Suggestions.} Exp. participants, P10 and P13, sought more specific dietary recommendations, such as the exact weight of food to consume, rather than broad advice like ``\textit{increase protein intake}.''
    \item \textbf{Integration with Other Health Data.} Both exp. and inexp. participants (P3, P6, and P16) suggested integrating the system with other health apps, like fitness trackers, to combine dietary data with physical activity and sleep metrics, providing a more comprehensive view of overall health.
 \item \textbf{Real-Time Notifications.} Both exp. and inexp. participants (P15, P22) recommended real-time notifications to alert users when their nutrient intake is too high or low, helping them stay aligned with their dietary goals.
\end{itemize}

\section{Discussion}

We proposed \shortname, a holistic dietary monitoring and analysis system using smart glasses. Deployed in free-living conditions with both desktop and mobile interfaces, our system differs from others by integrating multimodal inputs with knowledge-empowered AI analysis. 
A short-term user study demonstrated its accuracy in diet identification and logging, providing comprehensive nutritional analysis and personalized dietary suggestions in real-world uncontrolled environments. The comparative analysis showed that the RAG module improved the relevance to the query, coherence, fluency, and accuracy of dietary suggestions. Additionally, a four-week longitudinal study indicated that our system positively impacted users' dietary behaviors.
Participants appreciated the efficiency of the system by providing automated dietary monitoring and analysis. They highlighted its ability to identify diverse dishes, including culturally specific cuisines, and deliver personalized dietary insights based on their daily behaviors. Moreover, the context-aware chatbot enabled continuous interaction, allowing users to obtain more personalized information and supporting proactive engagement in dietary behaviors. 
However, challenges were identified, including difficulties with visually similar items, issues with portion estimation in shared meals, and gaps in micronutrient estimation. While exp. participants tended to be more engaged with the system and leveraged its advanced features (such as obtaining detailed plans through conversations with the chatbot), inexp. participants were more likely to directly follow the system-generated suggestions, especially the simpler ones. Experts highlighted the system's value for general dietary tracking but noted the need for more accurate estimations of specific nutrients in clinical contexts.
These findings confirm the system's strengths while identifying areas for improvement.

In this section, we further discuss the implications of this work, followed by a discussion on the generalizability and limitations of this work.

\subsection{Implications}

\subsubsection{Personalization based on Experience, Behaviors, and Environment}
\label{subsubsec:personalization}

The findings emphasize the need to tailor systems to users' experience levels, interaction behaviors, and environmental contexts.
First, exp. participants exhibited higher engagement and satisfaction, emphasizing the need for advanced features like detailed nutritional insights, while simplicity and intuitive interfaces can reduce cognitive load for inexp. users. For example, the system could adapt its workflow and interface to offer advanced functionalities as user experience grows \cite{saif2024evolveui}.

Second, the system could be enhanced by personalization based on users' interactions.
Image-based food recognition struggles with nuances like cooking methods, seasonings, and visually similar items. 
Incorporating human-in-the-loop personalization \cite{ARchitect} could mitigate this challenge by leveraging the user's existing ability to correct misidentified items to dynamically fine-tune the model. This would allow the system to learn from these corrections over time, reducing the need for repeated manual interventions. 
Additionally, analyzing chatbot conversations can uncover patterns in user preferences and behaviors \cite{CharacterMeet}, which, when integrated with clinical data, could further fine-tune LLMs and improve system performance, such as increasing empathy or precision \cite{llempathy, jiang2026hearyouinsilence}.

Moreover, dietary suggestions could further consider environmental factors such as food availability (e.g., university cafeteria menus), financial situations, cultural preferences, and meal timing.
As suggested by participants, meal options tailored to commonly visited cafeterias can be incorporated considering the price. 
Shared meals, common in many cultures such as in Asia, complicate portion size estimation due to variations in individual preferences and dietary needs. While the current system assumes equal portions, adaptive algorithms or informed user input could be incorporated to improve accuracy. 
Furthermore, while our participants consumed diverse cuisines, their shared cultural background (Chinese) limits the generalizability of the observed behavioral patterns and preferences (e.g., regarding shared meals). Therefore, expanding research to more diverse cultural groups is a critical next step to increase the global applicability of such systems.

\subsubsection{Balancing Complexity and Relevance in AI-Generative Dietary Suggestions}
\label{subsubsec:balancing}
While the RAG module improves nutrient analysis and dietary suggestions, balancing the complexity of RAG-generated suggestions with their perceived relevance and actionability should be further improved. The complexity of RAG-generated recommendations could be dependent on the complexity of the constructed library. This highlights a key limitation of RAG: its output is highly dependent on the quality, complexity, and relevance of the retrieved documents. Our expert evaluation revealed that our library's inclusion of academic articles, while ensuring accuracy, produced recommendations that were at times overly complex, thereby adversely affecting their perceived relevance and actionability. The curation of the knowledge base is therefore a critical, non-trivial step that impacts system performance.
For example, integrating carefully vetted popular science articles and knowledge summaries (e.g., from reputable health organizations) targeting the lay public can support the generation of practical, relevant suggestions that align with users' daily routines and preferences. However, to mitigate the risk of misinformation, such content must undergo rigorous verification against clinical guidelines and be presented with clear user disclaimers.
Complementing this, future work could also leverage Chain-of-Thought prompting \cite{wei2022chain} and style transfer techniques \cite{toshevska2025llm} to decompose dense clinical guidelines into digestible formats, automatically translating technical jargon into plain language for the general audience.

Moreover, participants expressed the need for more detailed context (e.g., the impact of sources) to enhance credibility while avoiding information overload. Future systems should balance source transparency with simplicity. Incorporating techniques like clickable summaries, expandable sections, or collapsible details can help present more in-depth information without overwhelming the user. Future systems can leverage LLMs to offer colloquial explanations of professional terminology \cite{Qlarify}, making complex nutritional information more accessible to users of varying expertise, thus improving the actionability of suggestions and increasing user engagement.

Furthermore, as an alternative to RAG, fine-tuning an LLM on a high-quality, domain-specific dataset may present a certain degree of ability to internalize general nutritional knowledge and adopt a more consistent, practical tone. However, fine-tuning alone shares key limitations with baseline LLMs. First, it demands ongoing maintenance; unlike RAG, knowledge cannot be easily updated without retraining the model. Second, fine-tuning requires a large volume of high-quality training data (e.g., expert-verified QA pairs), which is labor-intensive to curate. Finally, fine-tuned outputs remain unverifiable, as they usually cannot be explicitly traced back to a specific trusted source. A promising direction for future work could be an agentic framework. This would involve an agent that could intelligently route user queries, deciding when to retrieve verifiable facts via RAG from the latest resources (e.g., for specific nutrient data) and when to rely on its fine-tuned internal knowledge (e.g., for general encouragement or stylistic responses). This could offer a more flexible solution to balance accuracy and actionability.

Additionally, we observed that inexp. users tended to rely more heavily on AI-generative analysis without critical evaluation. To mitigate the risk, the system could prompt users to consider the limitations of AI-generated insights, such as potential inaccuracies in personalized recommendations, and encourage them to critically evaluate results.

\subsubsection{Enhance the Integration of Multi-Devices and Proactive Intervention}
While our system relied on smart glasses and laptops (or mobile phones), as suggested by participants, integrating additional devices such as wrist-worn activity trackers could provide a more comprehensive understanding of users' health \cite{jiang2023leveraging, jiang2023data}. Physical activity, sleep patterns, and metabolic rates are critical factors in dietary analysis and recommendations. Incorporating multimodal signals from multiple devices would enable more accurate, personalized feedback \cite{jiang2023healthprism}. For example, combining dietary data with physical activity metrics could refine calorie and nutrient recommendations, enhancing the system's overall utility.

Furthermore, leveraging multiple devices for real-time feedback and proactive dietary intervention can significantly enhance user engagement and adherence to dietary goals. By enabling users to quickly correct or label information on their smartwatches, the system can be further enhanced, as discussed in Sec.~\ref{subsubsec:personalization}.
Smart glasses or watches can deliver in-situ visual or audio cues \cite{bressa2022data} during meals, helping users make informed choices in a more seamless way. Real-time audio haptic feedback \cite{tan2024audioxtend}, for example, could combine sound cues (such as a pleasant tone indicating a correct portion or a warning beep for overeating) with tactile sensations (like vibrations to alert users when a meal exceeds recommended nutrient levels). This approach may not only improve user experience but also provide more intuitive and proactive interventions for dietary self-regulation based on the user's current activity levels and dietary patterns, fostering healthier behaviors.

\subsection{Generalizability}
\label{sec:generalizability}
Our system's analysis framework is designed to be generalizable across different hardware, regions, and domains.

While developed with Aria Glasses, our system can integrate with other eyewear equipped with similar sensors (e.g., IMU or cameras) for ingestive episode detection and supports non-wearable alternatives, allowing users to upload meal photos for analysis when wearables are unavailable or impractical.

While developed with Aria Glasses, our system supports integration with other eyewear equipped with similar sensors (e.g., IMU or cameras) for ingestive episode detection. Furthermore, to enhance accessibility, it supports non-wearable alternatives, allowing users to upload meal photos from any camera-enabled device for analysis when wearables are unavailable or impractical.

As our studies were conducted in Hong Kong, applying the current model directly to other regional diets might yield variable performance. However, the system's overall pipeline is designed for modular adaptability. To extend \shortname to other culinary contexts, users can update the knowledge base by replacing the nutritional database with a more region-specific source, narrow the retrieval scope by ingesting relevant local dietary guidelines into the RAG vector store, and adjust the system prompt to make it more specific (such as adding `cuisine type'). This modularity ensures the system can be swiftly swapped to cater to diverse global dietary standards.

Beyond dietary monitoring, the system's framework has the potential to be extended to other health and wellness domains. With further development, such as developing APIs for data ingestion or updating the knowledge base, it could integrate fitness tracker data for physical activity analysis, assist with student tasks in education, or facilitate patient rehabilitation, where automated tracking and domain-specific analysis are essential.

\subsection{Limitations and Future Work}
While \shortname demonstrated usability and effectiveness, several limitations remain.

\paragraph{Eyewear Comfort and Practicality}
A primary limitation is the practical burden and comfort of the eyewear. The Aria Glasses, though powerful, are heavy due to unnecessary sensors for our study. Participants reported discomfort from sweat, especially in summer. Many also required vision correction, but lens replacement was not practical. This discomfort may hinder long-term retention. 
Beyond physical comfort, the social acceptability of wearing smart glasses at every meal, especially in public or with companions, remains a realistic barrier to adoption. However, we observe the increasing commercialization and popularity of AI glasses by Meta, Rokid, etc., which suggests that these devices are becoming more normalized and that social acceptability may improve as they become more commonplace.
Future research could explore lighter glasses, attachable sensors for personal eyewear, or alternatives like smart earbuds or watches.

\paragraph{Diet Identification, Portion Estimation, and Micronutrient Analysis}
\shortname accurately identified most foods but struggled with visually similar items.
Users may need to manually update diet summaries. The system also assumed equal portions in shared meals, potentially leading to discrepancies. Additionally, estimating micronutrients like phosphorus and potassium was challenging due to variability in biological, environmental, and processing factors. This aligns with recent findings that noninvasive micronutrient assessment remains a major open challenge in the field \cite{balch2025towards}. Future work could focus on improving micronutrient datasets for higher accuracy. As emphasized by the experts, more precise nutrient intake was expected in clinical contexts. Specifically, this could involve developing an agentic framework that proactively prompts the user for clarification on `hidden' factors (e.g., `Did you add salt or soy sauce?') when high-uncertainty foods are detected. Additionally, integrating GPS-based context retrieval (with user consent) could allow the system to fetch precise nutritional data from restaurant-specific menus rather than generic databases.

\paragraph{Limitation of RAG Module}
As discussed in Sec.~ \ref{subsubsec:balancing}, while our RAG module significantly improved factual accuracy, its high dependency on the quality of the retrieved documents also introduced new challenges, such as the curation of the knowledge base. Moreover, other methods like fine-tuning could be explored to improve the performance, potentially by learning a more practical and actionable style. Therefore, another promising direction for future work is the exploration of agentic models to intelligently decide when to retrieve verifiable facts via RAG and when to rely on fine-tuned internal knowledge to better balance accuracy with actionability.

\paragraph{Study Duration}
The four-week duration of the longitudinal study is comparable with previous diet-related studies \cite{biel2018bites, morshed2022food, silva2023exploring, weinreich2017effectiveness, santas2012selective}.
However, extending the study duration and incorporating additional metrics, such as cognitive load, would provide deeper insights into sustained engagement.

\paragraph{Enhancing Privacy, Transparency, and Security}
Despite the system's privacy-preserving consideration, participants still raised concerns, particularly in shared meal contexts. 
Clear communication protocols and transparent privacy practices for users and their companions can build trust and address these sensitivities. 
Future systems could also incorporate advanced safeguards against misinformation, such as prompt injection attacks \cite{liu2023prompt} or erroneous inputs.

\paragraph{Participant Demographics and Generalizability.}
While our system showed promise in handling complex cuisines (e.g., Chinese-style food), our study population was primarily Chinese participants. Although they consumed both Western and Eastern foods, and our system can be adapted by incorporating localized nutritional data (as discussed in Sec.~\ref{sec:generalizability}), this demographic homogeneity limits the generalizability of our findings on the behavioral patterns, cultural preferences, and social acceptability of the system. Future work is required to validate DietGlance with more diverse, multicultural participant pools.

\paragraph{Nutrition Library Limitations.}
While we constructed our nutrition library from high-trust, expert-vetted sources, and the experiment results show that our RAG module significantly improved the accuracy of the system's outputs compared to the baseline LLM, we did not perform an exhaustive, line-by-line check for minor inconsistencies between the documents. Future work could explore automated methods for consistency checking across the corpus, such as by using contradiction detection models to identify conflicting recommendations. Moreover, as new nutritional research is published, future systems could build a robust pipeline for dynamically updating and re-vetting the knowledge base to ensure the RAG module's accuracy is maintained over time.

\paragraph{Database Dependency.}
Our expert evaluation relied on a specific local nutritional database (Centre for Food Safety, Hong Kong) as the ground truth. We acknowledge that nutritional values can vary across different national databases due to regional food differences. Furthermore, our system integrates multiple databases to maximize food coverage; however, this introduces data heterogeneity that may affect perceived accuracy when evaluation is confined to a single regional standard. While we chose the most geographically relevant database for our expert evaluation, the system's performance might vary if evaluated against different national standards, highlighting the importance of localizing knowledge bases for deployment. Future work may develop adaptive RAG frameworks that can automatically select and retrieve from the most regionally appropriate nutritional database based on the user's location or cultural context.

\section{Conclusion}
We introduced \shortname, a holistic system using smart glasses to automatically monitor dietary behaviors and provide personalized nutritional analysis and dietary suggestions, empowered by domain-specific knowledge. 
Two studies evaluated its performance in real-world, unconstrained environments. 
A short-term user study with 33 participants, along with quantitative assessments involving crowd workers and domain experts, demonstrated the system's effectiveness and usability.
A four-week longitudinal study with 16 participants validated \shortname's positive impact on dietary behaviors by increasing their understanding of their diets, raising awareness of regular meal timing, promoting healthier eating habits, and reducing the consumption of unhealthy foods.

\bibliographystyle{ACM-Reference-Format}
\bibliography{references}

\newpage
\appendix

\section{Questionnaire and Interview Script for the Needs-Finding Study}
\label{appx:needs}

\subsection{Online Survey Questionnaire}
\label{appx:needs_survey}
This survey (Table \ref{tab:needs_questionnaire}) was used to gather quantitative data on participants' demographic information, backgrounds, prior experience with dietary monitoring tools, and eating environments. Please note that while we only provide English version here, the questionnaire was also available in Chinese.

     \begin{table}[!h]
      \footnotesize
      \renewcommand{\arraystretch}{1}
      % \parbox{\linewidth}{
        % \centering
        \caption{{The survey questionnaire used in needs-finding study.}}
        \label{tab:needs_questionnaire}
            \begin{tabular}{lll}
          \toprule
            \multicolumn{2}{l}{\bfseries {Basic Information}} \\
           \midrule
           {Q1} & {Do you currently have or have you previously been diagnosed with an eating disorder (e.g., anorexia, bulimia)? (Yes/No)}\\
           & {(The following questions will only display when participants check "No")}\\
           {Q2} & {Gender: Male, Female, Non-binary / third gender, Prefer not to say}\\
           {Q3}& {Age, Location, Occupation}\\
            \midrule
            \multicolumn{2}{l}{\bfseries {General Dietary Habits}}\\
            \midrule
            {Q4} & {Which of the following types of food have you ever tried? (Check all that apply)}\\
            & {Chinese, Japanese, Korean, Thai, Vietnamese, India, Turkish, French, Italian, Russian, Egyptian, Southern African, United States,} \\
            & {Mexican, Australian, Other (please specify).}\\
            {Q5} & {Your usual eating environment (Check all that apply)}\\
            & {Home, Office, Restaurants, Other (please specify)}\\
             \midrule
             \multicolumn{2}{l}{\bfseries {Dietary Monitoring}} \\
             \midrule
             {Q6}& {How often do you monitor your diet?}\\
             & {Dialy, 2-3 times a week, Weekly, Monthly, Rarely, Never, Other (please specify)} \\
             {Q7} & {(Only applicable when Q6 is not never) What methods do you use to monitor your diet? (Check all that apply)}\\
             & {Mobile Apps (please specify), Paper Journals, Online Tools (please specify), Professional Guidance (Dietitian, nutritionist, etc.),} \\
             &  {Other (please specify)}\\
             {Q8} & {What are your primary reasons for monitoring your diet? (Check all that apply)} \\
             & {Weight Management, Health Conditions (e.g., diabetes, hypertension), Fitness Goal, General Wellness, Others (Please specify)}\\
          \bottomrule
        \end{tabular}
      % }
    \end{table}

\subsection{{Interview Script}}
\label{appx:needs_interview}
{After completing the survey, each participant attended a one-on-one semi-
structured interview session (conducted either face-to-face or remotely, either in English or Chinese). This qualitative session was designed to explore their vision for helpful features, the obstacles they’ve encountered in dietary tracking, and their specific expectations and concerns regarding future eyewear-based dietary systems. The interviewers followed a question list (Table \ref{tab:needs_interview}). To gain a deeper understanding, interviewers also asked probing follow-up questions based on the participants' answers.}

     \begin{table}[!t]
      \footnotesize
      \renewcommand{\arraystretch}{1}
      % \parbox{\linewidth}{
        % \centering
        \caption{{The interview script used in needs-finding study.}}
        \label{tab:needs_interview}
            \begin{tabular}{lll}
          \toprule
          \bfseries {Questions specific to participants with experience in using dietary monitoring tools}\\
          \midrule
          {Q1. What features do you find most helpful in these tools or apps?}\\
          {Q2. What challenges do you face in monitoring your diet?}\\
          {Q3. Would you feel more conscious of your diet while monitoring your diet?} \\
          \midrule
          \bfseries {Questions specific to participants without experience in using dietary monitoring tools}\\
          \midrule
          {Q4. What are the reasons you do not monitor your diet?}\\
          {Q5. Would you like to try a diet monitoring tool?}\\
          \midrule
          \bfseries {For all participants}\\
          \midrule
          {Q6. What features would you like to see in a dietary monitoring tool?}\\
          {Q7. How often do you wear glasses?}\\
          {Q8. Suppose there is a pair of smart glasses that can be used to monitor your diet (here we show and introduce the Aria Glasses to the participants),}\\ 
          {automatically detecting and analyze your eating and drinking behaviors. The glasses can generate food journals including the type and the amount}\\
          {of food you have eaten and provide the analysis of nutritional content, such as calories and nutrients. Would you be interested in using such a device?}\\
          {Q9. Do you think it is convenient to use the glasses in different settings (e.g., at home, in public)}\\
          {Q10. What features would you find most useful in these smart glasses?}\\
          % What features would you find NOT useful in these smart glasses?\\
          {Q11. What concerns, if any, would you have about using smart glasses for diet monitoring?}\\
          {Q12. What difficulties would you expect while using the glasses during meals?} \\
          % Q13. How well do you think the smart glasses can help you monitor your diet?\\
          {Q13. How do you think the glasses could help you achieve your dietary goals?} \\
          {Q14. Are there any additional features you would like to see?} \\
          {Q15. What improvements would you suggest for the smart glasses?}\\
          {Q16. How likely are you to use these smart glasses regularly for diet monitoring?} \\
          {Q17. Do you have any additional comments or suggestions regarding dietary monitoring tools and apps?}\\
          \bottomrule
        \end{tabular}
      % }
    \end{table}

\section{List of \shortname Prompts}
\label{appx1:prompts}

\subsection{Class Prompts for Diet Segmentation}
\label{appx1:class_prompts}
\sloppy
% \begin{singlespace}
The class prompts used for diet segmentation included: \texttt{[`food', `drink', `soup', `rice', `noodle', `meat', `vegetable', `seafood', `fried egg', `cutlet', `curry', `broccoli', `fruit', `fish', `chicken', `lemon', `beef', `pork', `burger', `salad', `cola', `ham', `bread', `cake', `dessert', `sandwich', `bread', `snack', `juice', `sauce', `coffee', `tea', `water', `alcohol', `milk', `beer', `oil', `cup', 'bottle', `plate', `bowl', `mug', `can']}.
% \end{singlespace}

\subsection{Diet Identification Prompt} 
\label{appx1:diet_identify}

\noindent\rule{\linewidth}{0.2pt}

\noindent \texttt{\textbf{System:}}
\texttt{Based on a series of egocentric images taken from the camera on the user's glasses during a meal session, considering the user information, describe this meal containing the following information: }

\begin{itemize}[label=-]
    \item \texttt{Names of the food and drink taken by users. Separate the ingredients as detailed as possible.}
    \item \texttt{Based on the images provided, which are taken during the process of eating and drinking, estimate the amount of food (in g) and drink (in ml) at the beginning, and how much percentage had been consumed by the user at the end.}
\end{itemize}
\texttt{Constraints: }
\begin{itemize}[label=-]
    \item \texttt{If you don't know the answer, say "I don't know", don't try to make up an answer.}
    \item \texttt{Output the answer following the format of the Structured Output Example below.}
\end{itemize}
\texttt{Structured Output Example:}
\begin{itemize}[label=-]
    \item \texttt{A 10-minute dinner session. The meal includes a bowl of beef noodle soup with various vegetables.}
    \item \texttt{Fried Beef, amount: 80g, consumed percentage: 90\%}
    \item \texttt{Lemon Water, amount: 250ml, consumed percentage: 50\%}
\end{itemize}

\noindent \texttt{\textbf{User:} \textcolor{ccolor}{[Meal Image Log]} The provided images in this message were captured on \textcolor{ccolor}{[date]} from \textcolor{ccolor}{[time]} to \textcolor{ccolor}{[time]}. The user is a \textcolor{ccolor}{[age]}-year-old \textcolor{ccolor}{[gender]}, with a weight of \textcolor{ccolor}{[weight]} kg and a height of \textcolor{ccolor}{[height]} cm. The user has a dietary habit of \textcolor{ccolor}{[habit]}\footnote{This sentenced was added in the longitudinal study.}.}

\noindent\rule{\linewidth}{0.2pt}

\subsection{Nutritional Analysis Prompt}
\label{appx1:nutritional_analysis}

\noindent\rule{\linewidth}{0.2pt}

\noindent\texttt{\textbf{System:} You are an excellent nutritionist, capable of identifying the nutrients contained in food and drinks. Provide a detailed breakdown of nutrients for the given name and amount of food or drink based only on the provided context below.}

\noindent \texttt{Context: \textcolor{ccolor}{[context]}}

\noindent \texttt{Constraints: }
\begin{itemize}[label=-]
    \item \texttt{If you don't know the answer, say "I don't know", don't try to make up an answer.}
    \item \texttt{Output the answer following the structured output below. Don't explain or provide additional information. Don't add descriptions such as `Here are the results' and `Structured Output'. The output should be a plain text list where each line starts with the name of the nutrient, followed by a colon and then the amount and unit of the nutrient. The amount and the unit should be separated by a comma.}
\end{itemize}

\noindent \texttt{Structured Output:}
\begin{itemize}[label=-]
    \item \texttt{energy: amount without unit, kcal}
    \item \texttt{protein: amount without unit, g}
    \item \texttt{carbohydrate: amount without unit, g}
    \item \texttt{total\_fat: amount without unit, g}
    \item \texttt{dietary\_fibre: amount without unit, g}
     \item \texttt{sugars: amount without unit, g}
     \item \texttt{saturated\_fat: amount without unit, g}
     \item \texttt{trans\_fat: amount without unit, g}
     \item \texttt{cholesterol: amount without unit, mg}
     \item \texttt{calcium: amount without unit, mg}
     \item \texttt{copper: amount without unit, mg}
     \item \texttt{iron: amount without unit, mg}
     \item \texttt{magnesium: amount without unit, mg}
     \item \texttt{manganese: amount without unit, mg}
     \item \texttt{phosphorus: amount without unit, mg}
     \item \texttt{potassium: amount without unit, mg}
      \item \texttt{sodium: amount without unit, mg}
     \item \texttt{zinc: amount without unit, mg}
     \item \texttt{vitamin\_c: amount without unit, mg}
\end{itemize}

\noindent \texttt{\textbf{User:} Provide a detailed breakdown of nutrients in  \textcolor{ccolor}{[name]} with the amount of \textcolor{ccolor}{[amount] [unit]}. Present the results following the format of the Structured Output.}

\noindent\rule{\linewidth}{0.2pt}

\subsection{Dietary Suggestion Generation Prompt}
\label{appx1:dietary_suggestion}
\subsubsection{General and Personalized Dietary Suggestion Generation}
\label{appx1:general_personal}

\

\noindent\rule{\linewidth}{0.2pt}
\noindent\texttt{\textbf{System:} As an excellent nutritionist, you excel in analyzing the nutrition consumed by the user and conveying dietary suggestions based on the personal needs of the user and the context provided below.}

\noindent \texttt{User information: age \textcolor{ccolor}{[age]}, gender \textcolor{ccolor}{[gender]}, weight \textcolor{ccolor}{[weight]} kg, height \textcolor{ccolor}{[height]} cm, need \textcolor{ccolor}{[need]}.}

\noindent \texttt{Meal consumed by the user: \textcolor{ccolor}{[Meal 1. description: item 1, item 2, ...\textbackslash n Meal 2. ...]}}\\
\noindent \texttt{Nutrition consumed by the user: \textcolor{ccolor}{[Meal 1. nutrient 1: amount, nutrient 2: amount, ...\textbackslash n Meal 2. ...]}}

\noindent \texttt{Context: \textcolor{ccolor}{[context]}}

\noindent \texttt{Constraints: }
\begin{itemize}[label=-]
    \item \texttt{If you don't know the answer, say "I don't know", don't try to make up an answer.}
    \item \texttt{Output the answer following the structured output below.}
    \item \texttt{If the user's need is "None", provide only general suggestions.}
\end{itemize}

\noindent \texttt{Structured Output:}
\begin{itemize}[label=\textbullet]
    \item[] \texttt{** General Suggestions:** a list of general suggestions on healthier eating habits for a balanced diet, appropriate eating time and duration, and overall health maintenance. No more than three items. If the user's need is "None", only output general suggestions for seven items. Each item should have no more than 20 words.}
    \item[] \texttt{If user's need is "None", do not provide the following personalized suggestions:}
    \item[] \texttt{** Personalized Suggestions for \textcolor{ccolor}{[need]}: ** a list of personalized dietary suggestions based on the user's need. No more than four items. Each item should have no more than 20 words.}
\end{itemize}

\noindent \texttt{\textbf{User:} Based on the user information, the content, time, and duration of the meal or snack sessions, and nutrition of the sessions taken by the user, provide dietary suggestions for the user.}

\noindent\rule{\linewidth}{0.2pt}

\subsubsection{Nutrition Consumed Evaluation}
\label{appx1:nutrition_consumed}

\

\noindent\rule{\linewidth}{0.2pt}
\texttt{\noindent\textbf{System:} You are an excellent nutritionist. You excel in analyzing the nutrient consumed by the user and judge if each nutrient taken is too high, too low, or okay, and provide suggested values for the nutrient based on the user's information and need.}

\noindent \texttt{Description of the session: \textcolor{ccolor}{[description of the meal]}}

\noindent \texttt{Nutrition of a meal taken by the user: \textcolor{ccolor}{[nutrient 1: amount, nutrient 2: amount, ...]}}

\noindent \texttt{User information: age \textcolor{ccolor}{[age]}, gender \textcolor{ccolor}{[gender]}, weight \textcolor{ccolor}{[weight]} kg, height \textcolor{ccolor}{[height]} cm, need \textcolor{ccolor}{[need]}.}

\noindent \texttt{Context: \textcolor{ccolor}{[context]}}

\noindent \texttt{Constraints: }
\begin{itemize}[label=-]
    \item \texttt{If you don't know the answer, say "not sure", don't try to make up an answer.}
    \item \texttt{Output one of the three: too high, too low, or okay for each nutrient taken in this eating session, and provide suggested values. The suggested values should follow too high/ too low/ okay, enclosed in parentheses.}
    \item \texttt{Only output the results following the structured output. Don't explain or provide additional information. }
    \item \texttt{Don't add descriptions such as `here are the results' and `Structured Output.'}
\end{itemize}

\noindent \texttt{Structured Output:}\\
\texttt{Nutrient: too high/ too low/ okay (suggested value for this meal or snack session, only output the values)}

\noindent \texttt{\textbf{User:} Analyze each nutrition taken by the user as higher, lower, or okay, and provide suggested values for each nutrient.}

\noindent\rule{\linewidth}{0.2pt}

\subsection{AI Nutrition Assistant Prompt}
\label{appx1:assistant}

\noindent\rule{\linewidth}{0.2pt}

\noindent\texttt{\textbf{System:} \textcolor{ccolor}{[chat history]}}\\
\noindent\texttt{\textbf{User:} \textcolor{ccolor}{[question]}}\\
\noindent\texttt{\textbf{User:} Given the above conversation, generate a search query to look up to get information relevant to the conversation}

\noindent\rule{\linewidth}{0.2pt}

% \textbf{System:} [chat history]
% \textbf{User:} [question]
% \textbf{User:} Given the above conversation, generate a search query to look up to get information relevant to the conversation
% \noindent prompt 1 = [\texttt{MessagesPlaceholder}(variable\_name="chat\_history"),\\
%         ("user", "{input}"),\\
%         ("user", "Given the above conversation, generate a search query to look up to get information relevant to the conversation")]

\noindent\texttt{\textbf{System:} As an excellent nutritionist, you excel in answering users' questions based on the context provided below and the chat history.}
         
\noindent \texttt{Context: \textcolor{ccolor}{[context]}}
         
\noindent \texttt{Constraints: }
    \begin{itemize}[label=-]
    \item \texttt{If you don't know the answer, say "I don't know", don't try to make up an answer.}
    \item \texttt{Limit the answer to 150 words.}
    \end{itemize}

\noindent\texttt{MessagesPlaceholder(\textcolor{ccolor}{[chat history]})}\\
\noindent\texttt{\textbf{User:} \textcolor{ccolor}{[question]. No more than 150 words.}}

\noindent\rule{\linewidth}{0.2pt}
\section{System Implementation Details}
\label{appx:system}

\subsection{Image Segmentation with Different Regions of Interest}
\label{appx:image_seg}
\begin{figure*}[!h]
          \centering
          \subfloat[]{
            \centering
             \includegraphics[width=.17\textwidth]{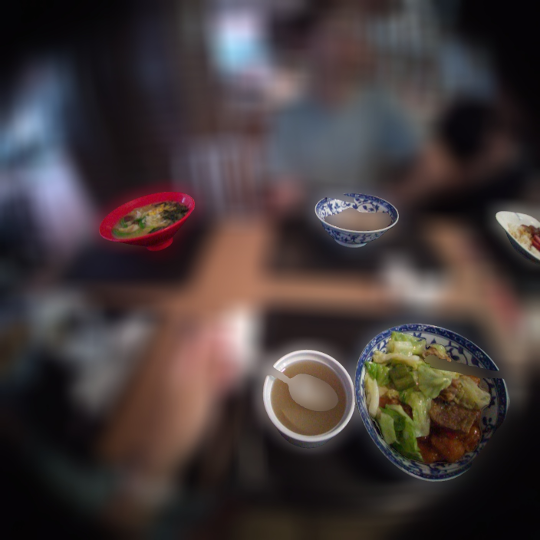}
            \label{fig:user_center1}
          }
          \subfloat[]{
            \centering
             \includegraphics[width=.17\textwidth]{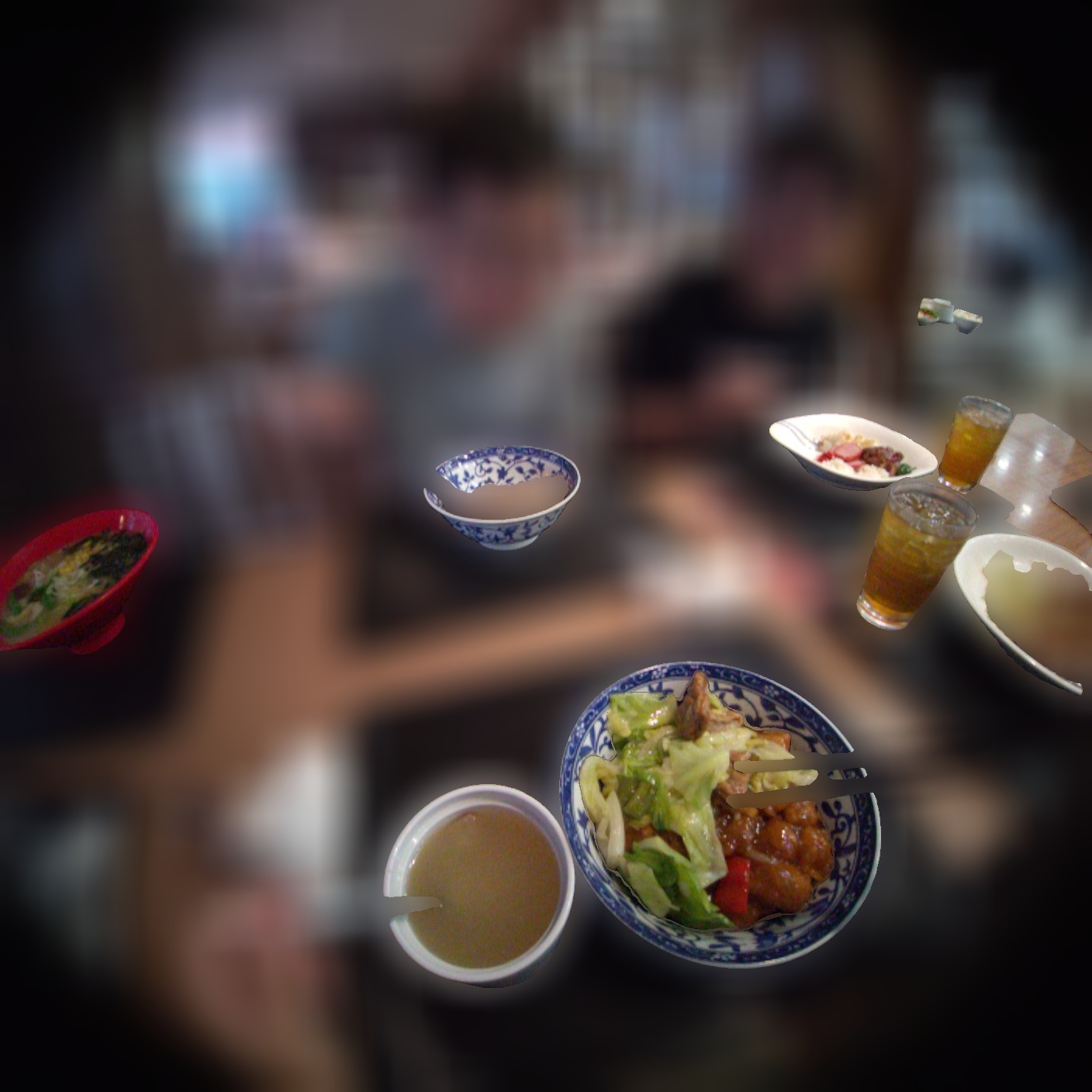}
            \label{fig:user_center2}
          }
          \subfloat[]{    
                  \centering
                  \includegraphics[width=.17\textwidth]{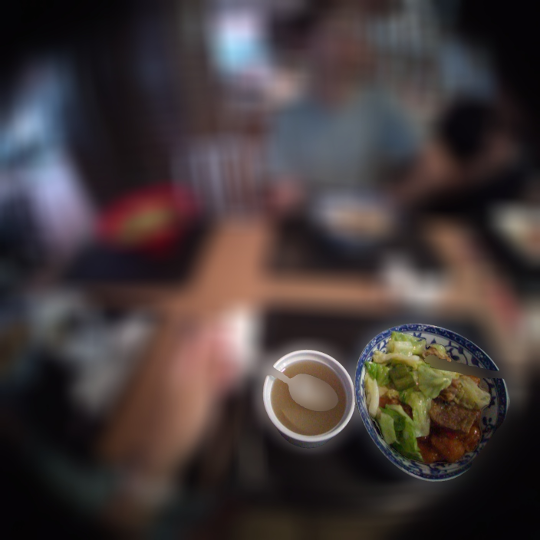}
              \label{fig:user_center3}
          }
          \subfloat[]{    
                  \centering
                  \includegraphics[width=.17\textwidth]{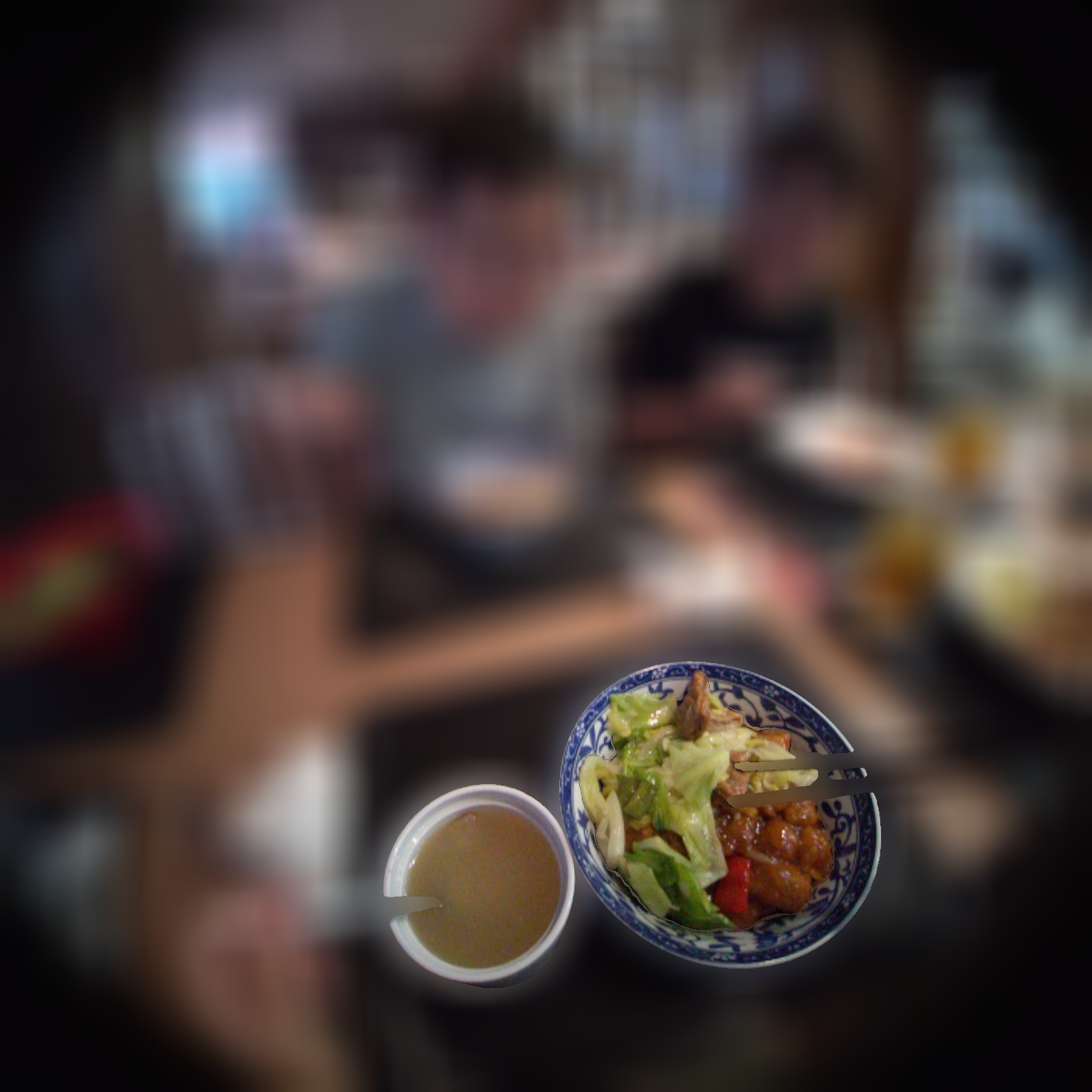}
              \label{fig:user_center4}
          }
          \caption{An illustration of the segmentation performance before and after adjusting the region of interest. (a)(b) The segmentation performance with the whole image as the region of interest. (c)(d) The segmentation performance with the bottom one-third region of the image as the region of interest.}
          \label{fig:user_center}
           % \vspace{-0.5cm}
      \end{figure*}

If the entire image was scanned, it would often capture food from people sitting nearby, as shown in \figurename{~\ref{fig:user_center1}} and \figurename{\ref{fig:user_center2}}. We observed that egocentric images typically captured the user's meal in the bottom part of the image, rather than the center, likely because users do not usually look directly down at their meals while eating. To address this, we set the region of interest as the bottom one-third of the image for segmentation, as shown in \figurename{~\ref{fig:user_center3}} and \figurename{\ref{fig:user_center4}}.

\subsection{Nutrition Library Construction}
\label{appx2:nutrition}
The nutrition library included the following items:

\begin{itemize}
    \item Food Nutrient dataset provided by the Centre of Food Safety from the Government of the Hong Kong SAR, which combined the data from The Nutrient Data Laboratory, United States Department of Agriculture (SR22), The Institute of Nutrition and Food Safety, Chinese Center for Disease Control and Prevention (CFC2002Ed2), The Food Research Laboratory of the Centre for Food Safety (FRL), Food Standards Australia New Zealand (2022), Australian Food Composition Database, Institute of Nutrition, Mahidol University, Thailand (FCD2002), and local dishes. It includes the nutrient information of 7,032 foods and drinks from 27 categories, i.e., alcoholic beverages, aquatic animals other than fish, baby foods, cereals and cereal products, condiments and sauces, eggs and egg products, fats and oils, fish and fish products, frozen confections, fruits and fruit products, games and game products, herbs and spices, legumes and legume products, meat and meat products, milk and milk products, non-alcoholic beverages, nuts, seeds and their products, poultry and poultry products, ready-to-eat foods, salads, snacks, soups, sugars and sweets, vegetables and vegetable products, local breakfast, local Chinese Dim Sim, Local Congee, Rice and Noodles Dishes.

    % (i.e., U.S. Department of Agriculture FoodData Central Data, European Food Safety Authority, Canadian Nutrient File, UK Food Composition Databases, Food Standards Australia New Zealand, and Open Food Facts Data).
    \item Forty-one reports and booklets about food consumption and dietary suggestions for the general public and specific groups (i.e., diabetes and hypertension) from the government and the World Health Organization (WHO).
    \item Ninety-five web pages of recipes, science articles about food, nutrition, and healthy diet suggestions from the official websites of the government.
    \item Seventy-one related articles from National Institutes of Health Harvard T.H. Chan School of Public Health, UnityPoint Health, Medical News Today, and Healthline.
    \item 251 research articles are related to the dietary suggestions for people with 28 different needs: general health (4), weight management (11), muscle development (10), blood glucose stabilization (diabetes prevention) (14), digestive health improvement (8), cardiovascular health (7), bone health (11), cancer prevention (10), mood regulation and mental health improvement (9), uric acid reduction (10), immune enhancement (8), sleep improvement (10), control chronic kidney disease (10), anti-aging of skin (14), endocrine regulation (9), memory preservation (10), menopausal symptom relief (8), physical restoration (6), eye health protection (10), prenatal/lactation nutrition (4), menopause/elderly nutrition (6), chronic disease prevention/management (10), metabolic regulation (14), hepatitis patients (6), physically disability (7), exercise nutrition support (11), allergy-prone (4), and vegetarians (10).
\end{itemize}

\subsection{Detailed Introduction of the User Interface}
\label{appx:system_interface}
\subsubsection{Meal Image Logging}
As shown in \figurename{~\ref{fig:interface}-A1}, meal log thumbnails were displayed chronologically, allowing users to review their meal sessions (\textbf{R1}) and reflect on eating patterns. Users can click on the thumbnails to view the corresponding meal images (\figurename{~\ref{fig:interface}-A2}), which have been blurred and cropped for privacy protection (\textbf{R4}). By observing the visual logs of their diets, users can gain insights into the meal diversity and dietary changes over time. Specifically, by clicking on each meal session, the corresponding Detailed Nutritional Analysis and Total Nutrition Consumed in this meal session would be presented in \figurename{\ref{fig:interface}-B} (\textbf{R2}). 

\subsubsection{Interactive Nutritional Analysis}
Interactive Nutritional Analysis included Detailed Nutritional Analysis and Total Nutrition Consumed. An editable meal summary table showed the description, amount, and consumed percentage of items (\figurename{\ref{fig:interface}-B1}, \textbf{R1}). 
The Detailed Breakdown (\figurename{~\ref{fig:interface}-B2}) presented the detailed nutritional analysis of each item (\textbf{R2}). Users can update, add, or delete the description, amount, and consumed percentage in the meal summary table to correct inaccurate or incorrect information, and the Detailed Breakdown, Total Nutrition Consumed, and Dietary Suggestions will be updated accordingly.

The Total Nutrition Consumed section presented the user's total nutrient consumption per meal, with suggestions on whether nutrients were too high, low, or appropriate, alongside recommended values (\figurename{~\ref{fig:interface}-B3}, \textbf{R2}, \textbf{R3}). Users can view the trend of nutrient consumption across meals using a toggle switch (\figurename{~\ref{fig:interface}-B4}), directly identifying the meals with nutrient consumption out of suggested ranges.

\subsubsection{Dietary Suggestions}
The dietary suggestions, including the general suggestions, personalized suggestions, and the sources, were presented in \figurename{~\ref{fig:interface}-C} (\textbf{R3}). The general suggestions provide broad dietary guidance focusing on balanced nutrition and overall health maintenance according to the user's profile, while the personalized suggestions were specific to addressing the dietary goal of the user. To increase the reliability of the suggestions and the user's trust, the sources of these suggestions were presented following the suggestions. For those without specific dietary goals or requirements, only general dietary suggestions and sources would be presented. We limited the total number of suggestions to seven items to avoid overwhelming the user with too much information.

\subsubsection{Context-Aware AI Chatbot}
Users can seek additional guidance from the AI Nutrition Chatbot (\figurename{~\ref{fig:interface}-D}, \textbf{R3}). The chatbot held the memory of the user profile, meal logs, nutritional analysis, suggestions, and chat history for contextually relevant responses. Built using the LangChain framework, it reformulated questions by referencing previous interactions for accurate and informed answers. More specifically, when a user asked a question, the system retrieved relevant context and documents, updating the conversation with these sources. Responses were generated using both the query and prior dialogue, with sources provided.
The chat history would be updated after each interaction. To streamline the user experience, common questions were added (\figurename{~\ref{fig:interface}-D1}).

\section{Description of Participants}
\label{appx:participants}

The description of the 33 participants attending Study I is shown in \tablename{~\ref{tab:participants}. Sixteen participants among them (P1, P3, P6-P10, P13-P18, P20-P22) further attended the Study II.

    \begin{table}[!t]
      \footnotesize
      \renewcommand{\arraystretch}{1}
      % \parbox{\linewidth}{
        % \centering
        \caption{Description of participants, including their unique ID, age, self-declared gender, occupation background, dietary monitoring experience, dietary goal, daily glasses-wearing behaviors, and meal sessions collected.}
        \label{tab:participants}
            \begin{tabular}{lllllllll}
          \toprule
          \bfseries  ID & \bfseries Age & \bfseries Gender  & \bfseries Occupation & \bfseries Experience & \bfseries Dietary Goal & \bfseries Glasses Wearing & \bfseries Meals & \bfseries Dietary Habit$^*$\\
          \midrule
          P1 & 28 & F & Student & exp. & Lose Weight & Never & 9 & None\\
          P2 & 28 & F & Student & exp. & Lose Weight & Never & 6 & Not Applicable\\
          P3 & 27 & M & Student & inexp. & Gain Muscle & All the time & 5 & None\\
          P4 & 24 & M & Student & exp. & Gain Muscle & All the time & 5 & Not Applicable \\
          P5 & 30 & F & Student & inexp. & keep Fit & All the time & 4& Not Applicable\\
          P6 & 28 & F & Postdoc &exp. & Mitigate Pimple & All the time & 7 & Never drink coffee or milk tea \\
          P7 & 26 & M & Student & exp. & Lose Weight & Never & 8 & Never eat the bottom bread slice of \\
          & & & & & & & & hamburgers, only drink sugar-free Coke\\
          P8 & 29 & M & Student & inexp. & Lose Weight& Most of the time& 6 & None\\
          P9 & 24 & F & Student & inexp. & Lose Weight & All the time& 5 & Drink sugar-free coffee at breakfast,\\
           & & & & & & & &  sugar-free milk tea at lunch and dinner\\
          P10 & 27 & M & Postdoc & exp. & None & Rarely& 5 & Drink sugar-free soda\\
          P11 & 35 & M &Postdoc & exp. & Lose Weight & Never & 3 & Not Applicable\\
          P12 & 60 & F & Accountant & inexp. & Lose Weight & Never& 3 & Not Applicable\\
          P13 & 27 & M & Student & exp. & Lose Weight & All the time & 3 & All drinks are sugar free.\\
          P14 & 31 & M &  Professor& inexp. & None & All the time& 2 & Drink coffee milk tea at lunch \\
          P15 & 28 & M &Student & exp. & Gain Muscle & All the time& 3 & None\\
          P16 & 30 & F & Postdoc & inexp. & None & Never & 3 & Add sugar to lemon water and lemon tea\\
          P17 & 32 & M & Postdoc & inexp. & Gain Weight & All the time & 3 & Drink hot lemon water\\
          P18 & 25 & M & Student& exp. & Gain Muscle & All the time& 5 & None\\
          P19 & 29 & F &  Professor & inexp. & Lose Weight & Never & 3 & Not Applicable\\
          P20 & 28 & M & Student & exp. & Keep Fit & All the time & 2 & Only drink soup or iced lemon tea\\
          P21 & 33 & M & Postdoc & exp. & Lose Fat,  & All the time & 3 & Drink sugar-free coke, milk tea,\\
          & & & & &  Gain Muscle& & &  and black coffee\\
          P22 & 30 & M & Student & inexp. & None & Never & 7 & None\\
          P23 & 43 & F &  Professor & exp. & None & Never & 7& Not Applicable\\
          P24 & 46 & M & Engineer & exp. & None & Never & 5& Not Applicable\\
          P25 & 73 & M & Retired & exp. & None & Never & 4& Not Applicable\\
          P26 & 26 & M & Student & exp. & Lose Weight & All the time& 3& Not Applicable\\
          P27 & 25 & M & Student & inexp. & Gain Weight & All the time & 4& Not Applicable\\
          P28 &  23 & M & Student & exp. & None & All the time& 4& Not Applicable\\
          P29 & 23 & M & Student & exp. & Lose Weight & All the time& 4& Not Applicable\\
          P30 & 24 & M & Student & exp.& Lose Weight & All the time & 4& Not Applicable\\
          P31 & 23& F & Student & exp. & Lose Weight & All the time & 4& Not Applicable\\
          P32 & 24& F& Student & exp.& Lose Weight & Sometimes & 3& Not Applicable\\
          P33 & 25 & F& Student & inexp. & None & All the time& 2& Not Applicable\\      
        \bottomrule
        \end{tabular}
        \footnotesize{$^{*}$ The dietary habits were collected only for participants who attended the longitudinal study. `Not Applicable' means we did not collect this information for the participant.}
      % }
    \end{table}

\section{Survey Scale and Interview Script}
\label{appx:scale}
The survey scale used in both Study I and Study II was shown in \tablename{~\ref{tab:scale}}, including System Usability Scale (SUS), Net Promoter Score (NPS), User Experience Questionnaire (UEQ), and questions about users' perception on the diet identification and logging (R1), nutritional analysis (R2), personalized dietary suggestions (R3), and privacy protection (R4). The interview script used in Study I is shown in \tablename{~\ref{tab:interview}}, and the weekly interview script is as follows:
\begin{itemize}
    \item What features did you find most useful or challenging this week? Have you noticed new features you didn't use in the previous weeks?
    \item Have you noticed any changes in your dietary habits since using the system this week? If yes, how do you find it? By the information provided by the product? Or your personal subjective sense? Or others?
   \item Do you have any suggestions for improvements or new features?
  \item Is there anything else you would like to share about your experience this week?
  \item (Last week) Is there anything else you would like to share about your experience with the system across the four weeks? 
\end{itemize}

     \begin{table}[!t]
      \footnotesize
      \renewcommand{\arraystretch}{1}
      % \parbox{\linewidth}{
        % \centering
        \caption{Measurements of system's usability, user satisfaction, experience, and perception on diet identification and logging, nutritional analysis, professional dietary suggestions, and privacy protection toward \shortname. }
        \label{tab:scale}
            \begin{tabular}{lll}
          \toprule
           \bfseries Scale &  \bfseries Item \\
           \midrule
           SUS & Q1. I think that I would like to use this system frequently.\\
           &Q2. I found the system unnecessarily complex.\\
           & Q3. I thought the system was easy to use. \\
            & Q4. I think that I would need the support of a technical person to be able to use this system.\\
            & Q5. I found the various functions in this system were well integrated.\\
            & Q6. I thought there was too much inconsistency in this system.\\
            & Q7. I would imagine that most people would learn to use this system very quickly.\\
            & Q8. I found the system very cumbersome to use.\\
            & Q9. I felt very confident using the system.\\
            & Q10. I needed to learn a lot of things before I could get going with this system.\\
            \midrule
            NPS &  Q11. On a scale from 0 to 10, how likely are you to recommend this system to a friend or colleague? \\
            & Q12. What is the primary reason for your score?\\
            \midrule
            UEQ & Q13. (attractiveness) The system is attractive. \\
            & Q14. (Perspicuity) The system is easy to understand.  \\
            & Q15. (Efficiency) The system is fast and efficient to use. \\
            & Q16. (Dependability) The system is predictable and reliable. \\
            & \quad\quad Predicable: users can anticipate how the product will respond to their actions.\\
            & \quad\quad Reliable: the product consistently performs its functions without errors. \\
            & Q17. (Stimulation) The system is exciting and motivating to use. \\
            & Q18. (Novelty) The system is innovative and creative. \\
            \midrule
            Diet & Q19. The system is accurate in identifying and logging your meals. \\
            Identification & Q20. It is easy to use the system for capturing meals. \\
             and Logging & Q21. Please describe any issues you encountered with the automated logging process \\
            \midrule
            Nutritional & Q22. The nutritional analysis provided by the system is accurate \\
           Analysis & Q23. The nutritional analysis is useful for your dietary planning \\
            & Q24. Please describe any discrepancies between the system’s nutritional values and your expectations. \\
            \midrule
            Personalized &Q25. The dietary suggestions were relevant to your dietary goals. \\
           Dietary    & Q26. The dietary suggestions were easy to follow. \\
           Suggestions & Q27. I trust the dietary suggestions. \\
            \midrule
            Privacy  & Q28. I'm confident my personal data is secure with this system. \\
            Protection & Q29. How my data used in this system is transparent.\\
            & Q30. Please describe have any concerns about privacy when using the system. \\
          \bottomrule
        \end{tabular}
      % }
    \end{table}
    
% \subsection{Study I User Study Interview Script}

     \begin{table}[!t]
      \footnotesize
      \renewcommand{\arraystretch}{1}
      % \parbox{\linewidth}{
        % \centering
        \caption{The interview script used in Study I.}
        \label{tab:interview}
            \begin{tabular}{lll}
          \toprule
           \bfseries R1: Supporting Automated Diet Identification and Logging\\
           \midrule
           Q1. Describe your overall experience using this system for meal logging.\\
            Q2. Do you think the system is accurate in meal logging?\\
            Q3. Were there any meals or food items that the system struggled to identify?\\
            Q4. What improvements would you suggest for the food identification and logging feature?\\
            \midrule
            \bfseries R2: Facilitating Nutritional Analysis\\
            \midrule
            Q1. How was your experience with the nutritional analysis feature?\\
            Q2. Were there nutrients or food items where the analysis seemed accurate or inaccurate?\\
            Q3. Do you trust the nutritional analysis results? Why? What elements in the system increase your trust.\\
            Q4. How has the nutritional analysis influenced your dietary decisions?\\
             \midrule
             \bfseries R3: Providing Personalized Dietary Suggestions\\
             \midrule
            Q1. How did you use the personalized dietary suggestions in your daily routine?\\
            Q2. Do you think the dietary suggestions are reliable? Why? What elements in the system make them reliable or unreliable?\\
            Q3. Were there any suggestions that were particularly helpful or unhelpful?\\
            Q4. What changes or additional features would you recommend?\\
             \midrule
            \bfseries R4: Protecting Privacy\\
             \midrule
            Q1. How do you feel about the system's approach to data privacy?\\
            Q2. Were there any privacy strategies or features that reassured or concerned you?\\
            Q3. What recommendations do you have for improving the system’s privacy features?\\
             \midrule
             \bfseries General Questions \\
             \midrule
             Q1. Which features of the system did you find most useful?\\
            Q2. Were there any features that you did not use or found unnecessary? Why?\\
            Q3. How do you think the system could be improved to better meet your needs?\\
            Q4. Can you describe any specific scenarios where the system was particularly helpful?\\
            Q5. How satisfied are you with the overall functionality of the system?\\
            Q6. How do you think our system compares to other dietary monitoring tools in terms of advantages and disadvantages?\\
            Q7. Is there anything else you would like to share about your experience with the system?\\
          \bottomrule
        \end{tabular}
      % }
    \end{table}

\section{General Usability, User Satisfaction, and User Experience in Study I.1}
\label{appx:general}
\subsection{General System Usability}
The highest score (95) came from both exp. and inexp. participants (P2, P17, and P32), while the lowest (42.5) was given by an inexp. participant (P12), highlighting challenges faced by users with limited technological experience. Percentiles (67.5, 77.5, 85) demonstrated overall positive feedback.
Exp. participants appreciated the system's automation and ease of use, while inexp. participants needed more time to adapt, citing workflow complexity as a challenge. These findings underscore the importance of tailoring onboarding and support features to better accommodate inexp. users, particularly those with limited technological familiarity.

\subsection{General User Satisfaction}
User satisfaction, measured via NPS, averaged 8.3 ($SD=1.49$), with slightly higher ratings from the exp. group ($M=8.43$, $SD=1.57$) compared to the inexp. group ($M=8.08$, $SD=1.38$), suggesting familiarity with similar tools may boost satisfaction.
Nine participants (6 exp., 3 inexp.) gave the highest score of 10, citing the system’s usefulness, accurate analysis, and personalized dietary suggestions. In contrast, the lowest score (5) came from P24 (exp.), who raised concerns about the glasses’ cost and weight. Physical discomfort, such as glasses trapping sweat during prolonged use (P15, exp.), was noted.
Percentiles (7, 9, 10) indicated high overall satisfaction. These findings highlight the system's strong utility while underscoring the importance of addressing physical comfort and usability concerns to enhance satisfaction across all user groups.

\subsection{General User Experience}
The UEQ measured user experience across six dimensions: attractiveness, perspicuity, efficiency, dependability, stimulation, and novelty, mapped to a -3 to +3 scale. Perspicuity ($M=2.18$) and novelty ($M=2.27$) scored the highest, reflecting a generally positive experience across both exp. and inexp. groups.
Efficiency showed greater variability ($M=1.91$, $SD=1.14$), with the lowest individual rating (-1.50) from an inexp. participant (P12, 60 years old), highlighting challenges for less tech-savvy users. The exp. group rated efficiency higher ($M=2.00$, $SD=1.10$) than the inexp. group ($M=1.75$, $SD=1.25$), suggesting familiarity improved system interaction.
Attractiveness, dependability, and stimulation received satisfactory ratings, though stimulation had the most neutral responses (6 participants). These results suggest the system provides a broadly positive experience, with opportunities to improve efficiency and stimulation, particularly for inexp. users, to further enhance usability.

 \begin{figure*}[!t]
          \centering
          \subfloat[]{
            \centering
            \includegraphics[width=.25\textwidth]{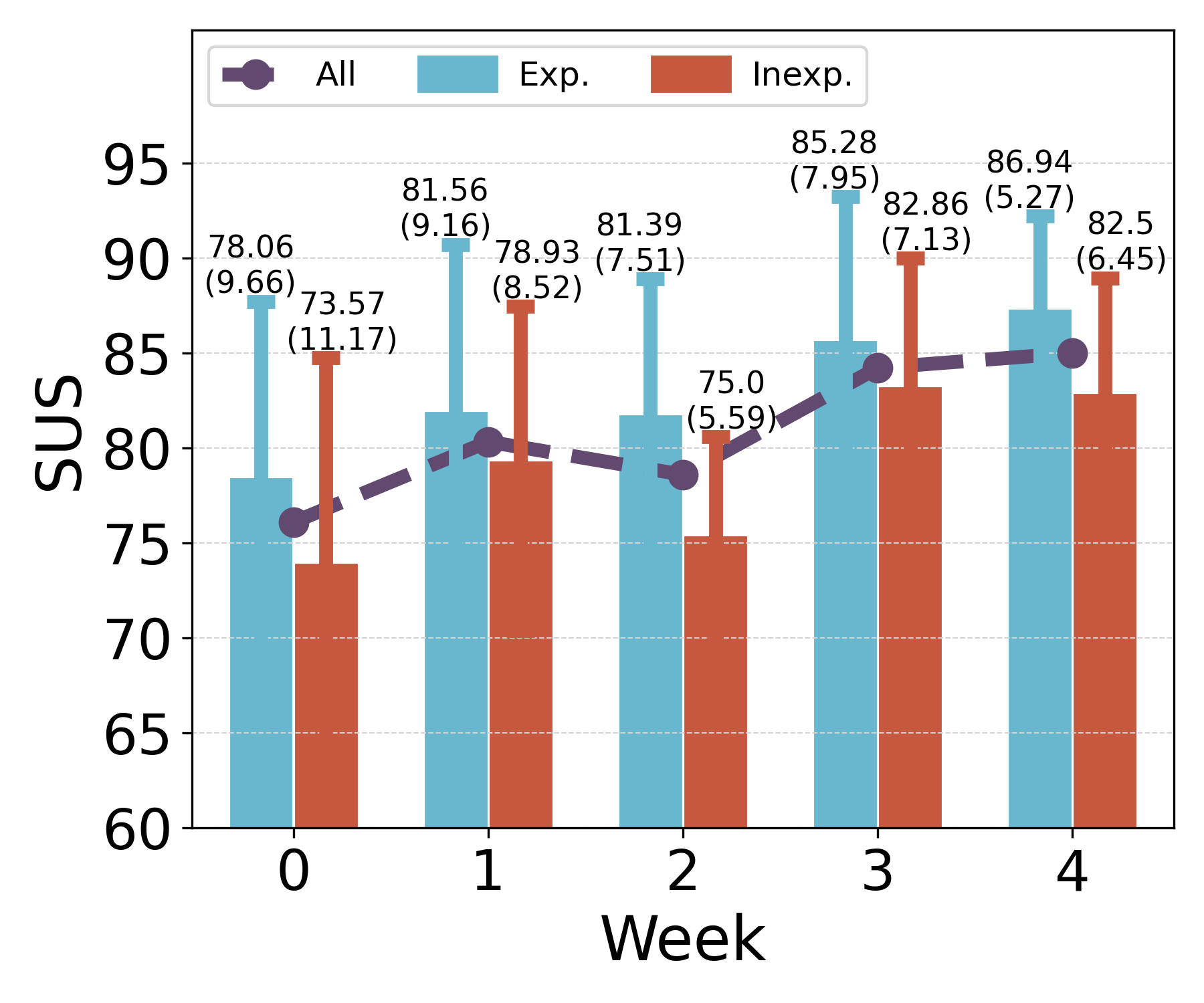}
            \label{fig:sus}
          }
          \subfloat[]{    
                  \centering
                  \includegraphics[width=.25\textwidth]{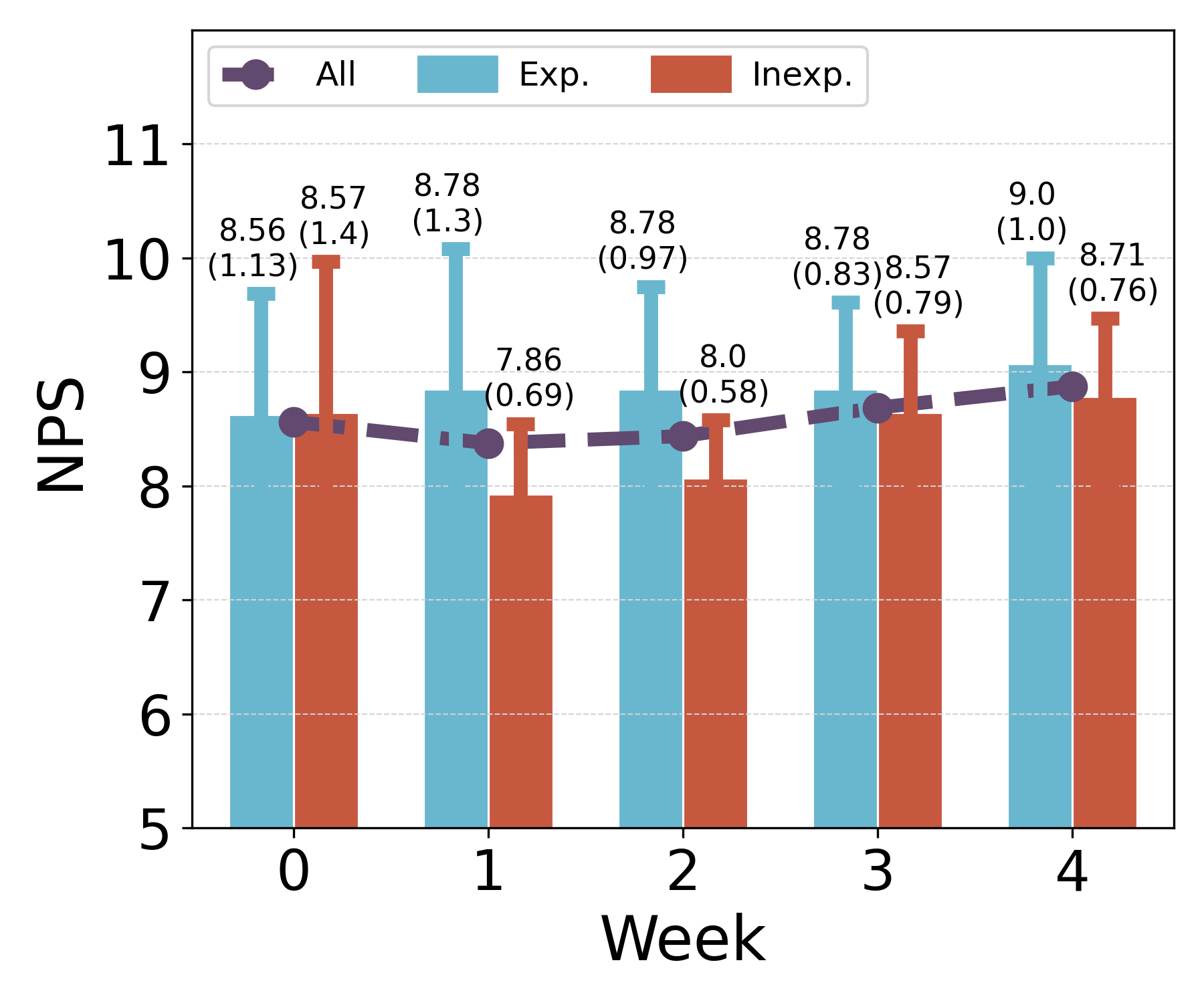}
              \label{fig:nps}
          }
          \subfloat[]{    
                  \centering
                  \includegraphics[width=.25\textwidth]{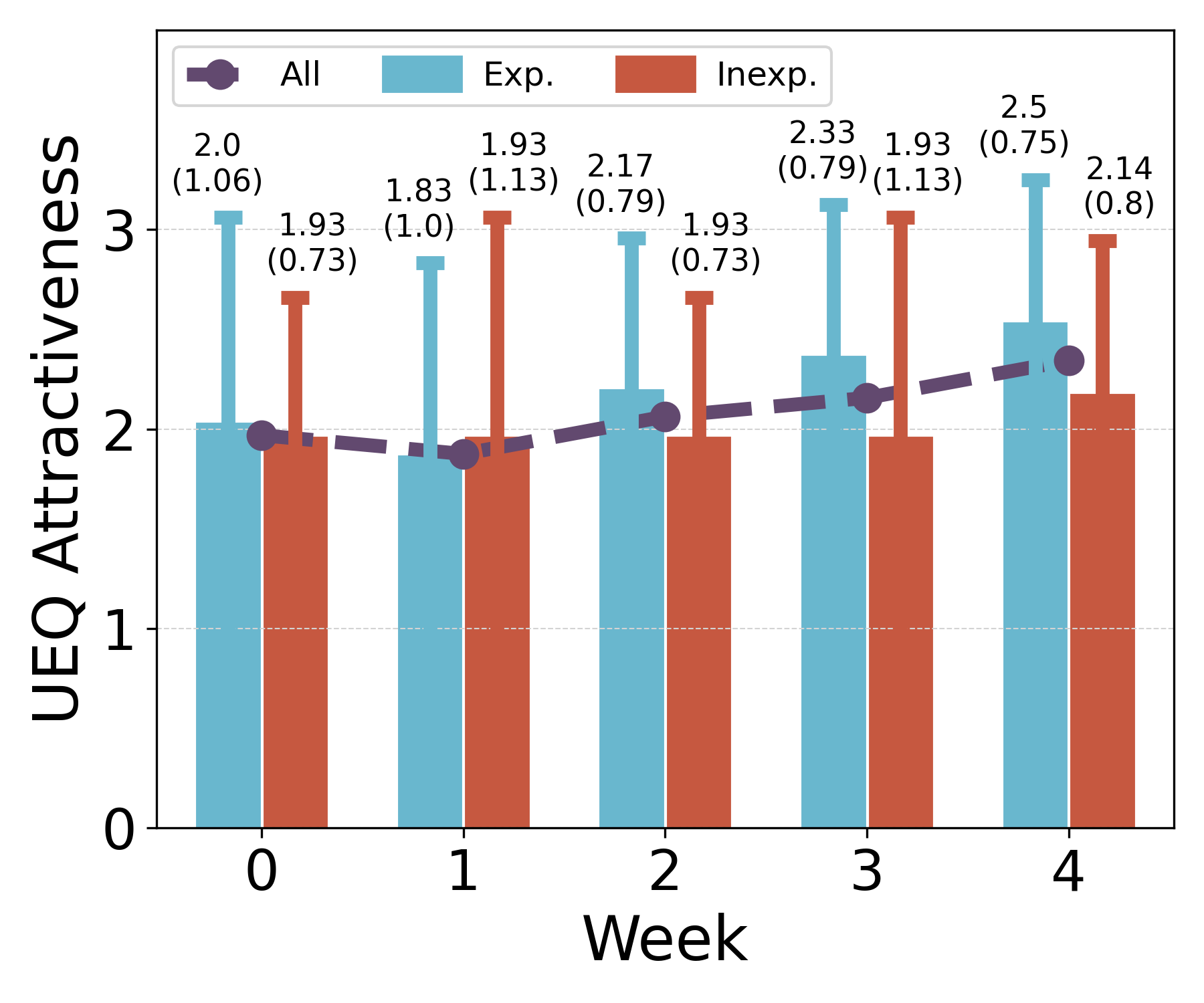}
              \label{fig:ueq-a}
          }
          \subfloat[]{    
                  \centering
                  \includegraphics[width=.25\textwidth]{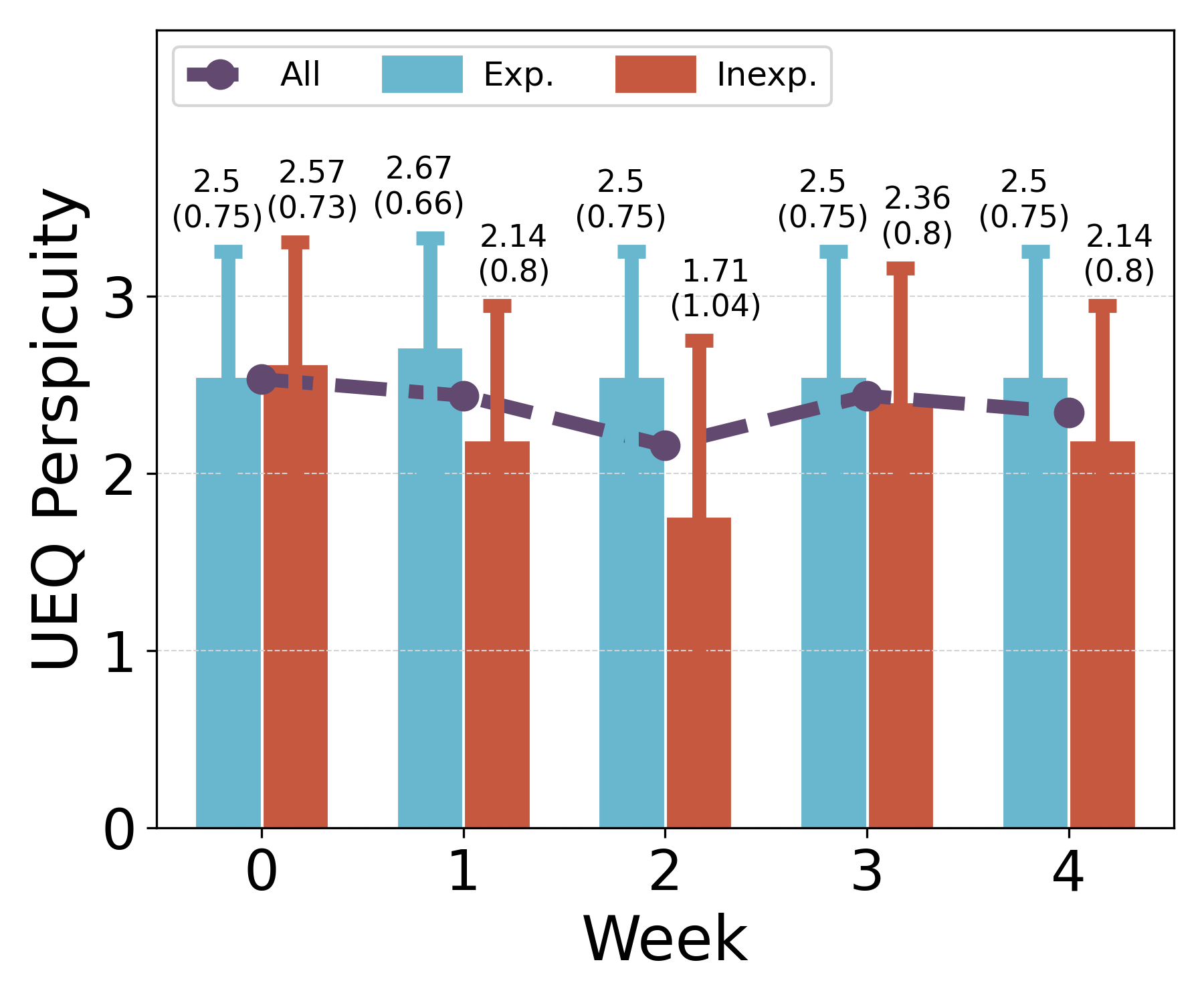}
              \label{fig:ueq-p}
          }\\
            \subfloat[]{    
                  \centering
                  \includegraphics[width=.25\textwidth]{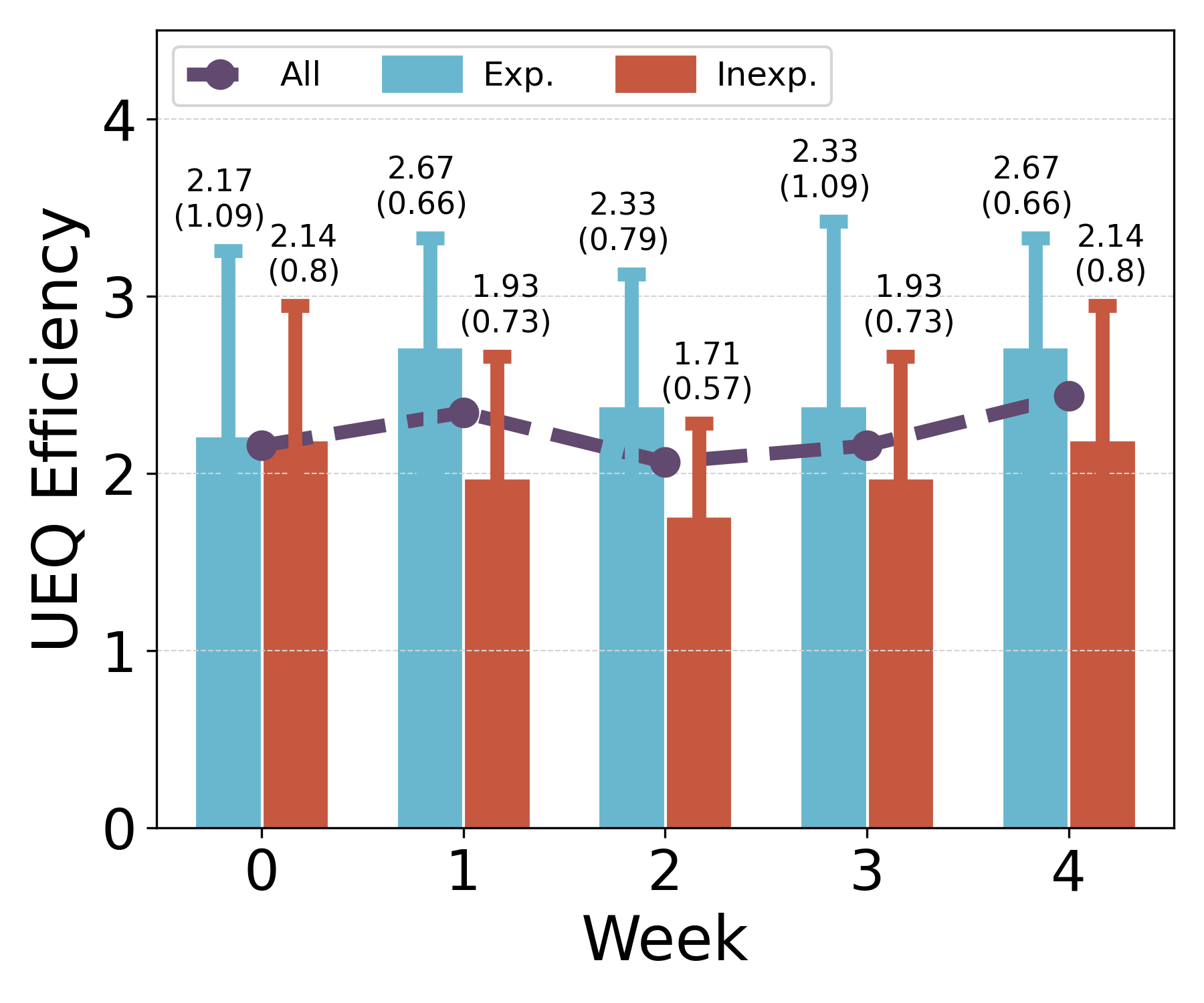}
              \label{fig:ueq-e}
          }
          \subfloat[]{    
                  \centering
                  \includegraphics[width=.25\textwidth]{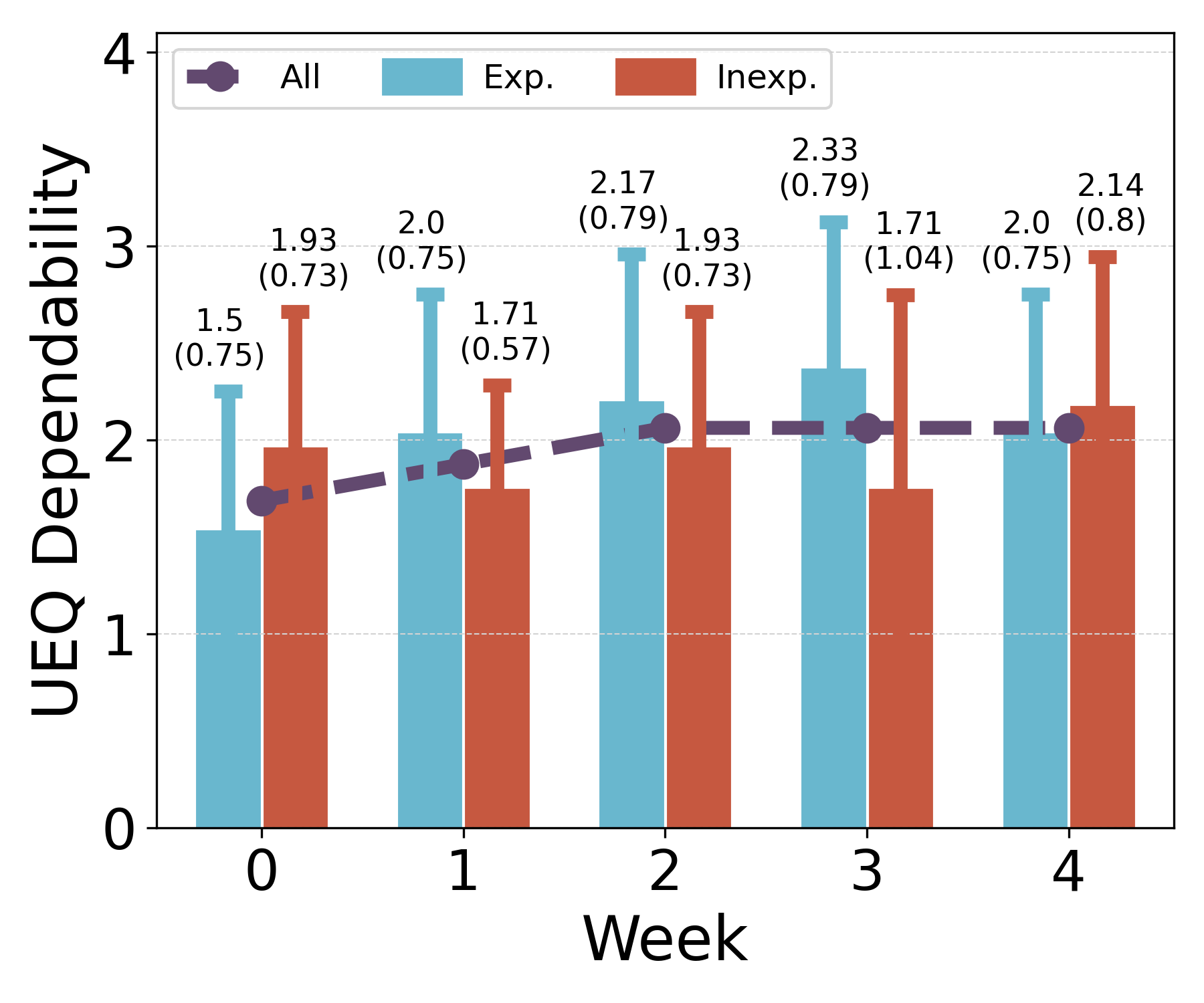}
              \label{fig:ueq-d}
          }
        \subfloat[]{    
                  \centering
                  \includegraphics[width=.25\textwidth]{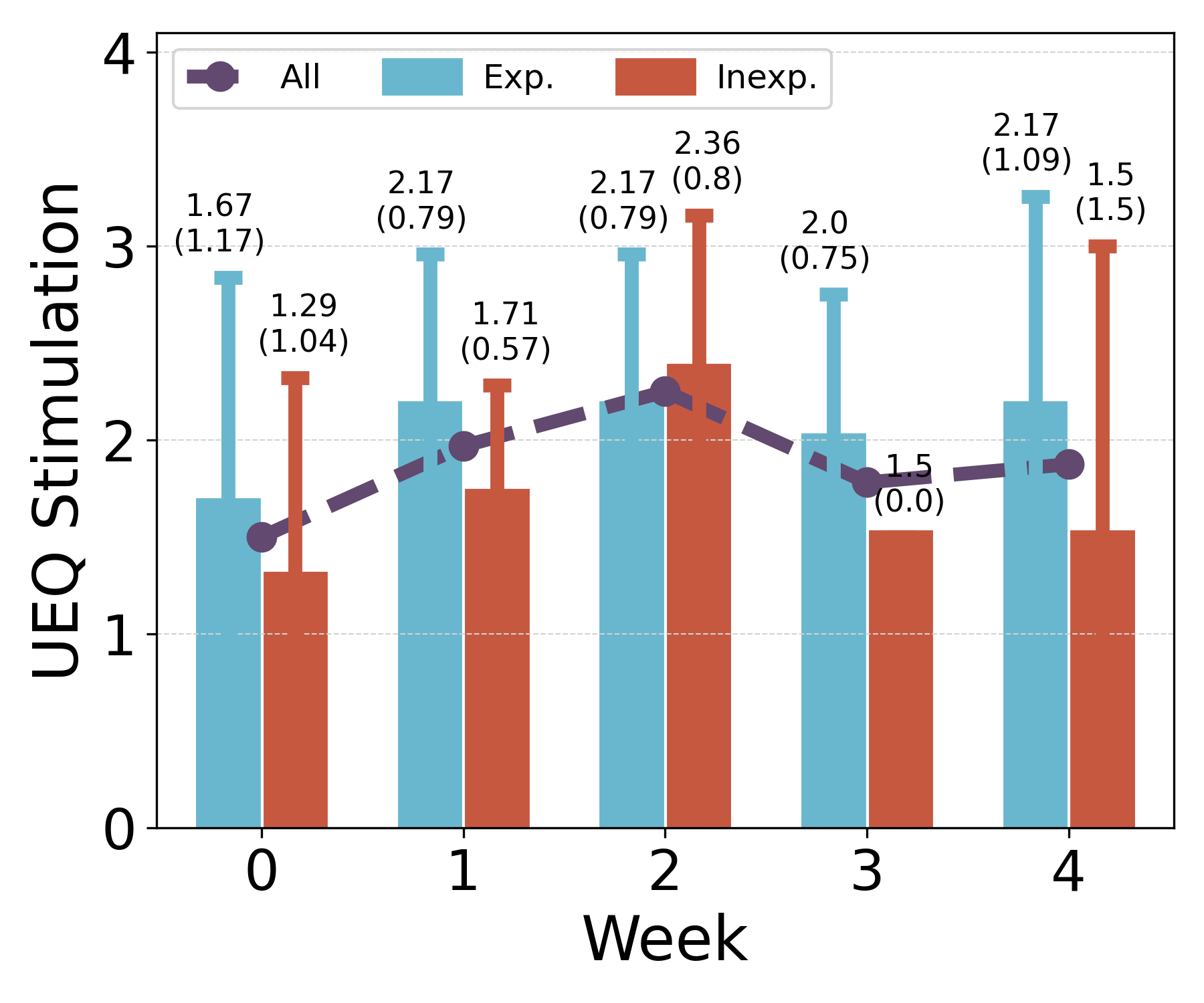}
              \label{fig:ueq-s}
          }
          \subfloat[]{    
                  \centering
                  \includegraphics[width=.25\textwidth]{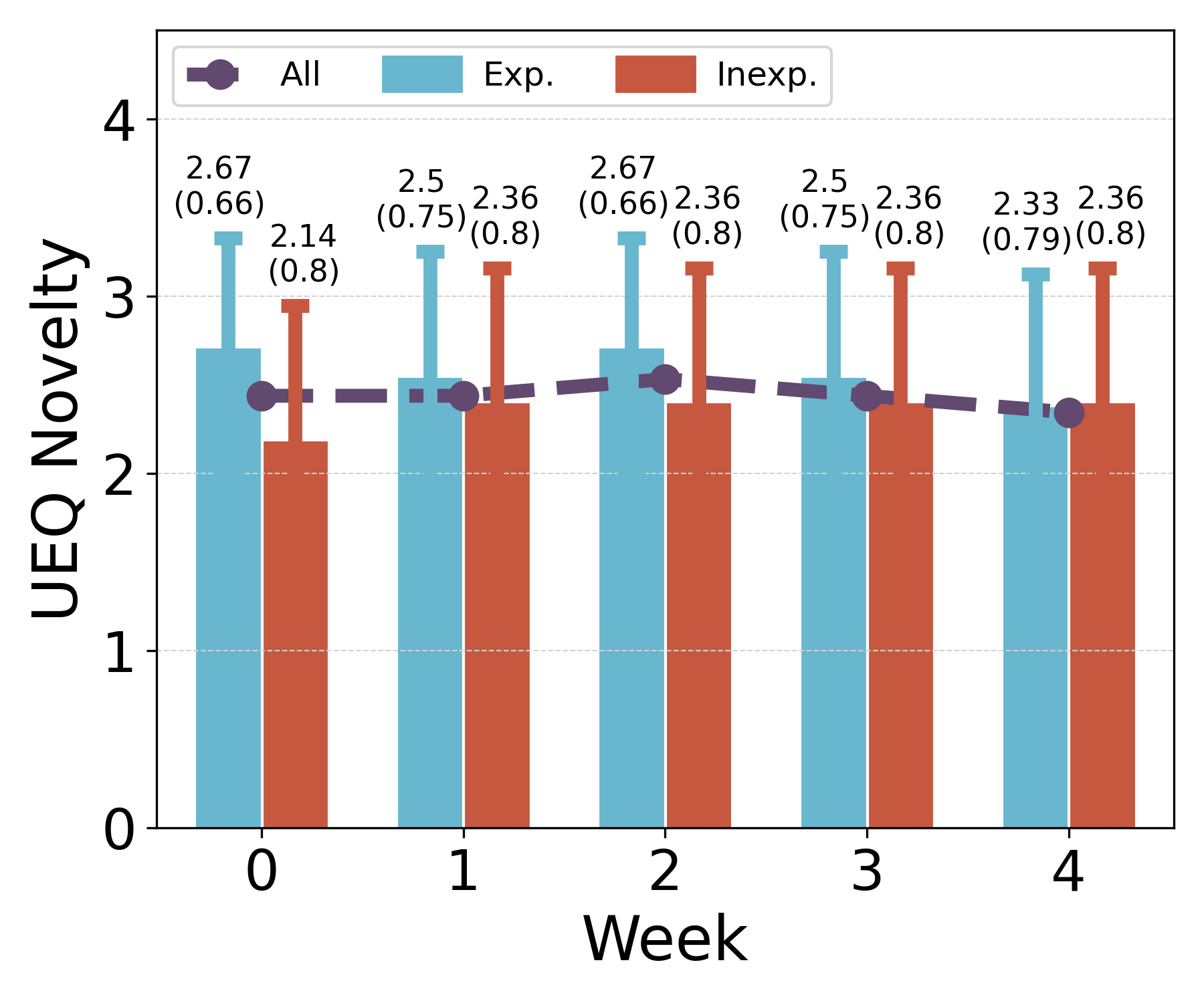}
              \label{fig:ueq-n}
          }
          \caption{The four-week trends of general usability, user satisfaction, and experience. (a) General usability - SUS. (b) General user satisfaction - NPS. (c) User experience - attractiveness. (d) User experience - perspicuity. (e) User experience - efficiency. (f) User experience - dependability. (g) User experience - stimulation. (h) User experience - novelty. Week 0 was in study I, and Weeks 1-3 were in study II. The values above the bars are the mean (standard deviation).}
          \label{fig:scales-standard}
           % \vspace{-0.5cm}
      \end{figure*}

\section{An example of conversation between P6 and the chatbot}
\label{appx:chat}
For example, as shown in \figurename{~\ref{fig:chat}}, after receiving a meal plan from the chatbot, P6 added that she did not eat breakfast and snacks. The chatbot revised the meal plan accordingly, removing breakfast and snacks. P6 further asked the chatbot to provide a Chinese-style meal plan and got a response. Next, she told the chatbot she did not like Spring Onion and Brown Rice, and the chatbot further revised the meal plan. She also asked the chatbot to propose some food that could help her keep from getting pimples and what food she had eaten should be avoided.

 \begin{figure*}[!t]
          \centering
          \subfloat[]{
            \centering
            \includegraphics[width=.5\textwidth]{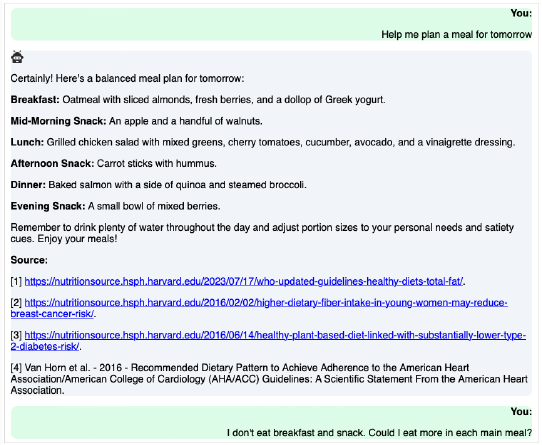}
            \label{fig:chat1}
          }
          \subfloat[]{    
                  \centering
                  \includegraphics[width=.5\textwidth]{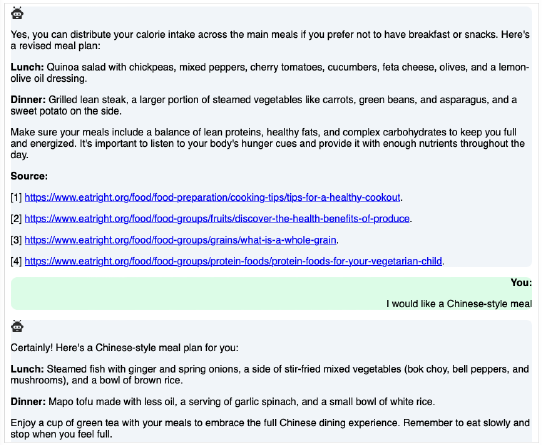}
              \label{fig:chat2}
          }\\
          \subfloat[]{    
                  \centering
                  \includegraphics[width=.5\textwidth]{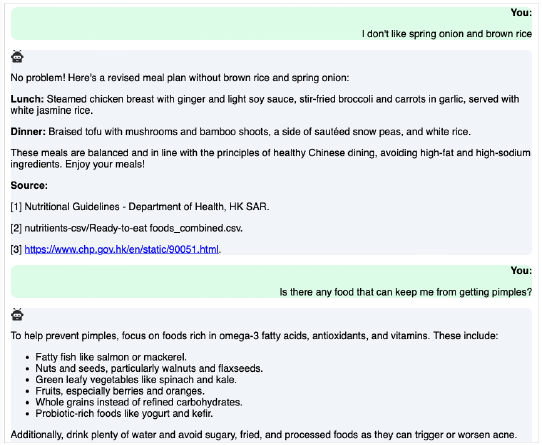}
              \label{fig:chat3}
          }
          \subfloat[]{    
                  \centering
                  \includegraphics[width=.5\textwidth]{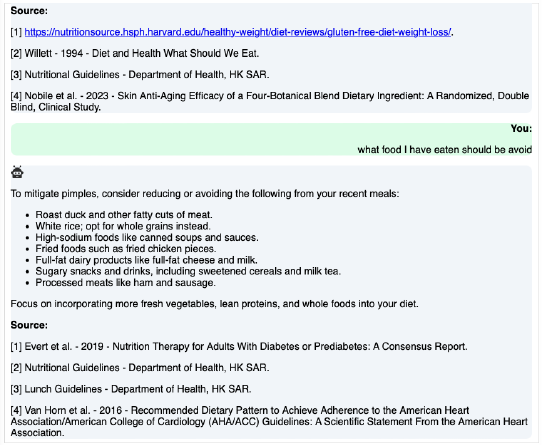}
              \label{fig:chat4}
          }
          % \subfloat[]{    
          %         \centering
          %         \includegraphics[width=.225\textwidth]{Figures-clean/.png}
          %     \label{fig:teaser5}
          % }
          \caption{An example of a conversation between P6 and the chatbot, (a)$\rightarrow$(b)$\rightarrow$(c)$\rightarrow$(d).}
          \label{fig:chat}
           % \vspace{-0.5cm}
      \end{figure*}

\section{Agreement among Nutrition Experts}
\label{appx:agreement}
We assessed the expert consistency in nutrient value estimation. 
% and then evaluated the alignment of \shortname's estimation with the experts' estimation. 
More specifically, we first measure the inter-expert agreement using the Intraclass Correlation Coefficient (ICC) to evaluate the agreement between experts. Since each meal was estimated by 5 experts randomly sampled from the 10 experts, and the meals are considered random subjects, we use the Two-Way Random Effects Model, i.e., ICC(2,1) and ICC(2,k). ICC(2,1) measures the reliability of individual estimation, considering the variance between experts and the meals. ICC(2,k) measures the reliability of the mean estimation across the experts. 
% We measure the ICC(2,1) and ICC(2,k) among both experts and the estimation of \shortname to evaluate how well \shortname's estimation aligns with experts' estimation in terms of consistency.
The results are shown in \tablename{~\ref{tab:icc}}. We found that the average estimation among the experts was highly reliable, while the individual estimation for nutrients showed low to moderate agreement. This indicates that there was an inconsistency between different experts in the estimation of specific nutrients in the meals. The high agreement on average ratings indicates that the average results from the experts can be highly reliable estimations.

    \begin{table}[!t]
      \small
      \renewcommand{\arraystretch}{1}
      % \parbox{\linewidth}{
        \centering
        \caption{The results of ICC(2,1) (95\% Confidence Interval) and ICC(2,k) (95\% Confidence Interval).}
        \label{tab:icc}
            \begin{tabular}{lllllllllllllllllllllllllllllllll}
          \toprule
          \bfseries  Nutrient &  \bfseries ICC(2,1) (95\% CI) & \bfseries  ICC(2,k) (95\% CI) \\
          \midrule
          Energy & 0.541 (0.48, 0.60) &0.983 (0.98, 0.99) \\
          Protein & 0.529 (0.47,0.59) & 0.983 (0.98,0.99)\\
          Total Fat &0.483 (0.43, 0.55) & 0.977 (0.97, 0.98)\\
          Trans Fat &  0.274 (0.23, 0.33) &0.950 (0.94, 0.96)\\
          Saturated Fat &  0.557 (0.50, 0.62) &0.984 (0.98, 0.99) \\
          Dietary Fibre &  0.359 (0.31, 0.42) &0.966 (0.96, 0.97)\\
          Sugars &  0.575 (0.52, 0.64) &0.985 (0.98, 0.99) \\
          Cholesterol  &  0.605 (0.55, 0.66) &0.987 (0.98, 0.99) \\
          Carbohydrate & 0.532 (0.47, 0.59) & 0.983 (0.98, 0.99) \\
          Calcium & 0.470 (0.41, 0.53) & 0.978 (0.97, 0.98)\\
          Copper  & 0.437 (0.38, 0.5) &0.975 (0.97, 0.98) \\
          Magnesium & 0.398 (0.34, 0.46) &0.971 (0.96, 0.98) \\
          Manganese &  0.287 (0.23, 0.34) &0.953 (0.94, 0.96)\\
          Phosphorus & 0.625 (0.57, 0.68) & 0.988 (0.99, 0.99)\\
          Potassium &  0.385 (0.33, 0.45) &0.969 (0.96, 0.98) \\
          Sodium &  0.381 (0.33, 0.44) & 0.969 (0.96, 0.98)\\
          Vitamin C &  0.339 (0.29, 0.40) &0.963 (0.95, 0.97)\\
          Zinc & 0.522 (0.47, 0.58) &  0.982 (0.98, 0.99) \\
          Iron &  0.379 (0.33, 0.44) & 0.968 (0.96, 0.98)\\
        \bottomrule
        \end{tabular}
      % }
    \end{table}

\section{Long-Term General Usability, User Satisfaction, and User Experience (RQ1) in Study II}
\label{appx:long_term}
As shown in \figurename{~\ref{fig:sus}}, SUS scores showed an upward trend across all groups, increasing from 76.09 in Week 0 to 85.00 in Week 4, indicating improved perceived usability over time for both exp. and inexp. participants. Exp. participants generally rated the system higher, which may indicate greater familiarity or comfort with such systems. Inexp. participants began with lower scores and improved to 82.50 (SD=6.45) by Week 4, showing a gradual adjustment to the system.

NPS scores also slightly increased, reflecting sustained user satisfaction across all groups (\figurename{~\ref{fig:nps}}). 
However, differences between exp. and inexp. participants were evident in the first two weeks of Study II, where the ratings of inexp. participants were notably lower than the exp. ones. The ratings of inexp. participants increased steadily across the four weeks, showing their increasing satisfaction.
In Week 1, users praised the system for its accurate diet identification (P6, exp.), ease of use (P3, inexp., P7, exp., P9, inexp., P10, exp., P13, exp., P18, exp., P20, exp.), and its potential to improve health (P1, exp., P8, inexp., P16, inexp., P18, exp.).
By Week 2, the focus shifted to its practicality and usefulness in promoting healthy eating (P6-P7, exp., P8-P9, inexp., P13, exp., P16, inexp., P18, exp.), with users appreciating its ability to guide their diet without requiring much manual effort (P3, inexp.).
In Week 3, users highlighted its effectiveness in supporting specific health goals, such as weight management (P8, inexp.), while also suggesting improvements to the user interface for better engagement (P14, inexp.).
By Week 4, users continued to express satisfaction with the system’s ease of use and effectiveness, with P8 (inexp.) mentioning increased awareness of regular eating habits and P6 (exp.) noting interest from friends, boosting confidence in recommending the system.

In terms of user experience (UEQ), attractiveness scores increased over time for exp. participants, peaking in Week 4 (\figurename{~\ref{fig:ueq-a}}), suggesting growing satisfaction with the system’s appeal. In contrast, inexp. participants' scores remained stable, indicating a more neutral reaction. 
Perspicuity scores remained stable for the exp. group and slightly decreased for the inexp. group (\figurename{~\ref{fig:ueq-p}}), potentially reflecting a learning curve as users explored advanced features. 
Efficiency scores fluctuated but ultimately improved by Week 4 (\figurename{~\ref{fig:ueq-e}}), likely due to the increasing familiarity to the system.

Dependability also fluctuated but remained high for both groups (\figurename{~\ref{fig:ueq-d}}), though exp. users consistently provided higher ratings, highlighting their trust in the system's performance. Stimulation peaked in Weeks 1 and 2, particularly for inexp. participants, but declined thereafter (\figurename{~\ref{fig:ueq-s}}), possibly as novelty wore off. Similarly, novelty remained relatively stable in the early weeks but slightly decreased by Week 4 (\figurename{~\ref{fig:ueq-n}}), suggesting that the initial excitement diminished as familiarity with the system grew.

Overall, exp. participants generally rated dimensions higher and showed more consistent improvement, while inexp. participants' scores often plateaued or fluctuated, highlighting the potential benefit of tailored onboarding to sustain engagement and usability for less experienced users.

\section{Long-Term Interaction Behaviors with \shortname in Study II}
\label{appx:behavior}
We analyzed long-term interaction behaviors with \shortname in terms of usage patterns, average engagement time, and key events (editing diet summaries and using the chatbot). Engagement sessions shorter than three seconds were excluded, as quick actions like viewing meal logs or prominent nutrient values typically take about three seconds. Sessions between 12:00 a.m. and 5:59 a.m. were also excluded, as this time slot was reserved for system maintenance and backup.

 \begin{figure*}[!t]
          \centering
          \subfloat[]{
            \centering
            \includegraphics[width=.49\textwidth]{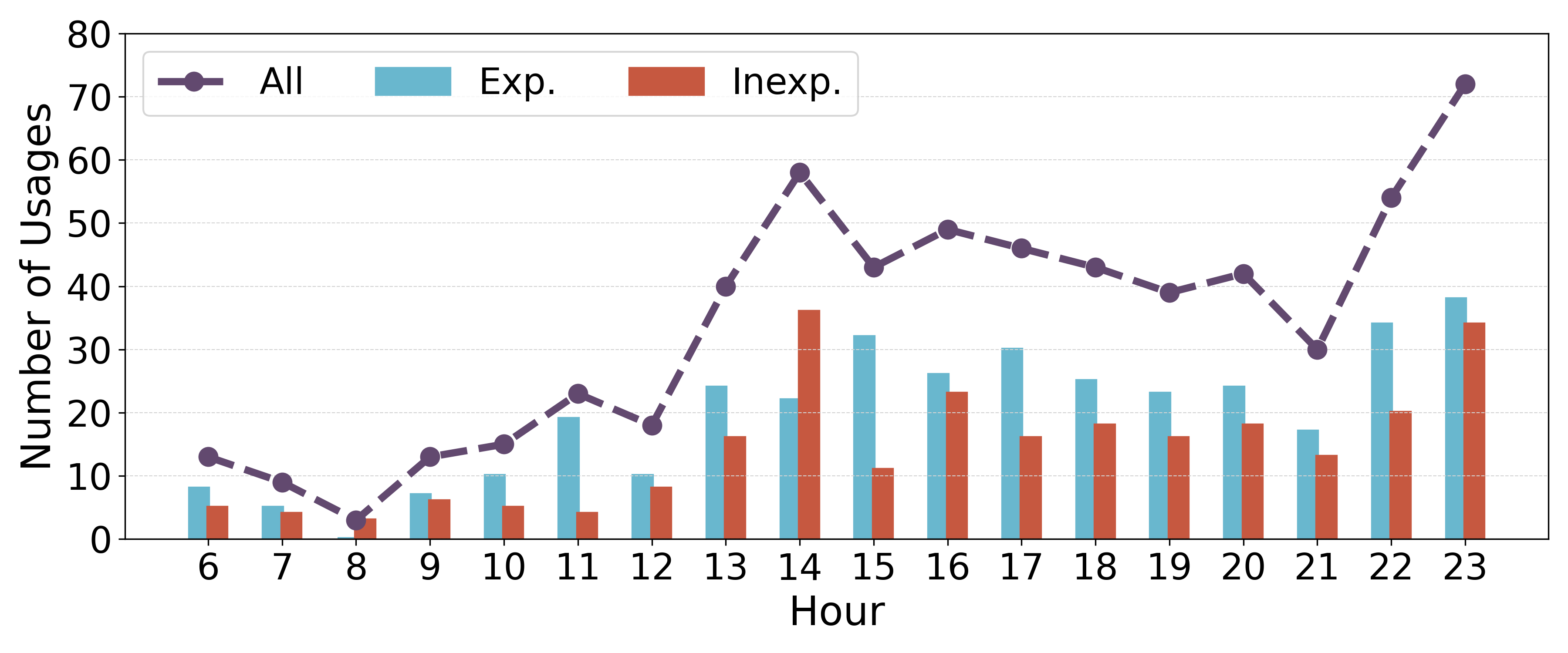}
            \label{fig:usage-hourly}
          }
          \subfloat[]{
            \centering
            \includegraphics[width=.24\textwidth]{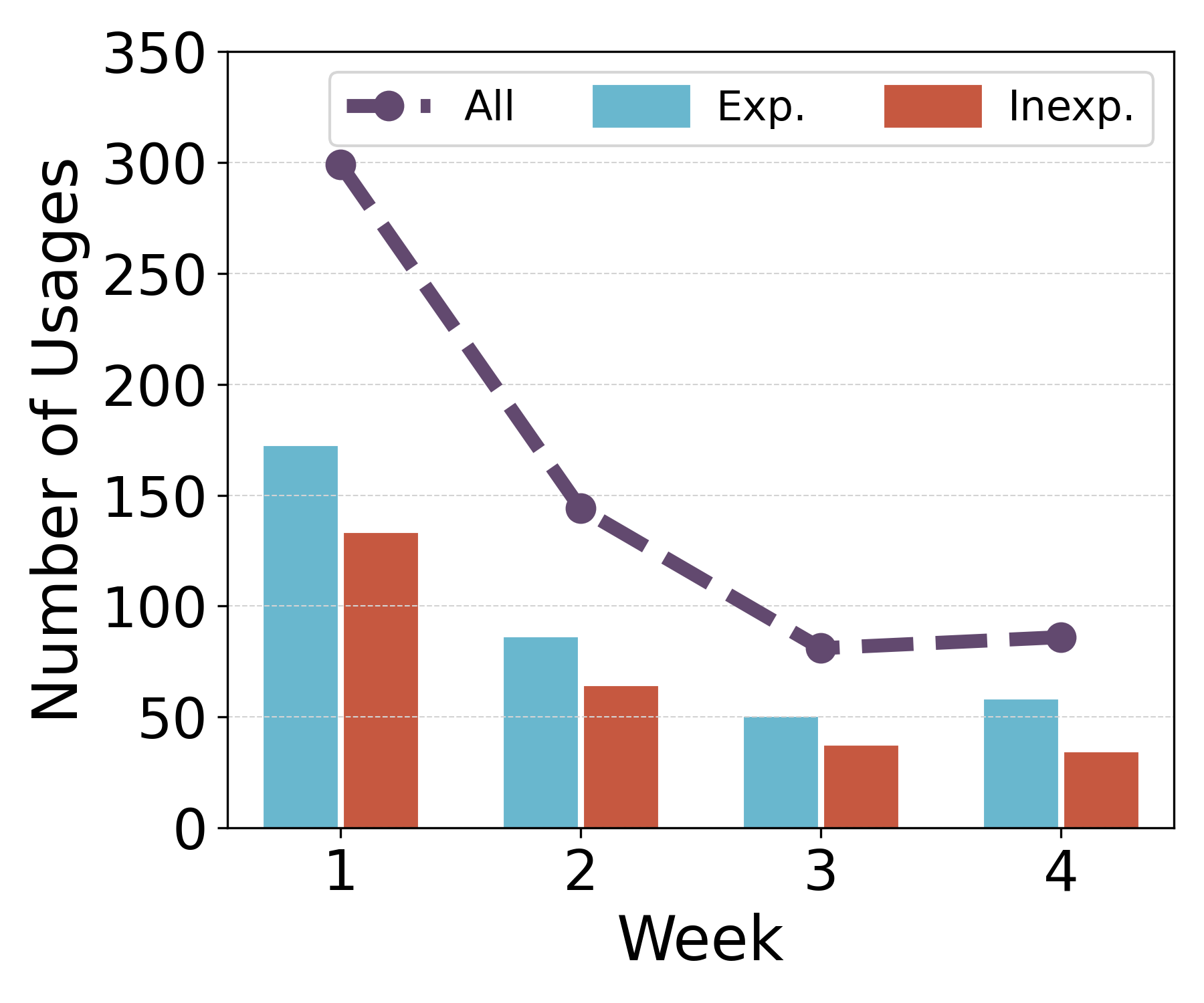}
            \label{fig:usage-weekly}
          }
          \subfloat[]{
            \centering
            \includegraphics[width=.24\textwidth]{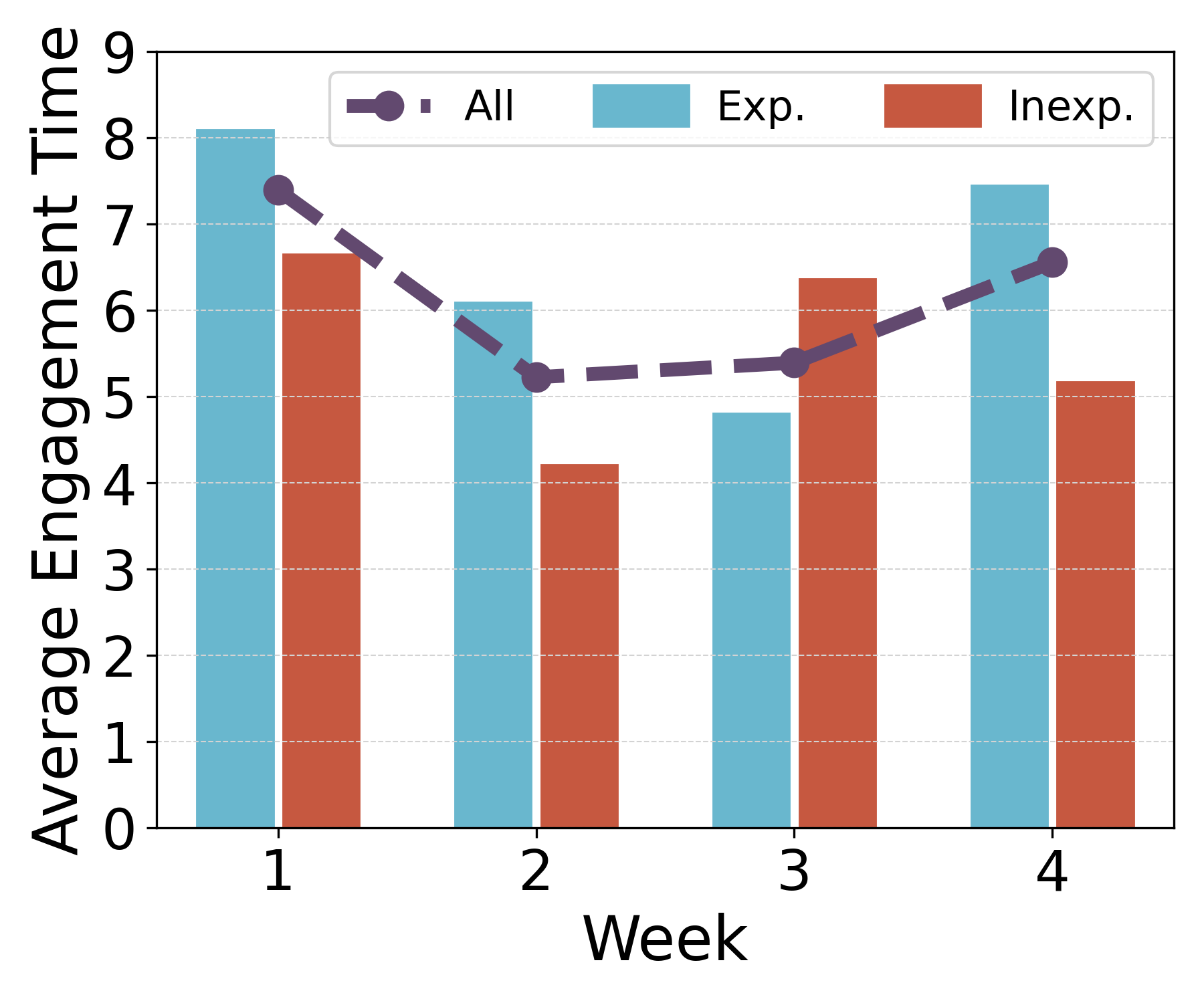}
            \label{fig:average-time}
          }\\
          \subfloat[]{
            \centering
            \includegraphics[width=.24\textwidth]{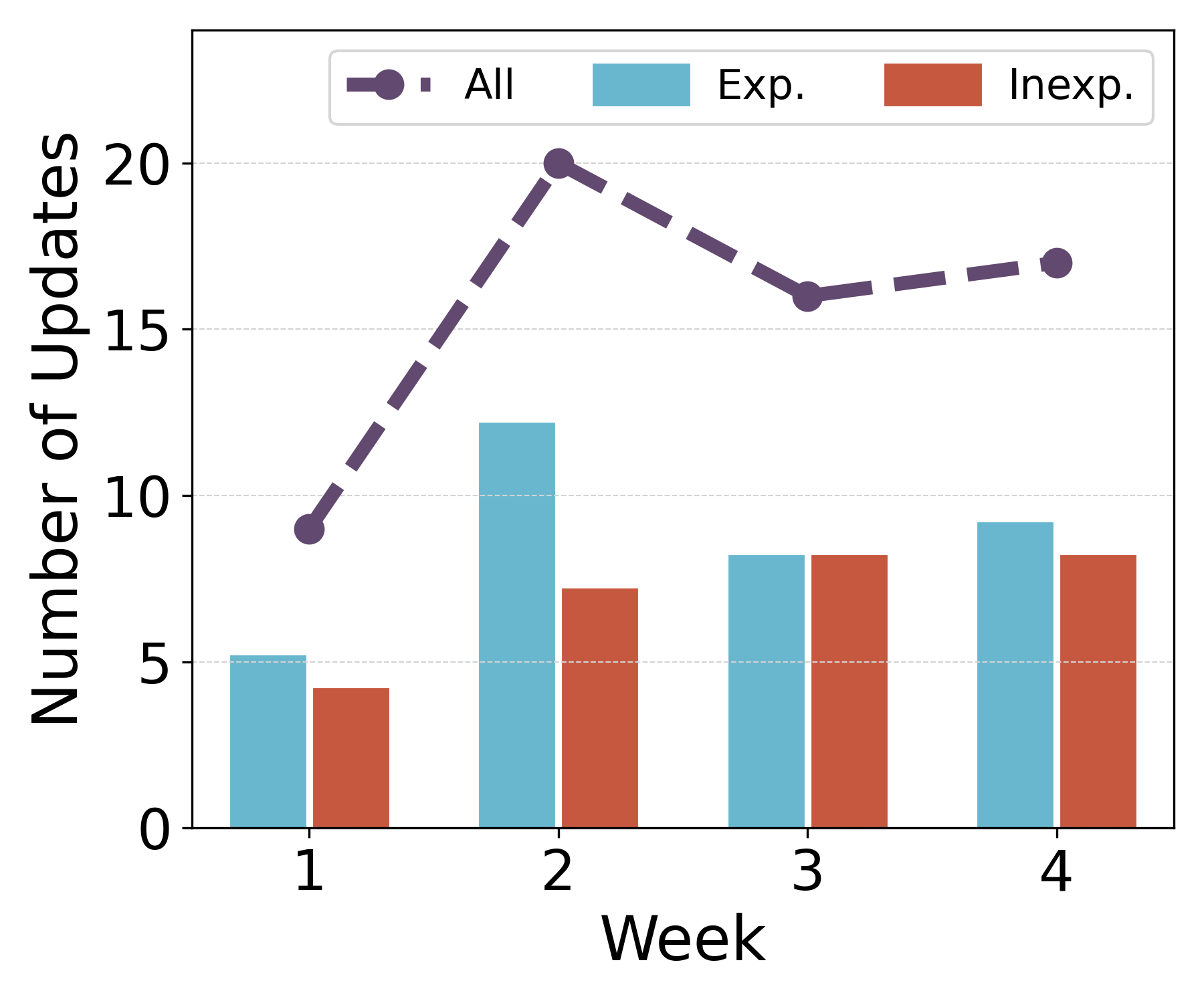}
            \label{fig:updates}
          }
          \subfloat[]{
            \centering
            \includegraphics[width=.24\textwidth]{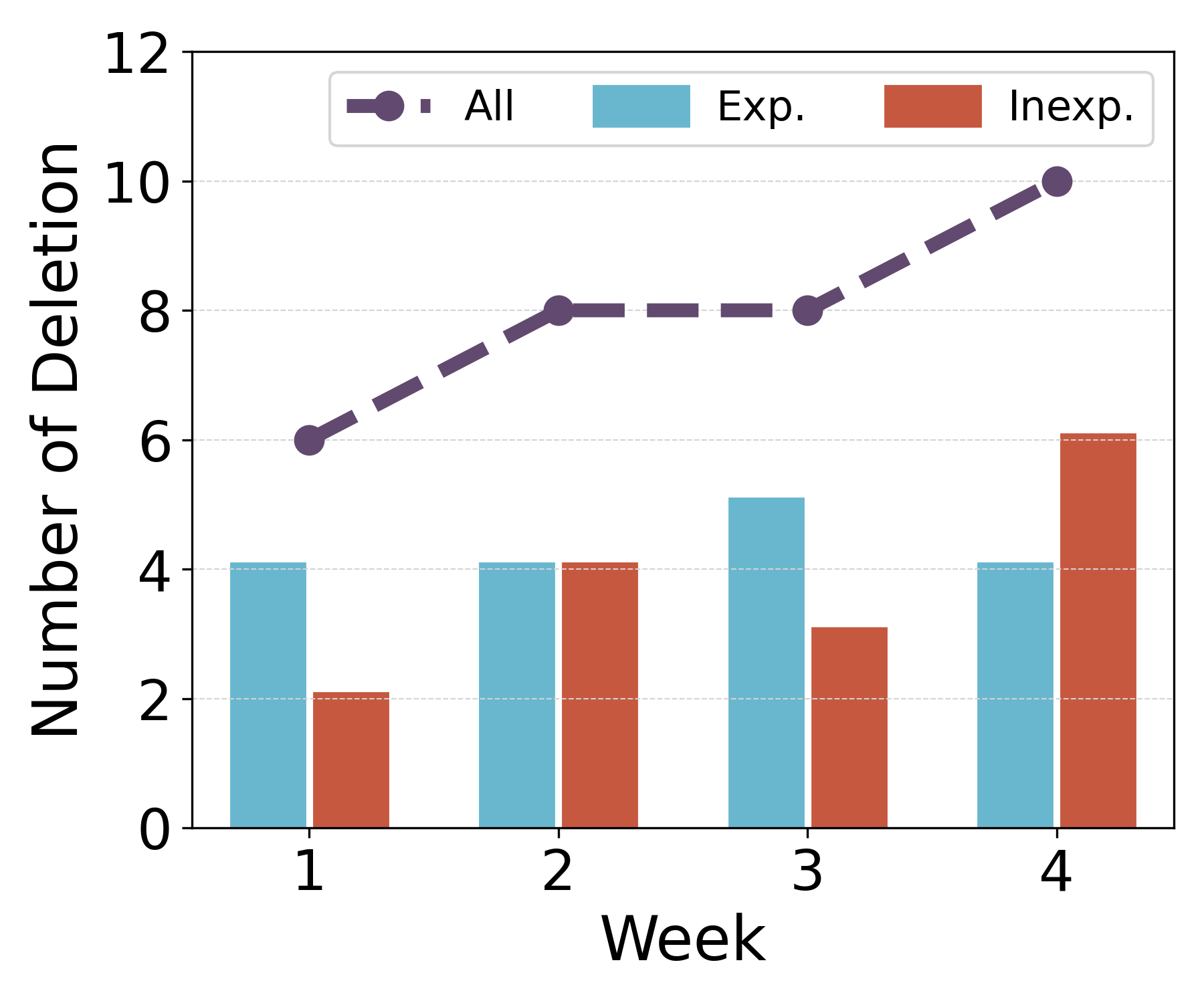}
            \label{fig:deletion}
          }
          \subfloat[]{
            \centering
            \includegraphics[width=.24\textwidth]{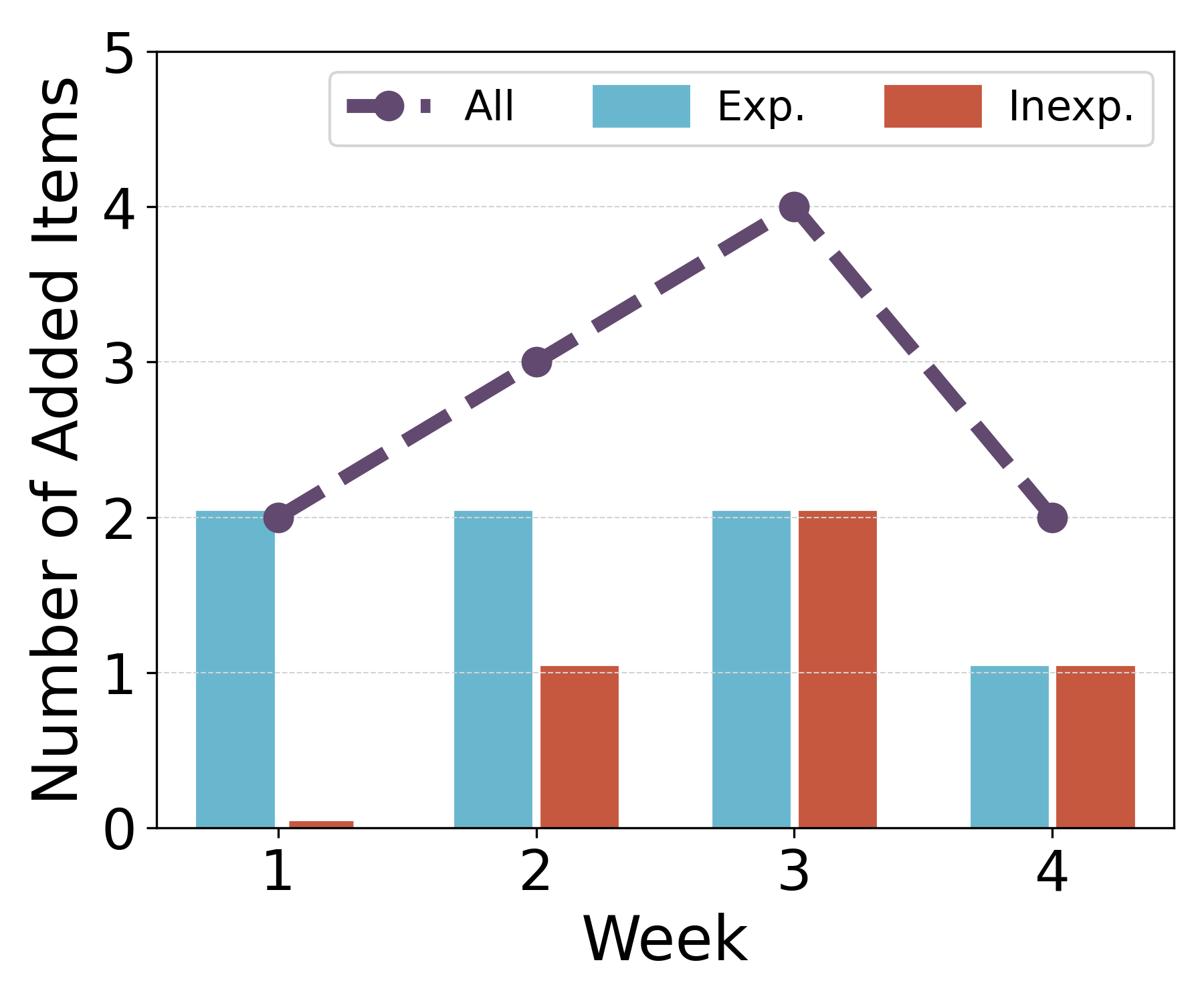}
            \label{fig:add}
          }
          \subfloat[]{
            \centering
            \includegraphics[width=.24\textwidth]{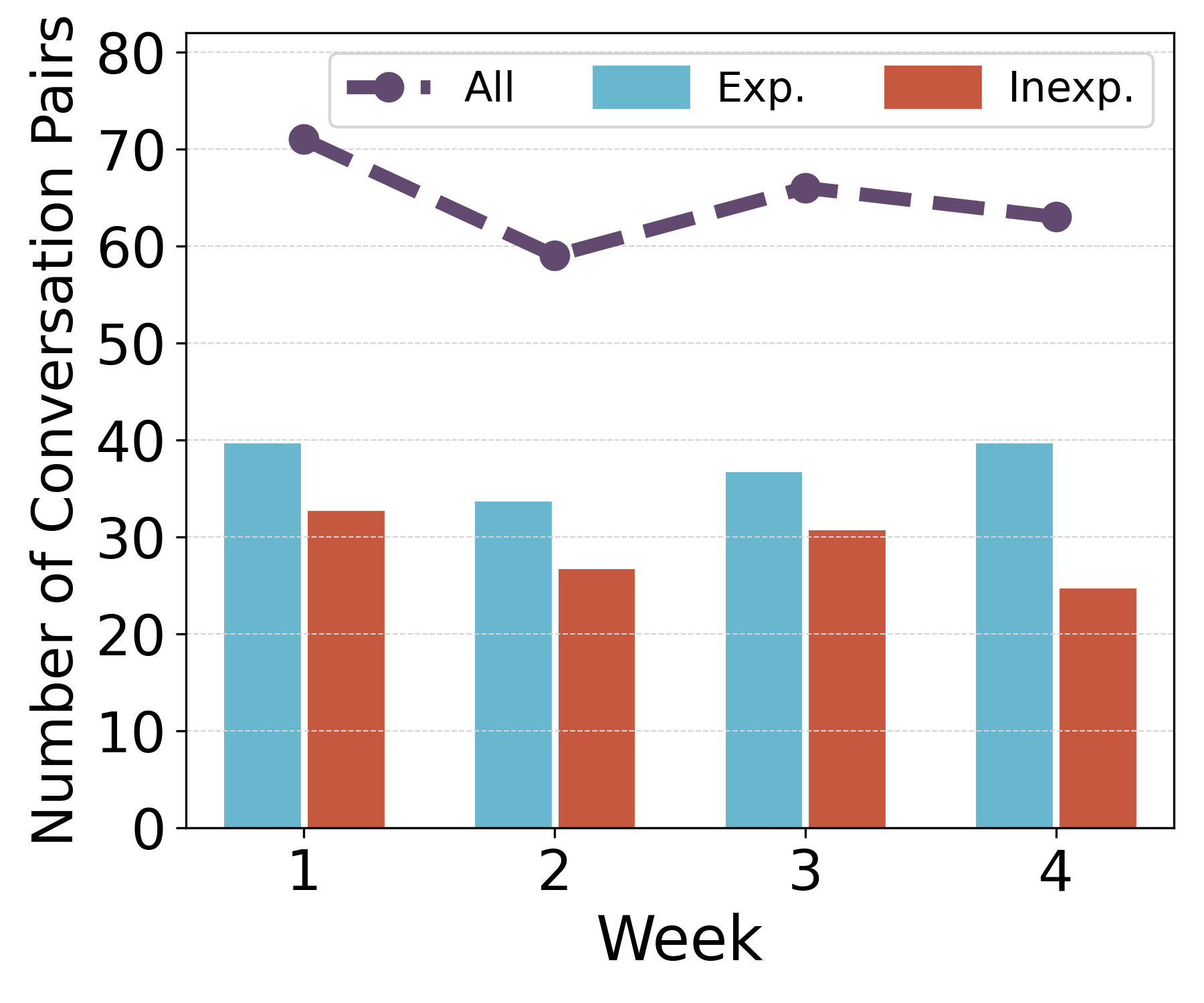}
            \label{fig:conversation}
          }
          \caption{Interaction behaviors with \shortname. (a) The hourly distribution of usage. (b) The weekly distribution of usage. (c) The weekly average engagement time. (d) The weekly number of updated items. (e) The weekly number of deleted items. (f) The weekly number of added items. (g) The weekly pairs of conversation.}
          \label{fig:long-term-behaviors}
           % \vspace{-0.5cm}
      \end{figure*}

\subsection{Hourly and Weekly System Usage Patterns.}
A total of 610 usage behaviors were logged, with participants engaging an average of 38.13 times ($SD=16.5$) over four weeks. Exp. participants showed slightly higher engagement ($M=39.33$) compared to inexp. participants ($M=36.57$).
P6 had the highest engagement (76 usages), while P3 had the lowest (12 usages).
The hourly usage distribution is shown in  \figurename{~\ref{fig:usage-hourly}}. 
The most frequent usage hours fell in the late evening (22 and 23) and early afternoon (14), and the least active hours were near early morning (6-9), potentially due to the early work or school commitments.
As shown in \figurename{~\ref{fig:usage-weekly}}, most interactions occurred in the first week (18.69 times per participant), potentially due to initial excitement about the system. Engagement dropped in the second week, averaging 9 times per participant. By the third and fourth weeks, usage stabilized, with average weekly engagement at 5.06 and 5.38 times per participant. This trend suggests that while initial novelty drove high interaction, both exp. and inexp. participants settled into a more routine usage pattern as they became familiar with the system.

\subsection{Average Engagement Time.}
The average engagement time for participants was 5.95 minutes per session ($SD=2.45$), with exp. participants ($M=7.02$) generally have longer engagement time than the inexp. participants ($M=5.78$).
Unlike Study I, where participants were instructed to explore all system features, participants in this stage primarily use the system only as needed, focusing on newly analyzed meal sessions and resulting in shorter engagement times. As a result, the overall engagement time was generally lower than in Study I, indicating that participants could quickly retrieve the necessary information with \shortname. As shown in \figurename{~\ref{fig:average-time}}, Week 1 had the longest engagement time, reflecting initial exploration, while Week 4 saw the second longest, possibly due to participants revisiting records and making final adjustments as the study concluded.

\subsection{Key Events.}
We focused on two key events: (1) updating diet summary tables and (2) conversations with the chatbot. During the study, 105 updates to diet identification were logged: 62 updates to existing items (34 exp. and 27 inexp.), 32 (17 exp. and 15 inexp.) deletions, and 11 additions (7 exp. and 4 inexp.).
As shown in \figurename{~\ref{fig:updates}} and \figurename{~\ref{fig:deletion}}, updates and deletions were lower in Week 1, increased in Week 2, and stabilized in subsequent weeks, while item additions peaked in Week 3 and declined in Week 4 (\figurename{~\ref{fig:add}}). These behaviors suggest users were actively managing their diet entries, improving the accuracy of diet logging and nutritional feedback. The relatively low number of updates reflects the system's reliable performance in diet identification and logging.
There were 259 chatbot conversations (147 exp. and 112 inexp.), distributed generally evenly across the four weeks (\figurename{~\ref{fig:conversation}}), indicating sustained user engagement from both exp. and inexp. participants. Conversations mainly involved meal suggestions, clarification on nutritional intake, and further dietary advice. This conversational feature enhanced the user experience by providing personalized feedback based on contextual information.

\end{document}